\documentclass[11pt,draftclsnofoot, onecolumn]{IEEEtran}

\usepackage{graphicx,cite,epsfig,amssymb,amsmath,url,stfloats,latexsym}
\usepackage{array}
\usepackage{arydshln}
\usepackage{amsfonts}
\usepackage{pgfplots}
\usepackage{algorithm}
\usepackage{algorithmic}
\usepackage{booktabs} 
\usepackage{epstopdf}
\usepackage{amsfonts,amsthm}
\usepackage{multirow}
\usepackage{mathrsfs}
\usepackage{xcolor}
\usepackage{amsfonts}
\usepackage{graphicx}
\usepackage{tabularx}
\usepackage{array}
\usepackage{arydshln}
\usepackage{multicol}
\usepackage{multirow} 
\usepackage{mathtools}
\usepackage[font=scriptsize]{subfig}
\usepackage[font=small]{caption}
\usepackage{textcomp}
\usepackage{stfloats}
\usepackage{booktabs}
\usepackage{verbatim}
\usepackage{graphicx}
\usepackage{cite}
\usepackage{url}
\usepackage{enumitem}

\usepackage{amssymb}
\usepackage{amsthm,extpfeil}

\definecolor{mygray}{gray}{.9}
\graphicspath{{figures/}}
\newcolumntype{C}[1]{>{\PreserveBackslash\centering}p{#1}}
\newcolumntype{R}[1]{>{\PreserveBackslash\raggedleft}p{#1}}
\newcolumntype{L}[1]{>{\PreserveBackslash\raggedright}p{#1}}

\newtheorem{theorem}{Theorem}

\usepackage[flushleft]{threeparttable}

\begin{document}

\title{\mbox{}\vspace{0.2cm}\\
\textsc{\huge Finite Field Multiple Access} 
\vspace{0.5cm}}

\vspace{0.5cm}
\author{\normalsize
Qi-yue~Yu, {\it IEEE Senior Member},
Jiang-xuan~Li,
and Shu~Lin, {\it IEEE Life Fellow} 

\thanks{Qi-yue~Yu (email: yuqiyue@hit.edu.cn) and Jiang-xuan~Li (email: 21S005095@stu.hit.edu.cn) are at Harbin Institute of Technology, China. Shu Lin (email: drshulin8@gmail.com) is at University of California, Davis.
}
\thanks{This paper was presented in part at \textit{IEEE Information Theory Workshop (ITW) 2024}, Nov. 2024.} 
}

\maketitle

\begin{abstract}
In the past several decades, various techniques have been developed and used for multiple-access (MA) communications. With the new applications for 6G, it is desirable to find new resources, physical or virtual, to confront the fast development of MA communication systems. For binary source transmission, this paper introduces the concept of element-pair (EP), and the Cartesian product of $J$ distinct EPs can form an EP code. EPs are treated as virtual resources in finite fields to distinguish users. 
This approach allows for the reordering of channel encoding and multiplexing modules, allowing superimposed signals to function as codewords decodable by a channel code, thereby effectively addressing the finite blocklength (FBL) challenge in multiuser transmissions.
We present methods for constructing symbol-wise EP codes with the unique sum-pattern mapping (USPM) property using finite fields.
Based on the orthogonal EP code constructed over GF($2^m$), we develop a time-division mode of finite-field multiple-access (FFMA) systems over a Gaussian multiple-access channel (GMAC), including both sparse-form and diagonal-form structures. Based on the diagonal-form (DF) structure, we introduce a specific configuration referred to as polarization-adjusted DF-FFMA, which achieves both power gain and coding gain across the entire blocklength. The proposed FFMA is then applied to network layer and forms network FFMA systems for pure digital networks. Simulation results demonstrate that, compared to popular complex-field MA systems, the proposed FFMA systems can offer superior error performance in a GMAC.
\end{abstract}


\newpage
\section{Introduction}
\IEEEPARstart{M}{ultiple} access (MA) is one of the most important techniques for wireless communications. During the past several decades, various MA techniques have been developed for mobile communications to support various users and services \cite{FAdachi1, YChen_2018}.

For the next generation of wireless communications, a range of exciting applications are being explored, including ultra-massive machine-type communications (um-MTC) for massive connectivity, the full digital world (e.g., digital twins), vehicle-to-everything (V2X), wireless data centers, immersive extended reality (XR), wireless brain-machine interfaces, emergency rescue communications, and more \cite{6G, 6G_white}.
To support these applications, there is a need to explore specialized MA techniques. In this paper, we focus on investigating MA techniques tailored to two specific applications: um-MTC and digital networks.
\begin{itemize}
  \item
   In um-MTC communications, the communication network is required to support a very large number of devices simultaneously. With increasing number of devices, the connection density of an um-MTC system is around $10^{6} \sim 10^{8}$ devices/km$^2$ \cite{6G, UMA_2022, UMA_PZ}. The required MA technique for this scenario should simultaneously support massive users (or devices) with short packet traffic and achieve an acceptable {per-user probability of error (PUPE)} \cite{MIT_2017}.
  \item
  For the digital network scenario, the digital twin is expected to map between the digital world and the physical world, which can help us predict and reflect the physical world \cite{6G}. However, to design MA systems for digital networks, few \textit{physical resources} can be used to distinct users. Evidently, the conventional MA techniques, e.g., FDMA (frequency division multiple access), SDMA (space division multiple access), are not suitable for a pure digital network.
\end{itemize}

Currently, sourced and unsourced random access (RA) techniques are popular for the um-MTC communications, since it can support massive users with low latency and power consumption 
\cite{SourcedRA_1, Yu_UDAS, SourcedRA_2, SourcedRA_3, SourcedRA_4}. 
The sourced RA is used to serve device-oriented scenarios, e.g., monitoring the status of sensors, enabling the base station (BS) to be aware of both the messages and users (or devices) identities \cite{SourcedRA_1}.
For a sourced RA system, each user is assigned a unique signature to identify the active user(s) at the BS. Typically, we can use different resources, sequences and/or codebooks to identity different users \cite{Yu_UDAS}. The classical sourced RA system utilizes time slots to distinguish users, 
operating in a {slotted ALOHA} scheme, which simplifies both the active user detection (AUD) and multiuser detection (MUD) algorithms \cite{ALOHA_1}.

In the case of unsourced RA (URA), the BS is only concerned with transmitting messages and is not interested in the identities of the users (or devices). 
In \cite{MIT_2017}, the author presented a random-access code for a $J$-user Gaussian multiple-access channel (GMAC). In usual, an URA system is assumed that the total number of users is a large number, and the active duration is set to be the length of a frame, in which the active users may generate collision. Many works have been done on the theoretical bound analyses \cite{Capacity_GMC_2017, Capacity_GMC_2020, Capacity_GMC_2021, Capacity_GMC_Yury}. 
Besides the asymptotic limit analyses, the key issue of realizing URA systems is to design suitable MA codes. 
The popular MA codes are 
\textit{Bose–Chaudhuri–Hocquenghem (BCH) MA code} \cite{MIT_2017_2, BCH_MA}, 
\textit{coded compressed sensing (CCS) MA code} \cite{CCS_1, CCS_2}, 
\textit{polar MA code} \cite{Polar_1, Polar_2, Polar_3, Polar_4}, 
\textit{interleave division multiple-access (IDMA) MA code} \cite{IDMA, IDMA_1}, 
\textit{sparse vector code} \cite{SVC_1, SVC_2} and etc.

In fact, for a massive random access scenario, the reliable performance of sourced/unsourced RA is limited by the blocklength \cite{Capacity_GMC_2017, Capacity_GMC_2021, Capacity_GMC_Yury}.
Let the blocklength, or the number of degrees of freedom (DoF), be a constant denoted by $N$, which is also called the \textit{finite blocklength (FBL) constraint}.
Suppose there are totally $J$ users. 
We take the classical slotted ALOHA scheme as an example.
To separate users without ambiguity, the number of time slots is set to be equal to the number of users $J$. Thus, the blocklength of each time slot is equal to $N/J$. 
According to \cite{Capacity_GMC_2021}, we know that the maximal rate of each time slot is equal to
\begin{equation*}
C(P) - \sqrt{\frac{V(P)}{(N/J)}} Q^{-1}(\epsilon) + \frac{1}{2} \frac{\log (N/J)}{(N/J)} + {\mathcal O}(\frac{1}{(N/J)})
\end{equation*}
where $C(P)$ is the channel capacity, $P$ is the maximal power constraint and the noise variance is set to be $1$. $Q(\cdot)$ is the $Q$-function, and $\epsilon$ is the average error probability.
$V(P) = \frac{P(P+2)}{2(1+P)^2}$ is the dispersion of the Gaussian channel.
Obviously, the maximal rate of each time slot decreases with the increased number of users $J$, especially $J$ is a large value.
To support massive users with short packet traffic, the payload of each user is always a short packet, i.e., $K = 10 \sim 100$ bits. 
Under the FBL constraint, designing MA codes to approach the asymptotic limit becomes extremely challenging.

For the digital networks scenario, it is difficult to utilize the current physical resources, e.g., frequency, space, and etc., to identify individual users. Thus far, there has been no specific MA technique developed for pure digital networks. It is appealing to find \textit{virtual resources} to distinct users and realize MA.

\subsection{Complex-field MA and Finite-field MA}

In this paper, we propose an MA technique to provide some solutions to what needed for supporting massive message transmissions from multiusers, and solving the FBL challenge in multiuser transmission.
In addition, the proposed MA can be implemented in pure digital networks, providing a feasible solution.
The proposed MA technique is devised based on \textit{element-pair (EP) coding} which provides virtual resources for MA communications. 
Since EP codes are constructed based on finite fields (FFs), we refer the proposed MA technique based on EP-coding as \textit{finite-field MA (FFMA)} technique (or coding). An MA communication system using FFMA-coding is referred to as an FFMA system for simplicity.

\begin{figure}[t]
  \centering
  \includegraphics[width=0.3\textwidth]{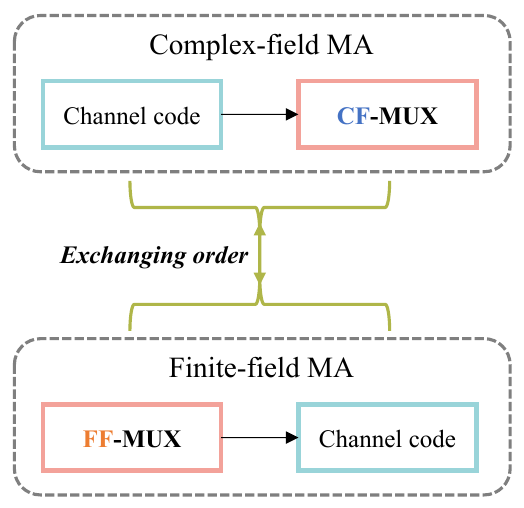} 
  \caption{Complex-field MA and finite-field MA, where CF-MUX and FF-MUX stand for complex-field multiplex module and finite-field multiplex module, respectively.} 
  \label{f.CFMA}
  \vspace{-0.15in}
\end{figure}

We now compare classical MA techniques with our proposed FFMA systems, as shown in Fig. \ref{f.CFMA}.
The current MA techniques separate users based on various physical resources, such as time, frequency, space, power, and more. We term such an MA system as \textit{complex-field MA (CFMA)}, where physical signals are typically represented as complex numbers.
In a CFMA system, each user is assigned different physical resources, enabling the users to be multiplexed. 
For simplicity, we refer to this multiplexing process as a \textit{complex-field multiplex module (CF-MUX)}.

\begin{itemize}
  \item 
  In a CFMA system, the transmitter first performs channel coding, followed by a CF-MUX. Consequently, at the receiving end, multiuser detection (MUD) is required before channel decoding can take place. If MUD and channel decoding are performed separately, we may not achieve the channel decoding gain needed to enhance MUD performance. An attractive MA technique is IDMA \cite{IDMA}, which utilizes a turbo structure for message exchange between MUD and channel decoding. However, similar to the slotted ALOHA case, each user's channel codeword blocklength remains $N/J$, which still presents a multiuser FBL challenge.
  \item 
  In an FFMA system, the transmitter can first apply the proposed element-pair (EP) coding for MUX, followed by channel coding. Consequently, the channel codeword blocklength in our proposed FFMA system is $N$, fully utilizing the entire \textit{degree of freedom (DoF)}. As a result, channel decoding can occur before multiuser separation at the receiving end, enabling simultaneous attainment of coding gain from the multiuser transmission process. 
  Notably, channel coding might be unnecessary for the proposed FFMA systems, as EP coding inherently provides error correction capabilities, a topic that will be explored further.
\end{itemize}

\subsection{Three challenges of FFMA}
  As is commonly known, one of the classical problems in mathematical theory involves the exchange of the order of functions. 
  For example, given two functions $f(t)$ and $g(t)$, the question is about the relationship between $f(g(t))$ and $g(f(t))$, i.e., whether $f(g(t)) = g(f(t))$, or $f(g(t)) \neq g(f(t))$, or possibly $f(g(t)) \le g(f(t))$, or $f(g(t)) \ge g(f(t))$.
  Similarly, in communication systems, a key challenge lies in the order of fundamental modules, such as channel coding and MUX. In our proposed FFMA system, we aim to exchange the order of the channel coding and MUX modules. However, this introduces three main challenges, which we will analyze one by one.

\begin{figure}[t]
  \centering
  \includegraphics[width=0.9\textwidth]{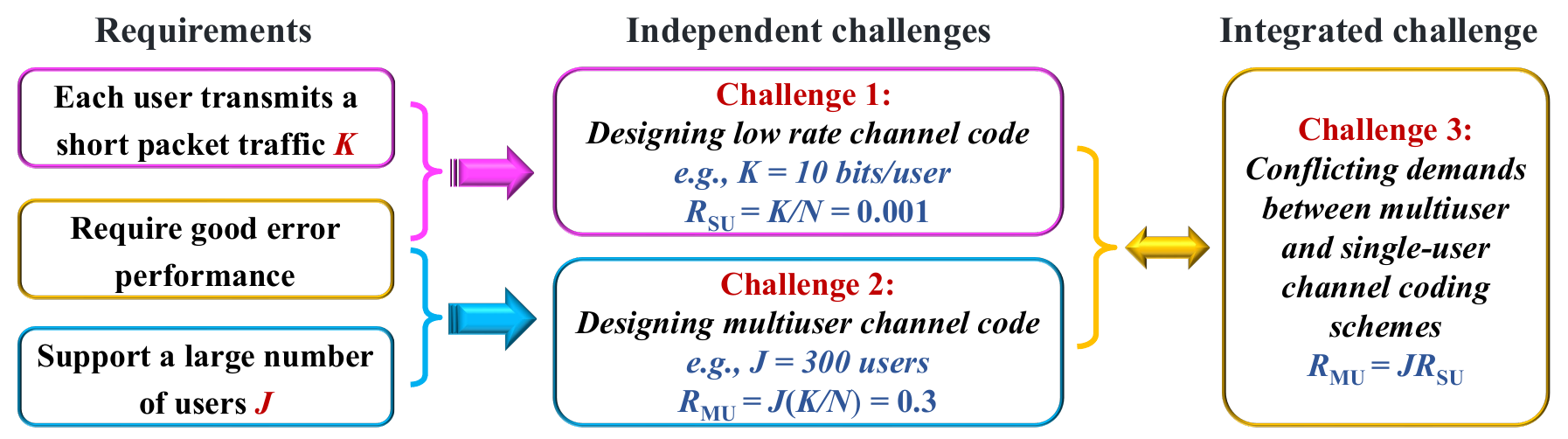} 
  \caption{Three challenges of FFMA tehcnique, which are designing low-rate channel codes, designing multiuser channel codes, and
  conflicting demands between multiuser and single-user channel codes. 
  In this example, we set the number of DoF to $N = 10000$, the number of bits to $K = 10$, and the number of users to $J = 300$.
  } 
  \label{f.Problems}
  \vspace{-0.25in}
\end{figure}

\subsubsection{Designing low-rate channel codes}

In coding theory, designing low-rate channel codes is a challenging problem. In the um-MTC scenario, each user transmits a small packet of data, for example, $K = 10$ bits/user. The total number of degrees of freedom (DoF), denoted by $N$, is typically large, e.g., $N = 10000$. 
This means that the channel code must be designed with a rate of $R_{\rm SU} = K/N = 10/10000=0.001$, where the subscript ``SU'' refers to ``single-user''.
Currently, designing such low-rate channel codes is still an open problem in the field.

\subsubsection{Designing multiuser channel codes}
There are fewer papers focusing on uplink \textit{multiuser channel codes}, as different users are located at different positions, and their information sequences are independent. This results in a lack of cooperation among users, preventing them from obtaining coding gain. Although there have been works on MA codes, as mentioned earlier, these studies primarily focus on multiuser detection, rather than on achieving coding gain through multiuser channel codes.

The development of multiuser channel codes is an attractive research topic, as it has the potential to provide both coding gain and user separation. Clearly, the design of multiuser channel codes remains a challenging research problem.

\subsubsection{Conflicting demands between multiuser and single-user channel codes}

As mentioned earlier, the total number of DoF is typically large. According to the FBL condition, to achieve a higher coding gain, it is desirable to utilize all the DoF in constructing a multiuser channel code. For instance, LDPC codes can be employed to construct such long multiuser channel codes. However, when designing LDPC codes, both the blocklength and rate must be carefully considered. A challenge arises, as there is a conflict in designing LDPC codes that effectively support multiuser transmission.

For example, consider a binary $(N, JK)$ LDPC code ${\mathcal C}_{gc}$ designed for error control in multiuser transmission. 
When $N = 10000$, $K = 10$ and $J = 300$, we observe the following data rates for single-user (SU) and multiuser (MU) cases:
\begin{itemize}
  \item
  Single-user rate: $R_{\rm SU} = \frac{K}{N} = \frac{1}{1000}$,
  corresponding to a low-rate $0.001$ LDPC code.
  \item
  Multiuser rate: $R_{\rm MU} = \frac{JK}{N} = \frac{3000}{10000}$,
  corresponding to a $0.3$ LDPC code.
\end{itemize}
Evidently, the multiuser rate is significantly higher than the single-user rate. 
According to Shannon's theory, a low-rate channel code generally provides better BER performance compared to a high data rate code. Therefore, the required $E_b/N_0$ (energy-per-bit to noise power spectral density ratio) for the single-user case should be lower than that for the multiuser scenario.

On the other hand, if each user has the same data rate $R_{\text{SU}}$, the multiuser rate $R_{\text{MU}}$ can be expressed as $R_{\text{MU}} = J \cdot R_{\text{SU}}$. 
For a given channel code, to achieve the same BER performance, the required $E_b / N_0$ in dB units for the multiuser (MU) and single-user (SU) cases can be expressed as follows:
\begin{equation} \label{e.EbN0}
  \begin{aligned}
    \left(\frac{E_b}{N_0}\right)_{\rm SU,dB} = 
    \left(\frac{E_b}{N_0}\right)_{\rm MU,dB} + 10 \lg (J)
  \end{aligned}.
\end{equation}
which indicates that the required $E_b / N_0$ for the single-user case is greater than that for the multiuser case. 
Consequently, as the number of incoming users increases, the required $E_b/N_0$ correspondingly rises.
This highlights a contradiction regarding the required $E_b / N_0$ in the single-user scenario.

To address the aforementioned issue, this paper integrates ideas from signal processing into channel coding and decoding. 
Specifically, an LDPC code can be conceptualized as a cascade of \textit{repetition code} and \textit{single parity-check (SPC) code}. 
Repetition codes, in turn, can be seen as \textit{time diversity}, which can be achieved by increasing transmit power directly.
Consequently, we incorporate power allocation into our coding and decoding framework to tackle the identified problem. Power allocation technique can adjust channel capacity, effectively implementing \textit{polarization adjusted} \cite{PAC_1, PAC_2, PAC_3}.
We propose a \textit{polarization-adjusted FFMA (PA-FFMA)} for multiuser transmission. 
When the number of users is one, i.e., $J = 1$, the PA-FFMA simplifies to a \textit{polarization-adjusted LDPC (PA-LDPC) code}. 
Importantly, the proposed PA-FFMA can be viewed as a type of multiuser channel code, obtaining both power and coding gains from the entire DoF. Hence, PA-FFMA can provide well-behaved BER than that of the current CFMA systems.

\vspace{-0.1in}
\subsection{Structure}
The work presented in this paper is organized into five parts.
In the first part, we define the concept of EP-coding and present methods for constructing \textit{symbol-wise} codes based on finite fields, including both \textit{prime fields} and \textit{extension fields} of prime fields.
The second part details the encoding of an EP code and introduces the concepts of \textit{finite-field multiplex module (FF-MUX)} and \textit{multiuser code}, which are generally used in conjunction with an EP encoder.
In the third part, we introduce the FFMA technique based on the \textit{orthogonal uniquely decodable EP (UD-EP)} code ${\Psi}_{\rm o, B}$ over GF($2^m$) for MA communication systems, which represents a type of \textit{multiuser channel code}. We describe both the sparse-form (SF) and diagonal-form (DF) FFMA structures, along with their applications in source RA systems.
The fourth part presents a \textit{polarization-adjusted FFMA (PA-FFMA)} system based on the diagonal-form structure. We define \textit{irregular} and \textit{regular} PA-FFMA and introduce the \textit{bifurcated minimum distance (BMD)} detection algorithm for the PA-FFMA systems.
In the fifth part, we apply the proposed FFMA to the network layer, forming \textit{network FFMA} systems suitable for pure digital networks.

The rest of this paper is organized as follows. 
Section II introduces the {concept} of EP-coding and gives constructions of symbol-wise EP codes over finite fields. 
Bounds on the number of UD-EP codes that can be constructed for a given finite field are derived. 
Section III presents encoding of EP codes. Section IV presents an FFMA system over GF($2^m$) by using an orthogonal UD-EP code $\Psi_{\rm o, B}$ for a massive MA scenario. 
Section V presents PA-FFMA systems based on the diagonal-form structure.
In Section VI, a network FFMA system over GF($p^m$) is presented for a pure digital network scenario. In Section VII, we compare the proposed FFMA systems with several classical CFMA techniques. 
Section VIII concludes the paper with some remarks. 

In this paper, the symbol $\mathbb B = \{0, 1\}$ and $\mathbb {C}$ are used to denote the binary-field and complex-field. 
The symbols $\lceil x \rceil$ and $\lfloor x \rfloor$ denote the smallest integer that is equal to or larger than $x$ and the largest integer that is equal to or smaller than $x$, respectively.
The notation $(a)_q$ stands for modulo-$q$, and/or an element in GF($q$).


\section{EP Codes over Finite Fields}

Suppose GF($q$) is a finite-field with $q$ elements, where $q$ is a prime number or a power of the prime number. Let $\alpha$ denote a primitive element of GF($q$), and the powers of $\alpha$, i.e., $\alpha^{-\infty} = 0, \alpha^0=1, \alpha, \alpha^2, \ldots, \alpha^{(q - 2)}$, give all the $q$ elements of GF($q$). 
For a binary source transmission system, a transmit bit is either $(0)_2$ or $(1)_2$.
Hence, we can use two different elements in a finite-field GF($q$) to express the bit information, 
i.e., $(0)_2 \mapsto \alpha^{l_{j,0}}$ and $(1)_2 \mapsto \alpha^{l_{j,1}}$, 
where $\alpha^{l_{j,0}}, \alpha^{l_{j,1}} \in$ GF($q$) and $l_{j,0} \neq l_{j,1}$. We define the selected two elements $(\alpha^{l_{j,0}}, \alpha^{l_{j,1}})$ as an \textit{element pair (EP)}.

In this section, we construct \textit{additive inverse element pair (AIEP) codes} based on finite fields. The first part of this section presents a method for constructing AIEP codes based on the \textit{prime fields}, and the second part presents a class of \textit{orthogonal} EP codes which are constructed based on the \textit{extension fields} of a prime field. 
Note that, $C_j$ and $\Psi$ stand for an EP and an EP code, respectively.
The subscripts and superscripts of $C_j$ and $\Psi$ are used to describe the structural property.

\subsection{AIEP Codes Based on Prime Fields}

Let GF($p$) = $\{0, 1, \ldots, p-1\}$ be a prime-field with $p > 2$.  
Partition the $p - 1$ nonzero elements GF($p$) into $(p - 1)/2$ 
\textit{mutually disjoint element-pairs (EPs)}, 
each EP consisting of a nonzero element $j$ in $\text{GF}(p)\backslash 0$ and 
its additive inverse $p - j$ (or simply $- j$). 
We call each such pair $(j, p - j)$ an \textit{additive inverse EP (AIEP)}. 
The partition of $\text{GF}(p)\backslash \{0\}$ into AIEPs, denoted by $\mathcal P$, is referred to as 
\textit{AIEP-partition}.
When $p = 2$, there is only one EP $C_{\rm B} =(0, 1)$ referred as the base EP, and the subscript ``B'' stands for binary (or $p=2$).

Let $C_1, C_2, \ldots, C_{J}$ be $J$ AIEPs in $\mathcal P$, 
where $J$ is a positive integer less than or equal to $(p - 1)/2$, i.e., $1 \le J \le (p - 1)/2$.
For $1 \le j \le J$, let the AIEP be $(j, p - j)$, i.e., $C_j = (j, p - j)$.
Denote $C_j$'s \textit{reverse order AIEP (R-AIEP)} by $C_{j,{\rm R}} = C_{p-j} = (p - j, j)$, where the superscript ``R'' of $C_{j,{\rm R}}$ stands for ``reverse order''.

The $J$ AIEPs $C_1, C_2, \ldots, C_J$ form a partition of a subset of $2J$ elements in GF$(p) \backslash \{0\}$, a sub-partition of $\mathcal P$. 
Let $\Psi_{\rm s}$ denote the set $\{C_1, C_2, \ldots, C_J\}$, i.e., ${\Psi}_{\rm s} = \{C_1, C_2, \ldots, C_J \}$, where the superscript ``s'' stands for ``symbol-wise''.
The $J$ R-AIEPs $C_{1,{\rm R}}, C_{2,{\rm R}}, \ldots, C_{J,{\rm R}}$ of 
$C_1, C_2, \ldots, C_J$ form a \textit{reverse order set} of $\Psi_{\rm s}$, 
denoted by ${\Psi}_{\rm s,R} = \{C_{1,{\rm R}}, C_{2,{\rm R}}, \ldots, C_{J,{\rm R}} \}$ or ${\Psi}_{\rm s, R} = \{C_{p-1}, C_{p-2}, \ldots, C_{p-J}\}$.

Let $(u_1, u_2, \ldots, u_j, \ldots, u_J)$  be a $J$-tuple over GF($p$) in which the $j$-th component $u_j$ is an element from $C_j$, where $1 \le j \le J$. The $J$-tuple $(u_1, u_2, \ldots, u_J)$ is an element in the \textit{Cartesian product} $C_1 \times C_2 \times \ldots \times C_J$ of the AIEPs in $\Psi_{\rm s}$. 
We view each $J$-tuple ${\bf u} = (u_1, u_2, \ldots, u_j, \ldots, u_J)$ in
$C_1 \times C_2 \times \ldots \times C_J$ as a $J$-user AIEP codeword, i.e.
\begin{equation}
  {\Psi}_{\rm s} \triangleq C_1 \times C_2 \times \ldots \times C_J,
\end{equation}
which forms a $J$-user AIEP code over GF($p$) with $2^J$ codewords.
It is noted that an AIEP can be treated as a set with two elements, which allows us to utilize the Cartesian product to define the AIEP code.
In this paper, ${\Psi}$ stands for an EP set and also an EP code.

The modulo-$p$ sum $w = \oplus_{j=1}^{J} u_j$ of the $J$ components in $(u_1, u_2, \ldots, u_{J})$
is called as the \textit{finite-field sum-pattern (FFSP)} of the $J$-tuple $(u_1, u_2, \ldots, u_{J})$, which is an element in GF($p$). 
The $J$-tuple $(p-u_1, p-u_2, \ldots, p-u_{J})$ is also an element in $C_1 \times C_2 \times \ldots \times C_{J}$. 
The FFSP of $(p-u_1, p-u_2, \ldots, p-u_{J})$ is $p - w = p -  \oplus_{j=1}^{J} u_j$. 
If the FFSP of $(u_1, u_2, \ldots, u_{J})$ is the zero element $0$ in GF($p$), the FFSP of $(p-u_1, p-u_2, \ldots, p-u_{J})$ is also the zero element $0$ in GF($p$), i.e., $w = 0$ and $p - w = 0$ (modulo $p$).

Let $(u_1, u_2, \ldots, u_{J})$ and $(u_1', u_2', \ldots, u_{J}')$ be \textit{any two} $J$-tuples in $\Psi = C_1 \times C_2 \times \ldots \times C_{J}$. 
If $\oplus_{j=1}^{J} u_j \neq \oplus_{j=1}^{J} u_j'$, 
then an FFSP $w$ uniquely specifies a $J$-tuple in $C_1 \times C_2 \times \ldots \times C_{J}$. 
That is to say that the mapping 
\begin{equation} \label{e.UDmap}
(u_1, u_2, \ldots, u_{J}) \Longleftrightarrow w = \bigoplus_{j=1}^{J} u_j,
\end{equation} 
is a one-to-one mapping. 
In this case, given the FFSP $w = \bigoplus_{j=1}^{J} u_j$, 
we can uniquely recover the $J$-tuple $(u_1, u_2, \ldots, u_{J})$ \textit{without ambiguity}. 
We say that the Cartesian product $\Psi_{\rm s} = C_1 \times C_2 \times \ldots \times C_{J}$ has a \textit{unique sum-pattern mapping (USPM) structural property}.

If we view each $J$-tuple $(u_1, u_2, \ldots, u_{J})$ in $C_1 \times C_2 \times \ldots \times C_{J}$ with the USPM structural property as a \textit{$J$-user codeword}, then we call ${\Psi}_{\rm s} = C_1 \times C_2 \times \ldots \times C_{J}$ a $J$-user \textit{uniquely decodable AIEP (UD-AIEP) code} over GF($p$), simply a $J$-user UD-AIEP code. 
It means that an UD-AIEP code is a special case of an AIEP code. 
In other words, if an AIEP code possesses the USPM structural property, the AIEP code is an UD-AIEP code. 
When this code is used for an MA communication system with $J$ users, the $j$-th component $u_j$ in a codeword $(u_1, u_2, \ldots, u_{J})$ is the symbol to be transmitted by the $j$-th user for $1 \le j \le J$. 
Note that, for a $J$-user UD-AIEP code ${\Psi}_{\rm s}$, we can replace any AIEP $C_j$ in ${\Psi}_{\rm s}$ by its R-AIEP $C_{j,{\rm R}}$, and still obtain a $J$-user UD-AIEP code,
i.e., ${\Psi}_{\rm s} = C_1 \times C_2 \times \ldots \times C_{j,{\rm R}} \times \ldots \times C_J$.

\begin{figure}[t]
  \centering
  \includegraphics[width=0.7\textwidth]{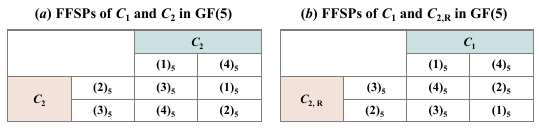} 
  \caption{A diagram in table-form of Example 1. 
  (a) The FFSPs of $C_1 = (1, 4)$ and $C_2 = (2, 3)$ in GF($5$); 
  (b) the FFSPs of $C_1 = (1, 4)$ and $C_{2,{\rm R}} = (3, 2)$ in GF($5$).
  } \label{Fig_GF5}
  \vspace{-0.2in}
\end{figure}

\textbf{Example 1:} 
Consider the prime field GF($5$). 
Using this prime field, two AIEPs $C_1 = (1, 4)$ and $C_2 = (2, 3)$ can be constructed. The R-AIEPs of $C_1$ and $C_2$ are $C_{1,{\rm R}} = (4, 1)$ and $C_{2,{\rm R}} = (3, 2)$, respectively.
The Cartesian product $C_1 \times C_2$ of $C_1$ and $C_2$ satisfies Eq. (\ref{e.UDmap}) as shown in Fig. \ref{Fig_GF5} (a). Hence, ${\Psi}_{\rm s} = C_1 \times C_2$ is a 2-user UD-AIEP code over GF($5$).

When we replace $C_2$ by $C_{2,{\rm R}}$, the Cartesian product 
$C_1 \times C_{2,{\rm R}}$ of $C_1$ and $C_{2,{\rm R}}$ still satisfies Eq. (\ref{e.UDmap}), thus $C_1 \times C_{2,{\rm R}}$ also forms a 2-user UD-AIEP code over GF($5$) as shown in Fig. \ref{Fig_GF5} (b).
$\blacktriangle \blacktriangle$

For a $J$-user UD-AIEP code $\Psi_{\rm s}$ over GF($p$) where $p > 2$, 
no codeword can have FFSP equal to the zero element $0$ of GF($p$). 
Since the FFSPs of the $2^J$ codewords in $\Psi_{\rm s}$ must be distinct elements in GF($p$), 
$2^{J}$ must be less than or equal to $p - 1$, 
i.e., $2^J \le p - 1$. 
Hence, the number of users $J$ for an UD-AIEP code over GF($p$) is upper bounded as follows:
\begin{equation} \label{e2.1}
  J \le \log_2 (p-1).
\end{equation}
Summarizing the results developed above, we have the following theorem.
\begin{theorem}
(\textbf{EP Upper bound over GF($p$)}) The Cartesian product of $J$ AIEPs over a prime field GF($p$) with $p > 2$ is a $J$-user UD-AIEP code if and only if the FFSPs of all its $2^J$ codewords are different nonzero elements in GF($p$) with $J$ upper bounded by $\log_2 (p-1)$.
\end{theorem}

There are totally $(p-1)/2$ AIEPs in GF($p$), and $\log_2 (p-1)$ of them can form an UD-AIEP code.
Owing to the feature of finite-field, we can totally have
$N_p = \lfloor \frac{p-1}{2\log_2 (p-1)}\rfloor$ UD-AIEP codes over GF($p$) for $p>3$, which form an UD-AIEP code set $\Xi$, i.e.,
 $ \Xi = \{ {\Psi}_{\rm s}(1), {\Psi}_{\rm s}(2), \ldots, 
   {\Psi}_{\rm s}(N_p)\},$
where ${\Psi}_{\rm s}(1), {\Psi}_{\rm s}(2), \ldots, {\Psi}_{\rm s}(N_p)$ are $\log_2 (p-1)$-user UD-AIEP codes.
It is noted that the UD-AIEP code set $\Xi$ is derived using a computer-based search algorithm.
Unfortunately, we have not yet discovered a mathematical method to provide a theoretical construction approach.

\begin{figure}[t]
  \centering
    \includegraphics[width=0.75\textwidth]{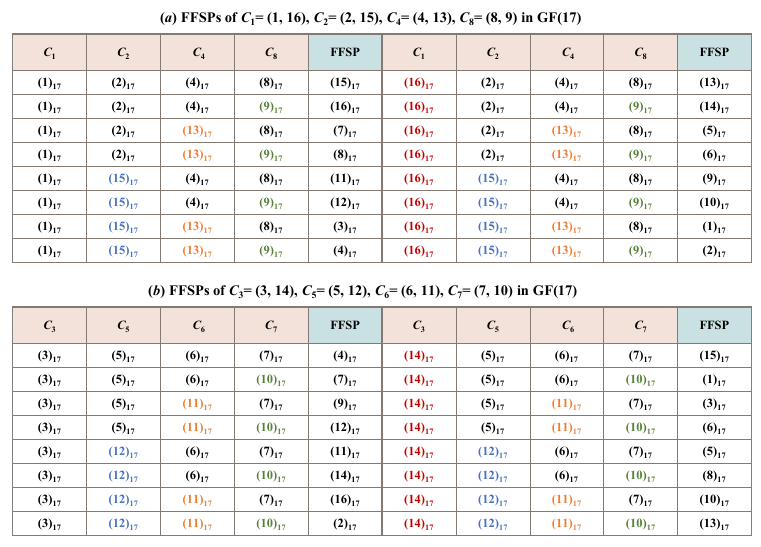}
  \caption{A diagram in table-form of Example 2. 
  (a) A $4$-user UD-AIEP code ${\Psi}_{\rm s}(1)$ over GF($17$) with $C_1 = (1,16), C_2 =(2,15), C_4 = (4,13), C_8=(8,9)$; and
  (b) a $4$-user UD-AIEP code ${\Psi}_{\rm s}(2)$ over GF($17$) with $C_3 = (3,14), C_5 =(5,12), C_6 = (6,11), C_7=(7,10)$.} \label{Fig_GF17}
  \vspace{-0.2in}
\end{figure}

\textbf{Example 2:} 
Suppose we use the prime field GF($17$) for constructing UD-AIEP codes. 
Eight AIEPs over GF($17$) can be constructed, which are 
$C_1 = (1, 16), C_2 = (2, 15), C_3 = (3, 14), C_4 = (4, 13), 
C_5 = (5, 12), C_6 = (6, 11), C_7 = (7, 10), C_8 = (8, 9)$. 
Since $\log_2 (17-1) = 4$, there are at most $4$ AIEPs whose Cartesian product satisfies the upper bound given by Theorem 1.

Based on the $8$ AIEPs, we can construct two $4$-user UD-AIEP codes.
One of these two $4$-user UD-AIEP code is formed by the following set of $4$ AIEPs:
\begin{equation*}
  {\Psi}_{\rm s}(1) = \{C_1 = (1, 16), C_2 = (2, 15), C_4 = (4, 13), C_8 = (8, 9)\},
\end{equation*}
whose Cartesian product 
${\Psi}_{\rm s}(1) = C_1 \times C_2 \times C_4 \times C_8$ 
gives a $4$-user UD-AIEP code as shown in Fig. \ref{Fig_GF17} (a). 
The other $4$-user UD-AIEP code is formed by the following set of $4$ AIEPs:
\begin{equation*}
  {\Psi}_{\rm s}(2) =\{C_3 = (3, 14), C_5 = (5, 12), C_6 = (6, 11), C_7 = (7, 10)\},
\end{equation*}
whose Cartesian product 
${\Psi}_{\rm s}(2) = C_3 \times C_5 \times C_6 \times C_7$ 
gives a $4$-user UD-AIEP code as shown in Fig. \ref{Fig_GF17} (b). 
All the AIEPs are different between ${\Psi}_{\rm s}(1)$ and ${\Psi}_{\rm s}(2)$.
$\blacktriangle  \blacktriangle$


\subsection{Orthogonal EP Codes Based on Extension Fields of Prime Fields}

In this subsection, we present a class of orthogonal EP codes constructed based on extension fields of prime fields.
Let $m$ be a positive integer with $m \ge 2$ and GF($p^m$) be the extension field of the prime field GF($p$). 
The extension field GF($p^m$) is constructed based on a primitive polynomial 
${\bf g}(X) = g_0 + g_1 X + g_2 X^2 + \ldots + g_m X^m$
of degree $m$ with coefficients from GF($p$) which consists of $p^m$ elements and contains GF($p$) as a subfield \cite{Shu2009}.

Let $\alpha$ be a primitive element in GF($p^m$). Then, the powers of $\alpha$, namely $\alpha^{-\infty} = 0, \alpha^0 = 1, \alpha, \ldots, \alpha^{(p^m - 2)}$, give all the $p^m$ elements of GF($p^m$). 
Each element $\alpha^{l_{j}}$, with $l_{j} = {-\infty}, 0,1, \ldots, p^m - 2$, in GF($p^m$) can be expressed as a linear sum of $\alpha^0 = 1, \alpha, \alpha^2, \ldots, \alpha^{(m - 1)}$ with coefficients from GF($p$) as 
\begin{equation} \label{e2.2}
    \alpha^{l_{j}} = a_{j,0} + a_{j,1} \alpha + a_{j,2} \alpha^2 + \ldots + 
    a_{j,m-1} \alpha^{(m-1)}.      
\end{equation}
From (\ref{e2.2}), we see that the element $\alpha^{l_{j}}$ can be uniquely represented by the $m$-tuple 
$(a_{j,0}, a_{j,1}, a_{j,2},\ldots, a_{j,m-1})$ over GF($p$), which is a linear combination of 
$\alpha^0, \alpha^1, \ldots, \alpha^{m-1}$, i.e., $\alpha^{l_{j}} = \oplus_{i=0}^{m-1} a_{j,i} \alpha^i$.

The field GF($p^m$) can form a vector space ${\mathbb V}_p(m)$ over GF($p$) of dimension $m$. 
Each vector in ${\mathbb V}_p(m)$ is an $m$-tuple over GF($p$). 
For $0 \le i < m$, it is known that $\alpha^i = (0, 0, \ldots, 1, 0,\ldots, 0)$ is an $m$-tuple with a 1-component at the $i$-th location and 0s elsewhere. 
The $m$ $m$-tuples $\alpha^0, \alpha^1, \alpha^2, \ldots, \alpha^{m-1}$ form an \textit{orthogonal (or normal) basis} of ${\mathbb V}_p(m)$. 
Hence, an element in GF($p^m$) can be expressed in three forms, namely \textit{power, polynomial and $m$-tuple forms}.

The sum of two elements 
$\alpha^{l_{j,0}} = \oplus_{i=0}^{m-1} a_{j_0,i} \alpha^i$
and
$\alpha^{l_{j,1}} = \oplus_{i=0}^{m-1} a_{j_1,i} \alpha^i$
is equal to
\begin{equation*}
\alpha^{l_{j,0}}+\alpha^{l_{j,1}} = (a_{j_0,0}+a_{j_1,0}) \alpha^0 + (a_{j_0,1}+a_{j_1,1}) \alpha^1 + \ldots
      +(a_{j_0,m-1}+a_{j_1,m-1}) \alpha^{m-1}.
\end{equation*}
The $m$-tuple representation of the sum $\alpha^{l_{j,0}} + \alpha^{l_{j,1}}$ is
\begin{equation*}
    \left((a_{j_0,0} + a_{j_1,0}), (a_{j_0,1} + a_{j_1,1}), \ldots, (a_{j_0,m-1} + a_{j_1,m-1})\right). 
\end{equation*}
For $0 \le i < m$, if $(a_{j_0,i}, a_{j_1,i})$ is an additive inverse pair over GF($p$) or a pair of zero elements $(0, 0)$, then $a_{j_0,i} + a_{j_1,i} = 0$ and $\alpha^{l_{j,0}} + \alpha^{l_{j,1}}  = 0$, 
i.e., $\alpha^{l_{j,0}}$ and $\alpha^{l_{j,1}}$ are additive inverse to each other. 
In this case, $(\alpha^{l_{j,0}}, \alpha^{l_{j,1}})$ forms an AIEP over GF($p^m$).

Now, we construct a type of \textit{orthogonal AIEP code}. 
If $(a_{j_0,i}, a_{j_1,i})$ is an nonzero AIEP over GF($p$), then 
\(
\alpha^i  (a_{j_0,i}, a_{j_1,i}) \triangleq (a_{j_0,i} \cdot \alpha^i, a_{j_1,i} \cdot \alpha^i),
\)
is an AIEP over GF($p^m$).

Let ${\Psi}_{\rm s} = \{C_1^{\rm s}, C_2^{\rm s}, \ldots, C_l^{\rm s}, \ldots, C_{L}^{\rm s}\}$ be a set of $L$ AIEPs over GF($p$), where $C_l^{\rm s} = (l, p - l)$ with $1 \le l \le L$.
For $0 \le i < m$,
\begin{equation} \label{e2.5}
\Psi_{{\rm o},i} = \{\alpha^i \cdot C_1^{\rm s}, \alpha^i \cdot C_2^{\rm s}, \ldots, 
\alpha^i \cdot C_l^{\rm s}, \ldots, \alpha^i \cdot C_{L}^{\rm s} \},
\end{equation}
is a set of $L$ AIEPs over GF($p^m$) with the USPM structural property,
where the subscript ``o'' stands for ``orthogonal''.
Hence, the Cartesian product
\begin{equation*} 
\Psi_{{\rm o},i} \triangleq (\alpha^i \cdot C_1^{\rm s}) \times (\alpha^i \cdot C_2^{\rm s}) \times \ldots \times 
(\alpha^i \cdot C_L^{\rm s})
\end{equation*}
of the $L$ AIEPs in $\Psi_{{\rm o},i}$ forms an $L$-user UD-AIEP code over GF($p^m$) with $2^L$ codewords,
each consisting of $2^L$ nonzero elements in GF($p^m$).
With $i = 0, 1, \ldots, m-1$, we can form $m$ $L$-user UD-AIEP codes over GF($p^m$),
$\Psi_{{\rm o},0}, \Psi_{{\rm o},1}, \ldots, \Psi_{{\rm o},m-1}$, which are \textit{mutually disjoint}, 
i.e., $\Psi_{{\rm o},k} \bigcap \Psi_{{\rm o},i} = \emptyset$ for $k \neq i$ and $0 \le k, i < m$.

If we represent an element in GF($p^m$) as an $m$-tuple over GF($p$), 
the two additive inverse elements in the pair 
$\alpha^i \cdot C_l^{\rm s} = (l \cdot \alpha^i, (p - l) \cdot \alpha^i)$ is a pair of two $m$-tuples with nonzero components, $l$ and $p - l$, at the $i$-th location, respectively,
and $0$s at all the other locations, 
i.e., $(0, 0, \ldots, l, 0, \ldots, 0)$ and $(0, 0,\ldots, p - l, 0, \ldots, 0)$. 
Hence, in $m$-tuple form, all the $L$ AIEPs in $\Psi_{{\rm o},i}$ have either $l$ and $p - l$ at the $i$-th location and $0$'s elsewhere with $1 \le l \le L$, i.e., 
\begin{equation*}
  \begin{aligned}
\alpha^i \cdot C_l^{\rm s} = \{(0, 0, \ldots, l, 0, \ldots, 0), (0, 0, \ldots, p-l, 0, \ldots, 0) \}.\\
  \end{aligned}
\end{equation*}

From $m$-tuple point of view, $\Psi_{{\rm o}, 0}, \Psi_{{\rm o}, 1}, \ldots, 
\Psi_{{\rm o}, m-1}$ are \textit{orthogonal} to each other, and they form $m$ orthogonal sets of AIEPs over GF($p$). 
Hence, $\Psi_{{\rm o}, 0}, \Psi_{{\rm o}, 1}, \ldots, \Psi_{{\rm o}, m-1}$ give $m$ orthogonal $L$-user UD-AIEP codes over GF($p^m$) (or over GF($p$) in $m$-tuple form). 
In $m$-tuple form, each codeword in $\Psi_{{\rm o},i}$ consists of $L$ $m$-tuples over GF($p$), 
each consisting of a single nonzero element from GF($p$) at the same location.
The union $\Psi_{\rm o} \triangleq \Psi_{{\rm o}, 0} \bigcup \Psi_{{\rm o}, 1} \bigcup \ldots 
\bigcup \Psi_{{\rm o}, m-1}$ forms an $Lm$-user orthogonal UD-AIEP code over GF($p^m$) with $2^{Lm}$ codewords over GF($p^m$) (or over GF($p$) in $m$-tuple form).
$\Psi_{\rm o}$ can be viewed as a \textit{cascaded} UD-AIEP code obtained by cascading the $m$ $L$-user UD-AIEP codes $\Psi_{{\rm o}, 0}, \Psi_{{\rm o}, 1}, \ldots, \Psi_{{\rm o}, m-1}$. 
We call $\Psi_{{\rm o}, 0}, \Psi_{{\rm o}, 1}, \ldots, \Psi_{{\rm o}, m-1}$ the \textit{constituent codes} of $\Psi_{\rm o}$.

Now, we consider a special case of the orthogonal UD-EP code $\Psi_{\rm o}$.
For the popular finite-field GF($2^m$) which is an extension field of the binary field GF($2$),
let ${\Psi}_{\rm o, B}$ be the orthogonal UD-EP code constructed over GF($2^m$).
In this case, there is only one base EP over GF($2$), not an additive inverse pair, 
defined by $C_{\rm B} =(0,1)$.
Hence, it is able to derive that
\begin{equation*}
  \begin{aligned}
  {\Psi}_{\rm o, B} 
  = \bigl\{\alpha^{j-1} \cdot C_{\mathrm{B}} \bigm| 1 \leq j \leq m\bigr\}
  = \bigl\{C_j^{\rm td} \bigm| 1 \leq j \leq m\bigr\},
  \end{aligned}
\end{equation*}  
where $C_j^{\rm td} = {\Psi}_{{\rm o},j-1} = \alpha^{j-1} \cdot C_{\rm B} = \alpha^{j-1} \cdot (0,1)$ 
for $1 \le j \le m$.
The subscript ``td'' of $C_j^{\rm td}$ stands for ``time division'', which will be introduced in a later section.

\textbf{Example 3:} 
For $p = 5$ and $m = 4$, consider the extension field GF($5^4$) of the prime field GF($5$). 
As shown in Example 1, using the prime field GF($5$), 
two AIEPs $C_1^{\rm s} = (1, 4)$ and $C_2^{\rm s} = (2, 3)$ can be constructed whose Cartesian product $C_1^{\rm s} \times C_2^{\rm s}$ is a 2-user UD-AIEP code over GF($5$). 
Based on this code, $8$ UD-AIEP codes over GF($5^4$) can be formed. 
They form $4$ orthogonal groups,
\begin{equation*}
  \begin{aligned}
    \Psi_{{\rm o},0} = \{\alpha^{0} \cdot C_1, \alpha^{0} \cdot C_2\} = 
                    \{(1 \cdot \alpha^{0}, 4\cdot \alpha^{0}), (2 \cdot \alpha^{0}, 3\cdot \alpha^{0})\},\\
    \Psi_{{\rm o},1} = \{\alpha^{1} \cdot C_1, \alpha^{1} \cdot C_2\} = 
                    \{(1 \cdot \alpha^{1}, 4\cdot \alpha^{1}), (2 \cdot \alpha^{1}, 3\cdot \alpha^{1})\},\\
    \Psi_{{\rm o},2} = \{\alpha^{2} \cdot C_1, \alpha^{2} \cdot C_2\} = 
                    \{(1 \cdot \alpha^{2}, 4\cdot \alpha^{2}), (2 \cdot \alpha^{2}, 3\cdot \alpha^{2})\},\\
    \Psi_{{\rm o},3} = \{\alpha^{3} \cdot C_1, \alpha^{3} \cdot C_2\} = 
                    \{(1 \cdot \alpha^{3}, 4\cdot \alpha^{3}), (2 \cdot \alpha^{3}, 3\cdot \alpha^{3})\}.\\
   \end{aligned}
\end{equation*}
The Cartesian products of these $4$ groups give $4$ orthogonal 2-user UD-AIEP codes 
$\Psi_{{\rm o},0}, \Psi_{{\rm o},1}$, $\Psi_{{\rm o},2}, \Psi_{{\rm o},3}$ over GF($5^4$). 
Their union gives an orthogonal $8$-user UD-AIEP code $\Psi_{\rm o}$ over GF($5^4$) with $2^{8} = 256$ EP codewords. 
$\blacktriangle  \blacktriangle$

\section{Encoding of EP Codes}

In this section, we first introduce the encoding of an EP code, which is realized based on a \textit{binary to finite-field GF($q$) transform function} denoted as ${\rm F}_{{\rm B}2q}$.
Then, we introduce the conceptions of \textit{finite-field multiplex module (FF-MUX)} for the AIEP codes $\Psi_{\rm s}$ constructed over GF($p$), and \textit{multiuser code (MC)} for the orthogonal UD-EP codes $\Psi_{\rm o, B}$ constructed over GF($2^m$).
Next, we investigate the orthogonal EP codes $\Psi_{\rm o}$ encoded by a channel encoder ${\mathcal C}_{gc}$, which is called orthogonal encoding of an error-correcting code over GF($p^m$) where $m \ge 2$.
The subscript ``gc'' of ${\mathcal C}_{gc}$ stands for ``globe channel code'', since the channel code ${\mathcal C}_{gc}$ is used through the MA transmission.

\subsection{Binary to Finite-field GF($q$) Transform Function}

Let the bit-sequence at the output of the $j$-th user be ${\bf b}_j = (b_{j,0}, b_{j,1},\ldots, b_{j,k}, \ldots, b_{j,K-1})$, where $K$ is a positive integer, $1 \le j \le J$ and $0 \le k < K$.
The EP encoder is to map each bit-sequence ${\bf b}_j$ uniquely into an element-sequence 
${\bf u}_j = (u_{j,0},u_{j,1},\ldots, u_{j,k},\ldots, u_{j,K-1})$ 
by a \textit{binary to finite-field GF($q$) transform function} ${\rm F}_{{\rm B}2q}$, 
i.e., $u_{j,k} = {\rm F}_{{\rm B}2q}(b_{j,k})$.
Similarly, we can decode the EP code by an inverse function ${\rm F}_{q2{\rm B}}$ which transforms \textit{from finite-field GF($q$) to binary field}, i.e., $b_{j,k} = {\rm F}_{q2{\rm B}}(u_{j,k})$.

Assume each user is assigned an EP, e.g., the EP of $C_j = (\alpha^{l_{j,0}}, \alpha^{l_{j,1}})$ is assigned to the $j$-th user for $1 \le j \le J$.
The subscript ``$j$'' of ``$l_{j,0}$'' and ``$l_{j,1}$'' stands for the $j$-th EP,
and the subscripts ``$0$'' and ``$1$'' of ``$l_{j,0}$'' and ``$l_{j,1}$'' represent the input bits are $(0)_2$ and $(1)_2$, respectively. 
For $1 \le j \le J$, we can set the $k$-th component $u_{j,k}$ of ${\bf u}_j$ as
\begin{equation} \label{F_b2q}
  u_{j,k} = {\mathrm F}_{{\mathrm B}2q}(b_{j,k}) \triangleq b_{j,k} \odot C_j = 
  \left\{
    \begin{aligned}
      \alpha^{l_{j,0}}, \quad b_{j,k} = (0)_2  \\
      \alpha^{l_{j,1}}, \quad b_{j,k} = (1)_2  \\
    \end{aligned},
  \right.
\end{equation}
where $b_{j,k} \odot C_j$ is defined as a \textit{switching function}.
If the input bit is $b_{j,k} = (0)_2$, the transformed element component is $u_{j,k} = \alpha^{l_{j,0}}$; 
otherwise, if the input bit is $b_{j,k} = (1)_2$, $u_{j,k}$ is equal to $u_{j,k} = \alpha^{l_{j,1}}$.
Let the input \textit{user block} of $J$ users of the $k$-th component denote by ${\bf b}[k]$, 
i.e., ${\bf b}[k] = (b_{1,k}, b_{2,k}, \ldots, b_{J,k})$, where $0 \le k < K$.
We also express the EP codeword $(u_{1,k}, u_{2,k}, \ldots, u_{J,k})$ in $\Psi$ as ${\bf u}[k] = (u_{1,k}, u_{2,k}, \ldots, u_{J,k})$.

\subsection{FF-MUX of an AIEP Code $\Psi_{\rm s}$}

A \textit{finite-field multiplex module (FF-MUX)} is used together with the AIEP code $\Psi_{\rm s}$ for multiplexing the element-sequences of $J$-user.
There are various FF-MUXs for different AIEP codes.

For the AIEP codes $\Psi_{\rm s}$ constructed over GF($p$), 
let ${\mathcal A}_{{\rm M}}$ be an FF-MUX over a prime field GF($p$), which is a $J \times T$ binary matrix, i.e., ${\mathcal A}_{{\rm M}} \in {\mathbb B}^{J \times T}$, given by
\begin{equation}
  {\mathcal A}_{{\rm M}} = \left[
  \begin{matrix}
    a_{1,1} & a_{1,2} & \ldots & a_{1,T}\\
    a_{2,1} & a_{2,2} & \ldots & a_{2,T}\\
    \vdots  & \vdots  & \ddots & \vdots \\
    a_{J,1} & a_{J,2} & \ldots & a_{J,T}\\
  \end{matrix}
  \right],
\end{equation}
where $1 \le T \le J$, and the subscript ``M'' stands for ``multiplex''.
Then, the {FFSP block} ${w}_{k}$ of the element-block ${\bf u}[k]$ is calculated as 
\begin{equation*}
  {w}_{k} = {\bf u}[k] \cdot {\mathcal A}_{{\rm M}},
\end{equation*}
which is a $1 \times T$ vector, i.e., $w_k = (w_{k,1}, \ldots, w_{k,t},\ldots, w_{k, T})$.
If each EP codeword ${\bf u}[k]$ in ${\Psi}_{\rm s}$ is mapped into a unique FFSP block $w_k$, we can uniquely recover ${\bf u}[k]$ from the FFSP block $w_k$
i.e., ${\bf u}[k] \leftrightarrow {w}_{k}$.
In this case, we must have $2^J \le p^T$ (or $J \le T \cdot \log_2 p$).
While, if $T=1$ and $J \le \log_2 (p-1)$, the AIEP code becomes as an UD-AIEP code with the USPM structural property.

To evaluate the performance of an AIEP code $\Psi_{\rm s}$, the \textit{loading factor (LF)} $\eta$ is defined as the ratio of the number of rows to the number of columns of the matrix ${\mathcal A}_{\rm M}$, given as  
\begin{equation}
\eta = {J}/{T},
\end{equation}
which is upper bounded by $\log_2 p$, because of $J \le T \cdot \log_2 p$.
Hence, the range of LF is $1 \le \eta \le \log_2 p$.
If $T = J$, the LF is equal to one, i.e., $\eta = 1$.

Note that, the LF of an UD-AIEP code over GF($p$) is equal to ${\log_2 (p-1)}$. Then, the LF of an AIEP code can be approximately upper bounded by an UD-AIEP code.
Although the AIEP code cannot improve the LF, it can support more users than an UD-AIEP code for a given prime field GF($p$).

\subsubsection{FF-MUX of an UD-AIEP code}

For an UD-AIEP code over GF($p$), the number of served users $J$ is equal to or smaller than $\log_2 (p-1)$, i.e., $J \le \log_2 (p-1)$. 
In this case, the FF-MUX ${\mathcal A}_{\rm M}$ is a $J \times 1$ vector, i.e.,
\begin{equation} \label{e.G_M_addition}
  {\mathcal A}_{{\rm M}} = \left[
    \begin{matrix}
      1, 1, \ldots, 1\\
     \end{matrix}
    \right]^{\rm T},
\end{equation}
and computing the FFSP as $w_k = {\bf u}[k] \cdot {\mathcal A}_{{\rm M}} = \bigoplus_{j=1}^{J} u_{j,k}$, which is a simple \textit{finite-field addition operation} to the EP codeword ${\bf u}[k]= (u_{1,k}, u_{2,k}, \ldots, u_{J,k})$.

To explain the multiplexing of an UD-AIEP code, let us consider the 2-user UD-AIEP code over GF($5$) given in Example 1 which consists of two AIEPs $C_1 = (1, 4)$ and $C_2 = (2, 3)$ whose FFSPs are shown in Fig. 1 (a). First, we assign $C_1$ and $C_2$ to users $1$ and $2$, respectively. 
The two input bits $(0)_2$ and $(1)_2$ of user-$1$ are mapped into two elements $(1)_5$ and $(4)_5$ in GF($5$), respectively. For user-$2$, the two input bits $(0)_2$ and $(1)_2$ are mapped into two elements $(2)_5$ and $(3)_5$ in GF($5$), respectively. The multiplex process is an addition operation, i.e., $w = u_1 \oplus u_2$, which is sent to the channel. Based on the FFSP table given by Fig. 1 (a), we can recover the bit information of the two users from an FFSP appearing in the FFSP table. For example, if the received FFSP of the two-users in GF($5$) is $(3)_5$, the nonbinary elements of users $1$ and $2$ are $(1)_5$ and $(2)_5$, respectively, which indicate the bits from users 1 and 2 are $(0)_2$ and $(0)_2$, respectively.

\subsubsection{FF-MUX of an AIEP code}
As presented early, there are a total of $(p-1)/{2}$ AIEPs in a finite-field GF($p$).
However, only $\log_2 (p-1)$ AIEPs can form UD-AIEPs.
If $p > 3$, we have $\frac{p-1}{2} > \log_2 (p-1)$.
This indicates that $\frac{p-1}{2} - \log_2 (p-1)$ AIEPs are unused.
These unused AIEPs can also be assigned to users by designing FF-MUX ${\mathcal A}_{{\rm M}} \in {\mathbb B}^{J \times T}$.

There are various methods to construct FF-MUX. 
In the following, we present a \textit{concatenated multiple UD-AIEP codes (C-UDEPs)} method, which consists of two steps:
  \begin{enumerate}
    \item
    Step one is to find all the UD-AIEP codes in the finite-field GF($p$), i.e.,
    \begin{equation*} 
     \Xi = \{ {\Psi}_{\rm s}(1), {\Psi}_{\rm s}(2), \ldots, {\Psi}_{\rm s}(N_p)\},
    \end{equation*}
    where ${\Psi}_{\rm s}(1), {\Psi}_{\rm s}(2), \ldots, {\Psi}_{\rm s}(N_p)$ are $\log_2 (p-1)$-user UD-AIEP codes, with $N_p = \lfloor \frac{p-1}{2\log_2(p-1)} \rfloor$.
    \item
    Step two is to make each output FFSP $w_{k,t}$ of ${w}_k$ for $1 \le t \le T$ belong to an UD-AIEP code in $\Xi$, i.e., $w_{k,t} \in {\Psi_{\rm s}}(t)$,
    where ${w}_k = (w_{k,1}, \ldots, w_{k,t}, \ldots, w_{k,T})$ and $T \le N_p$.  
  \end{enumerate}

\textbf {Example 4:} For the finite-field GF(17), its eight AIEPs have been given in Example 2, which provide two UD-AIEP codes 
${\Psi}_{\rm s}(1) = \{C_1 =(1, 16), C_2 =(2, 15), C_4 = (4, 13), C_8 = (8, 9)\}$ and ${\Psi}_{\rm s}(2) = \{C_3 =(3, 14), C_5 =(5, 12), C_6 = (6, 11), C_7 = (7, 10)\}$.
If there are $J = 8$ users, the UD-AIEP code cannot support $8$ users.
We can design a $8 \times 2$ binary FF-MUX ${\mathcal A}_{{\rm M}}$ as follows:
\begin{equation*}
  {\mathcal A}_{{\rm M}}^{\rm T} = \left[
  \begin{matrix}
    1 & 1 & 0 & 1 & 0 & 0 & 0 & 1\\
    0 & 0 & 1 & 0 & 1 & 1 & 1 & 0\\
  \end{matrix}
  \right].
\end{equation*}
With this FF-MUX, ${w} = {\bf u} \cdot {\mathcal A}_{\rm M}$ is a $1 \times 2$ vector in GF($17$).
The two components of ${w}$ are:
\begin{equation*}
  \begin{aligned}
  w_1 = u_1 \oplus u_2 \oplus u_4 \oplus u_8 \in {\Psi}_{\rm s}(1), \\
  w_2 = u_3 \oplus u_5 \oplus u_6 \oplus u_7 \in {\Psi}_{\rm s}(2). \\
  \end{aligned}
\end{equation*}
Since ${\Psi}_{\rm s}(1)$ and ${\Psi}_{\rm s}(2)$ are UD-AIEP codes, 
the LF of ${\mathcal A}_{\rm M}$ is ${\eta} = 4$, 
which is also equal to the LF of an UD-AIEP code over GF(17).

For example, if ${w} = (w_1, w_2) = (15, 4)_{17}$, following from the Fig. 2 (a) and (b),
we find that $u_1 = (1)_{17}, u_2 = (2)_{17}, u_4 = (4)_{17}, u_8 = (8)_{17}$, 
and $u_3 = (3)_{17}, u_5 = (5)_{17}, u_6 = (6)_{17}, u_7 = (7)_{17}$.
$\blacktriangle \blacktriangle$

\subsection{Multiuser Code of an Orthogonal EP Code $\Psi_{\rm o,B}$}

For the orthogonal UD-EP code $\Psi_{\rm o,B}$ over GF($2^m$), 
the FF-MUX ${\mathcal A}_{\rm M}$ is the same as Eq. (\ref{e.G_M_addition}), i.e.,
 ${\mathcal A}_{{\rm M}} = \left[1, 1, \ldots, 1 \right]^{\rm T}$,
resulting in a \textit{finite-field addition operation}.
Without loss of generality, we assume the FF-MUX to be the finite-field addition operation by default.

Now, we introduce the conception of \textit{multiuser code (MC)}, which is presented for an EP code constructed over an extension field GF($2^m$) where $m \ge 2$.
We take the orthogonal UD-EP code $\Psi_{\rm o,B}$ constructed over GF($2^m$) as an example.
The orthogonal UD-EP code $\Psi_{\rm o,B}$ is used to support an $m$-user FFMA. 
We assign $C_j^{\rm td}$ to the $j$-th user for $1 \le j \le m$, and the FFSP is calculated as 
\begin{equation} \label{e.w_tdma}
\begin{aligned}
w_k &\overset{(a)}{=} {\bf u}[k] \cdot {\mathcal A}_{{\rm M}} = u_{1,k} + u_{2,k} + \ldots + u_{m,k} \\
    &\overset{(b)}{=} b_{1,k} \alpha^0 + b_{2,k} \alpha^1 + \ldots + b_{m,k} \alpha^{m-1} 
     {=} {\bf b}[k] \cdot {\bf G}_{\rm M}^{\bf 1}
\end{aligned},
\end{equation}
in which step (a) indicates that the EP codeword ${\bf u}[k]$ is encoded by the FF-MUX ${\mathcal A}_{{\rm M}}$ that is a finite-field addition operation,
and step (b) is deduced based on (\ref{F_b2q}) and the orthogonal UD-EPs $C_j^{\rm td}$ for $1 \le j \le m$.
We call ${\bf G}_{\rm M}^{\bf 1}$ the \textit{full-one generator matrix} of a \textit{multiuser code} ${\mathcal C}_{\rm M}$, which is used to encode the input user block ${\bf b}[k]$ of $J$-user.

For the orthogonal UD-EP code $\Psi_{\rm o,B}$ over GF($2^m$), the generator matrix of the MC,
i.e., ${\bf G}_{\rm M}^{\bf 1} = [\alpha^0, \alpha^1, \ldots, \alpha^{m-1}]^{\rm T}$, is given as
\begin{equation*}
{\bf G}_{\rm M}^{\bf 1} = \left[ 
\begin{matrix}
\alpha^{0}\\
\alpha^{1}\\
\vdots\\
\alpha^{m-1}
\end{matrix}
\right] =
\left[ 
\begin{matrix}
1 & 0 & \ldots & 0\\
0 & 1 & \ldots & 0\\
\vdots & \vdots &\ddots &\vdots\\
0 & 0 & \ldots & 1\\
\end{matrix}
\right],
\end{equation*}
which is an $m \times m$ identity matrix. 
Thus, based on the orthogonal UD-EP code $\Psi_{\rm o, B}$ over GF($2^m$), the proposed system is \textit{a type of TDMA in finite-field (FF-TDMA)}, in which the outputs of the $m$ users completely occupy $m$ locations in an $m$-tuple (similarly to the $m$ time slots).

For an EP code constructed over an extension field of a prime field GF($p^m$) where $m \ge 2$, if its FF-MUX is a $J \times 1$ full one vector, i.e., ${\mathcal A}_{{\rm M}} = \left[1, 1, \ldots, 1 \right]^{\rm T}$, then its loading factor is only determined by the size of its full-one generator matrix ${\bf G}_{\rm M}^{\bf 1}$. For the orthogonal UD-EP code over GF($2^m$), its LF is equal to $\eta = J/m = 1$.

\subsection{Orthogonal Encoding of an Error-correcting Code}

In the following, we present an encoding of an error-correcting code over GF($p^m$) in a form to match the orthogonal UD-EP code $\Psi_{\rm o}$.
In a latter section, orthogonal UD-EP codes will be used in conjunction with error-correcting codes for error control in FFMA systems.

Suppose ${\bf w} = (w_0, w_1, \ldots, w_k, \ldots, w_{K-1})$ is a sequence over GF($p^m$),
where $K$ is a positive integer.  
For $0 \le k < K$, let $(u_{0,k}, u_{1,k},\ldots, u_{j,k},\ldots, u_{m-1,k})$ be the $m$-tuple representation of the $k$-th component $w_k$ of ${\bf w}$, i.e.,
\begin{equation*} 
  \begin{aligned}
w_k 
    &= u_{0,k} \alpha^0 + u_{1,k} \alpha^1 + \ldots + u_{j,k} \alpha^j + \ldots + u_{m-1,k} \alpha^{m-1},
  \end{aligned}
\end{equation*}
where $u_{j,k} \in$ GF($p$) and $ 0 \le j < m$.

Hence, $w_k$ can be viewed as the FFSP of the $m$ elements $u_{0,k}, u_{1,k},\ldots, u_{j,k},\ldots, u_{m-1,k}$. It is a one-to-one mapping between $w_k$ and the EP codeword ${\bf u}[k]$, 
where ${\bf u}[k] = (u_{0,k}, u_{1,k},\ldots, u_{j,k},\ldots, u_{m-1,k})$.

For $0 \le j < m$, we form the following element-sequence over GF($p$), i.e.,
${\bf u}_j = (u_{j,0}, u_{j,1}, \ldots, u_{j,k},\ldots, u_{j,K-1})$,                           
where $u_{j,k}$ is the $j$-th component of $w_k$. 
Define the following $K$ $m$-tuples over GF($p$) as
\begin{equation} \label{e2.11}
    {\bf u}_j \cdot \alpha^j \triangleq (u_{j,0} \alpha^j, u_{j,1} \alpha^j, \ldots, 
    u_{j,k} \alpha^j, \ldots, u_{j,K-1} \alpha^j).   
\end{equation}

Then, the $K$-tuple $\bf w$ over GF($p^m$) can be decomposed into the following ordered sequence of $K$ $m$-tuples over GF($p$),
\begin{equation} \label{e2.12}
    {\bf w} \triangleq  
    {\bf u}_0 \cdot \alpha^0 + {\bf u}_1 \cdot \alpha^1 + \ldots + 
    {\bf u}_j \cdot \alpha^j + \ldots + {\bf u}_{m-1} \cdot \alpha^{m-1},  
\end{equation}
which is the \textit{orthogonal $m$-tuple decomposition of $\bf w$}.
This orthogonal form indicates a sequence ${\bf w}$ over GF($p^m$) can be decomposed into $m$ symbol-sequences ${\bf u}_0, {\bf u}_1, \ldots, {\bf u}_{m-1}$.
In other words, ${\bf w}$ is the FFSP sequence of the $m$ element sequences ${\bf u}_0, {\bf u}_1, \ldots, {\bf u}_{m-1}$.

Let ${\bf G}_{gc}$ be the generator matrix of a $p^m$-ary $(N, K)$ linear block code ${\mathcal C}_{gc}$ over GF($p^m$) with $m \ge 2$. Let ${\bf g}_0, {\bf g}_1, \ldots, {\bf g}_{K-1}$ be the $K$ rows of ${\bf G}_{gc}$, each an $N$-tuple over GF($p^m$). 
Let ${\bf w} = (w_0, w_1, \ldots, w_k, \ldots, w_{K-1})$ be a message over GF($p^m$) whose orthogonal decomposition is given by (\ref{e2.12}). 
Then, we encode $\bf w$ into a codeword $\bf v$ in ${\mathcal C}_{gc}$ using the generator ${\bf G}_{gc}$, i.e.,
\begin{equation*} 
  {\bf v} = {\bf w} \cdot {\bf G}_{gc} = (v_0, v_1, v_2, \ldots, v_{N-1}).
\end{equation*}

The orthogonal $m$-tuple decomposition of $\bf v$ is given by (\ref{e2.13}), 
\begin{equation} \label{e2.13}
  \begin{aligned}
{\bf v} =& {\bf w} \cdot {\bf G}_{gc} 
        = w_0 {\bf g}_0 + w_1 {\bf g}_1 + \ldots + w_{K-1} {\bf g}_{K-1}\\
        =& \left(u_{0,0}  \alpha^0 \oplus u_{1,0} \alpha^1 \oplus \ldots \oplus 
                     u_{m-1,0} \alpha^{m-1} \right) {\bf g}_0 +
               \left(u_{0,1}  \alpha^0 \oplus u_{1,1} \alpha^1 \oplus \ldots \oplus 
                     u_{m-1,1} \alpha^{m-1} \right) {\bf g}_1 + \ldots +\\
               &\left(u_{0,K-1} \alpha^0 \oplus u_{1,K-1} \alpha^1 \oplus \ldots \oplus 
                     u_{m-1,K-1} \alpha^{m-1} \right) {\bf g}_{K-1} \\
        =& \left(u_{0,0} \alpha^0 {\bf g}_0 + u_{0,1} \alpha^0 {\bf g}_1 + \ldots 
                    + u_{0,K-1} \alpha^0 {\bf g}_{K-1}\right) \oplus
           \left(u_{1,0} \alpha^1 {\bf g}_0 + u_{1,1} \alpha^1 {\bf g}_1 + \ldots 
                    + u_{1,K-1} \alpha^1 {\bf g}_{K-1}\right) \oplus \ldots \oplus\\
          &\left(u_{m-1,0} \alpha^{m-1} {\bf g}_0 + u_{m-1,1} \alpha^{m-1} {\bf g}_1 + \ldots 
                    + u_{m-1,K-1} \alpha^{m-1} {\bf g}_{K-1}\right) \\
        =& ({\bf u}_0 \cdot {\bf G}_{gc}) \alpha^0 \oplus 
           ({\bf u}_1 \cdot {\bf G}_{gc}) \alpha^1 \oplus \ldots \oplus 
           ({\bf u}_{m-1} \cdot {\bf G}_{gc}) \alpha^{m-1}
        {=} {\bf v}_0 \alpha^0 \oplus {\bf v}_1 \alpha^1 \oplus \ldots \oplus {\bf v}_{m-1}\alpha^{m-1}.    
  \end{aligned}
\end{equation}
where ${\bf v}_j = {\bf u}_j {\bf G}_{gc}$ is the codeword of ${\bf u}_j$ for $0 \le j < m$. The codeword $\bf v$ in orthogonal form is referred to as \textit{orthogonal encoding} of the message $\bf w$, indicating the codeword $\bf v$ is the FFSP of the codewords ${\bf v}_0, {\bf v}_1, \ldots, {\bf v}_{m-1}$.

Note that the orthogonal encoding is also applicable to the linear bock code ${\mathcal C}_{gc}$ constructed over GF($p$). In this case, ${\mathcal C}_{gc}$ is an $(N, Km)$ linear block code over GF($p$).

\section{An FFMA System for a Massive MA Communication Scenario}
This section presents an uplink FFMA system in a GMAC, based on the orthogonal UD-EP code $\Psi_{\rm o, B}$ constructed over the extension field GF($2^m$) of the binary field GF($2$), i.e., 
\begin{equation*}
  \Psi_{\rm o,B} = \{C_1^{\rm td}, C_2^{\rm td}, \ldots, C_j^{\rm td}, \ldots, C_J^{\rm td} \},
\end{equation*}
where $C_j^{\rm td} = \alpha^{j-1} \cdot C_{\rm B}$ for $1 \le j \le J$.
The parameter $m$ indicates the number of finite-field time-slots which can be viewed as \textit{virtual resource blocks (VRBs)}. The VRB will be further discussed in the following section.
The number of users $J$ that the system can support is equal to or smaller than $m$, i.e., $J \le m$.
The UD-EP $C_j^{\rm td}$ is assigned to the $j$-th user. 
For a massive MA case, $m$ is a large number.
The block diagram for such a system is shown in Fig. \ref{Fig_UL}.

\begin{figure*}[t]
  \centering
  \includegraphics[width=0.95\textwidth]{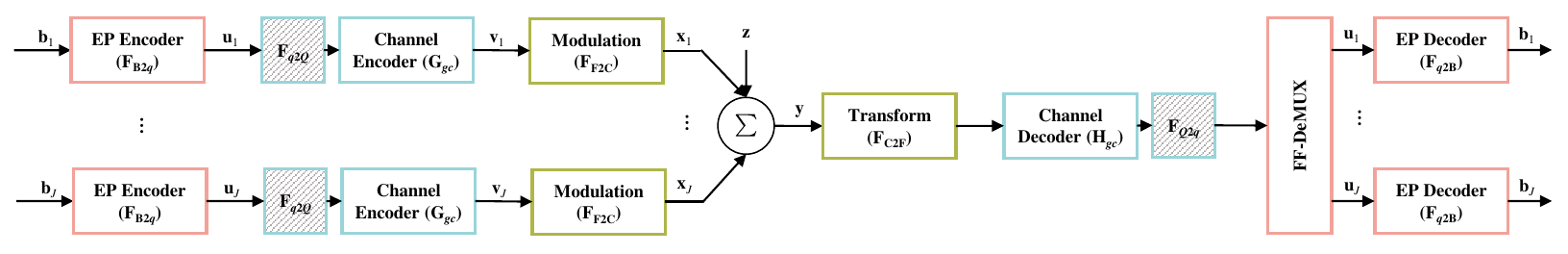}
  \label{Fig_UL}
  \caption{A block diagram of an FFMA system in a GMAC,
  where ${\rm F}_{{\rm B}2q}$ and ${\rm F}_{q2{\rm B}}$ stand for ``binary to finite-field GF($q$) transform'' and ``finite-field GF($q$) to binary transform''; 
  ${\rm F}_{q2Q}$ and ${\rm F}_{Q2q}$ are ``finite-field GF($q$) to finite-field GF($Q$) transform'' and ``finite-field GF($Q$) to finite-field GF($q$) transform''; 
  $\rm F_{F2C}$ and $\rm F_{C2F}$ stand for ``finite-field to complex-field transform'' and ``complex-field to finite-field transform''.} 
  \label{Fig_UL}
  \vspace{-0.2in}
\end{figure*}

\subsection{Transmitter of a Sparse-form FFMA System}

Let ${\bf b}_j = (b_{j,0}, b_{j,1},\ldots, b_{j,k}, \ldots, b_{j,K-1})$ be the bit-sequence at the output of the $j$-th user, where $1 \le j \le J$ and $0 \le k < K$.
The transmitter maps each bit-sequence ${\bf b}_j$ uniquely into an element-sequence 
${\bf u}_j = (u_{j,0},u_{j,1},\ldots, u_{j,k},\ldots, u_{j,K-1})$ by ${\rm F}_{{\rm B}2q}$, 
i.e., $u_{j,k} = {\rm F}_{{\rm B}2q}(b_{j,k})$, 
which is determined by the EP $C_j^{\rm td} = \alpha^{j-1} \cdot C_{\rm B}$.
For $0 \le k < K$, the $k$-th component $u_{j,k}$ of ${\bf u}_j$ is represented by its corresponding $m$-tuple representation over GF($2$), i.e., $u_{j,k} =(u_{j,k,0}, u_{j,k,1},\ldots, u_{j,k,i},\ldots, u_{j,k,m-1})$, where $0 \le i < m$.
Then, the $m$-tuple form of $u_{j,k}$ is
\begin{equation} \label{e.u_j_k}
{u}_{j,k} = {\rm F}_{{\rm B}2q}(b_{j,k}) = b_{j,k} \odot C_j^{\rm td} = (0,\ldots, 0, b_{j,k}, 0,\ldots, 0),
\end{equation} 
and the $i$-th component of $u_{j,k}$ is 
\begin{equation} \label{e.u_j}
u_{j,k,i} =
\left\{
  \begin{matrix}
    b_{j,k}, & i = j-1\\
    0,       & i \neq j-1 \\
  \end{matrix}. \right.
\end{equation}
For a large $m$, $u_{j,k}$ is a {sparse vector} with only one element.

Next, the element-sequence ${\bf u}_j$ is encoded into a codeword ${\bf v}_j$ of an $(N_Q, K_Q)$ linear block code ${\mathcal C}_{gc}$ of length $N_Q$ over GF($Q$) specified by a $K_Q \times N_Q$ generator matrix ${\bf G}_{gc}$.
Since $Q$ can be different from $q$, a \textit{finite-field GF($q$) to finite-field GF($Q$) transform function} ${\rm F}_{q2Q}$ is required. In this paper, we only consider two cases.

In the first case, $Q$ is equal to $q=2^m$, i.e., $Q=q=2^m$, and the block code ${\mathcal C}_{gc}$ is constructed based on the same field GF($2^m$) as that of the UD-EP code $\Psi_{\rm o,B}$. 
The transform function ${\rm F}_{q2Q}$ is used to transform the \textit{$m$-tuple form} of ${u}_{j,k}$ into the \textit{power form} of ${u}_{j,k}$. 
In this case, set $K_Q = K$, and the generator matrix ${\bf G}_{gc}$ becomes as a $K \times N_Q$ matrix.
Then, we encode ${\bf u}_j$ by the generator matrix ${\bf G}_{gc}$, 
and obtain the codeword ${\bf v}_j$ of ${\bf u}_j$ as
\begin{equation*} \label{e5.3}
{\bf v}_j = {\bf u}_j {\bf G}_{gc} = (v_{j,0}, v_{j,1},\ldots, v_{j,n_q}, \ldots, v_{j,N_Q-1}),   
\end{equation*}
where $v_{j,n_q} \in$ GF($2^m$) and $0 \le n_q < N_Q$.
Express each element $v_{j,n_q}$ in ${\bf v}_j$ into an $m$-tuple over GF($2$), i.e., 
$v_{j,n_q} = (v_{j,n_q,0}, v_{j,n_q,1},\ldots, v_{j,n_q,i}, \ldots, v_{j,n_q,m-1})$.
Then, the codeword ${\bf v}_j$ for ${\bf u}_j$ becomes an $m N_Q$-tuple over GF($2$).
If we set $N = mN_Q$, then ${\bf v}_j$ is an $N$-tuple over GF(2), 
i.e., ${\bf v}_j = (v_{j,0}, v_{j,1},\ldots, v_{j,n}, \ldots, v_{j,N-1})$, 
where $v_{j,n} \in {\mathbb B}$ and $0 \le n < N$.
We say that the codeword ${\bf v}_j$ for ${\bf u}_j$ in binary-form.
Hence, in this case, encoding of the bit-sequence ${\bf b}_j$ to its binary-form codeword ${\bf v}_j$ is binary.

In the second case, we set $Q=2$ and $q=2^m$.
In this case, the function of ${\rm F}_{q2Q}$ is the same as the transform function ${\rm F}_{{\rm B}2q}$. Hence, the transform function ${\rm F}_{q2Q}$ in encoding can be removed.
Let $K_Q = K \cdot m$ and $N_Q = N$.
Then, the generator matrix ${\bf G}_{gc}$ is a $mK \times N$ matrix over GF($2$). 
The encoded codeword ${\bf v}_j$ is a vector over GF($2$).

In summary, the overall encoding may be viewed as a two-level concatenated encoding, with the EP encoding as the inner encoding and the channel code ${\mathcal C}_{gc}$ encoding as the outer encoding. We refer to this encoding as $\Psi_{\rm o,B}$/${\mathcal C}_{gc}$-encoding.

Generally, the generator matrix ${\bf G}_{gc}$ is either in systematic form ${\bf G}_{gc,sym}$ or non-systematic form ${\bf G}_{gc,nonsym}$. Both forms of ${\bf G}_{gc}$ can be used in the proposed FFMA systems.
Nevertheless, a systematic form of the generator ${\bf G}_{gc,sym}$ may provide more flexibility in designing FFMA systems. This will be shown later in this section.

Suppose the $mK \times N$ generator matrix over GF($2$) is in systematic form, defined by ${\bf G}_{gc,sym}$. Then, the codeword ${\bf v}_j = (v_{j,0}, v_{j,1}, \ldots, v_{j,n}, \ldots, v_{j,N-1})$ is of the following form
\begin{equation*}
  \begin{array}{ll}
  {\bf v}_j = {\bf u}_j \cdot {\bf G}_{gc,sym} 
  = ({\bf u}_j, {\bf v}_{j,\rm red}) = (u_{j,0}, u_{j,1}, \ldots, u_{j,K-1}, \textcolor{blue}{{\bf v}_{j,\rm red}}),
  \end{array}
\end{equation*}
where $u_{j,k}$ is a sparse vector given by (\ref{e.u_j_k}),
and ${\bf v}_{j, \rm red}$ is the parity block (or called redundancy), and the subscript ``red'' stands for ``redundancy''.

In systematic form, the codewords of $m$ users can be arranged in an $m \times N$ codeword matrix
${\bf V} = [{\bf v}_1, {\bf v}_2, \ldots, {\bf v}_j, \ldots, {\bf v}_m]^{\mathrm T}$,
where $1 \le j \le J$ and $J = m$.
The codeword matrix ${\bf V}$ can be divided into two sections, information section ${\bf U}$ and parity section ${\bf E}$, as shown in (\ref{e.TM_sparse}).

\begin{small}
\begin{equation} \label{e.TM_sparse}
  \begin{aligned}
  {\bf V} =
  \left[
  \begin{matrix}
  {\bf v}_1 \\
  {\bf v}_2 \\
   \vdots   \\
  {\bf v}_{J} 
  \end{matrix}
  \right] = [
  {\bf U},{\bf E}
  ]
  =\left[
  \begin{array}{cccc:c}
  u_{1,0} & u_{1,1} & \ldots & u_{1,K-1} & \textcolor{blue}{{\bf v}_{1, \rm red}}\\
  u_{2,0} & u_{2,1} & \ldots & u_{2,K-1} & \textcolor{blue}{{\bf v}_{2, \rm red}}\\
  \vdots  & \vdots  & \ddots &  \vdots   & \vdots  \\
  u_{J,0} & u_{J,1} & \ldots & u_{J,K-1} & \textcolor{blue}{{\bf v}_{J, \rm red}}\\
  \end{array}
  \right] 
  {=} 
  \left[
  \begin{array}{c:c:c:c}
    \textcolor{red}{b_{1,0}}, 0,\ldots,0    & \ldots & \textcolor{red}{b_{1,K-1}},0, \ldots, 0  & \textcolor{blue}{{\bf v}_{1,\rm red}}\\
    0, \textcolor{red}{b_{2,0}}, \ldots, 0  & \ldots & 0, \textcolor{red}{b_{2,K-1}}, \ldots, 0 & \textcolor{blue}{{\bf v}_{2,\rm red}}\\
    \vdots                 & \ddots &  \vdots                 & \vdots\\
    0, 0, \ldots, \textcolor{red}{b_{J,0}}  & \ldots & 0, 0, \ldots, \textcolor{red}{b_{J,K-1}} & \textcolor{blue}{{\bf v}_{J,\rm red}}\\
  \end{array}
  \right].
  \end{aligned}
\end{equation}
\end{small}

The information section ${\bf U}$ of ${\bf V}$ is a $1 \times K$ array, i.e.,
${\bf U} = [{\bf U}_0, {\bf U}_1, \ldots, {\bf U}_k, \ldots, {\bf U}_{K-1}]$,
and ${\bf U}_k$ for $0 \le k < K$ is an $m \times m$ matrix.
When $J = m$, $J$ information bits $b_{1,k}, b_{2,k}, \ldots, b_{J,k}$ from the $J$ users are lying on the main diagonal of $ {\bf U}_k$.
When $J < m$, the $J$ information bits $b_{1,k}, b_{2,k}, \ldots, b_{J,k}$ are still lying on the main diagonal of ${\bf U}_k$, and the rest rows and/or columns of ${\bf U}_k$ are all zeros.
The parity section ${\bf E}$ of ${\bf V}$ is a $J \times (N-mK)$ matrix which consists of all the parity bits formed based on the generator matrix.

To support massive users with short packet transmission scenario, $m$ may be very large. 
In this case, ${\bf V}$ is a sparse matrix, and we call ${\bf V}$ a \textit{sparse codeword matrix}, and the corresponding FFMA system is referred to as \textit{sparse-form FFMA (SF-FFMA)}.

Then, each codeword ${\bf v}_j$ is modulated with BPSK signaling and mapped to a complex-field signal sequence ${\bf x}_j \in {\mathbb C}^{1 \times N}$, i.e., 
${\bf x}_j = (x_{j,0}, x_{j,1}, \ldots, x_{j, n}, \ldots, x_{j, N-1})$. 
For $0 \le n < N$, the $n$-th component $x_{j, n}$ is given by
\begin{equation}  \label{e.x_j}
{x}_{j,n} = {\rm F}_{\rm F2C}(v_{j,n}) =  2 {v}_{j,n} - 1,
\end{equation}
where $x_{j,n} \in \{-1, +1\}$ for $0 \le n < N$. 
The mapping from ${\bf v}_j$ to ${\bf x}_j$ is regarded as 
\textit{finite-field to complex-field transform (F2C)}, 
denoted by $\rm F_{F2C}$. 
Then ${\bf x}_j$ is sent to a GMAC.

\subsection{Receiver of a Sparse-form FFMA System}

At the receiving end, the received signal sequence ${\bf y} \in {\mathbb C}^{1 \times N}$ is the combined outputs of the $J$ users plus noise, i.e.,
\begin{equation} 
{\bf y} = \sum_{j=1}^{J} {\bf x}_j + {\bf z} = {\bf r} + {\bf z},
\end{equation}
where ${\bf z} \in \mathbb{C}^{1 \times N}$ is an AWGN vector
with ${\mathcal N}(0, N_0/2)$.
The sum in ${\bf y}$ is called \textit{complex-field sum-pattern (CFSP)} signal sequence, i.e., ${\bf r} = (r_0, r_1, \ldots, r_n, \ldots, r_{N-1}) \in \mathbb{C}^{1 \times N}$, 
which is the sum of the $J$ modulated signal sequences ${\bf x}_1, {\bf x}_2,\ldots, {\bf x}_J$. 
For $0 \le n < N$, the $n$-th component $r_n$ of ${\bf r}$ is given by 
\begin{equation} \label{e5.7}
    r_{n} = \sum_{j=1}^J x_{j,n} = 2 \sum_{j=1}^J v_{j,n} - J.              
\end{equation}

To separate the superposition signals, the classical method is to utilize \textit{multiuser interference cancellation} algorithms, e.g., successive interference cancellation (SIC) algorithm.
In our paper, we present an alternative solution, where the MUI is regarded as a form of MASK (amplitude shift keying) signal.
So that, we do not separate the superposition signals of $J$ users in the complex-field. We only need to map each CFSP signal $r_n$ to a unique FFSP symbol $v_n$, i.e., $r_n \mapsto v_n$.  
Then, the bit-sequences of $J$ users can be recovered by decoding the EP codeword, since the orthogonal UD-EP code $\Psi_{\rm o, B}$ satisfies the USPM structural property.

Thus, the first step of the detection process is to transform the received CFSP signal sequence 
${\bf r} = (r_0, r_1, \ldots, r_n, \ldots, r_{N-1})$ into its corresponding FFSP codeword sequence 
${\bf v} = (v_0, v_1, \ldots, v_n, \ldots, v_{N-1})$ 
by a \textit{complex-field to finite-field (C2F)} transform function $\rm F_{C2F}$, 
i.e., ${v}_n = {\rm F_{C2F}}({r}_n)$ for $0 \le n < N$. 
It is important to find the transform function $\rm F_{C2F}$, otherwise, the system is ineffective.
It is able to find the following facts of the CFSP ${r}_n$ and FFSP ${v}_n$.
\begin{enumerate}
\item
The value of CFSP $r_{n}$ is determined by the number of users who send ``$+1$'' and the number of users who send ``$-1$''. Thus, the maximum and minimum values of $r_{n}$ are $J$ and $-J$, respectively. 
The set of $r_{n}$'s values in ascendant order is $\Omega_r = \{-J, -J+2, \ldots, J-2, J\}$, 
in which the difference between two adjacent values is $2$. 
The total number of $\Omega_r$ is equal to $|\Omega_r| = J+1$. 
\item
Since $v_{n} \in {\mathbb B}$, the possible values of $v_{n}$ are $(0)_2$ and $(1)_2$. 
Then, $v_{n}$ is uniquely determined by the number of $(1)_2$ bits coming from the $J$ users, 
i.e., $v_{1,n}, v_{2,n}, \ldots, v_{J,n}$. 
If there are odd number of bits $(1)_2$ from the $J$ users, i.e., $v_{1,n}, v_{2,n}, \ldots, v_{J,n}$, 
then $v_{n} = (1)_2$; otherwise, $v_{n} = (0)_2$. 
Since the values of $r_{n}$ are arranged in ascendant order, 
the corresponding FFSP set $\Omega_v$ of $\Omega_r$ is $\Omega_v = \{0, 1, 0, 1, \ldots\}$, 
in which $(0)_2$ and $(1)_2$ appear alternatively. 
The number of $|\Omega_v|$ is also equal to $J+1$, i.e., $|\Omega_v| = |\Omega_r|$.
\item
Let $C_J^\iota$ denote the number of users which send ``+1''. 
The values of $\iota$ are from $0$ to $J$. 
When $\iota = 0$, it indicates that all the $J$ users send $(0)_2$, 
i.e., $v_{j,n} = (0)_2$ for all $1 \le j \le J$, 
thus, $v_{n} = \bigoplus_{j=1}^J v_{j,n} = (0)_2$. 
If $\iota$ increases by one, the number of $(1)_2$ bits coming from $J$ users increases by one accordingly. Therefore, the difference between two adjacent values is $2$, and the bits $(0)_2$ and $(1)_2$ appear alternatively.
\end{enumerate}

\begin{figure}[t] 
  \centering
  \label{Fig_MapTable}
  \includegraphics[width=0.7\textwidth]{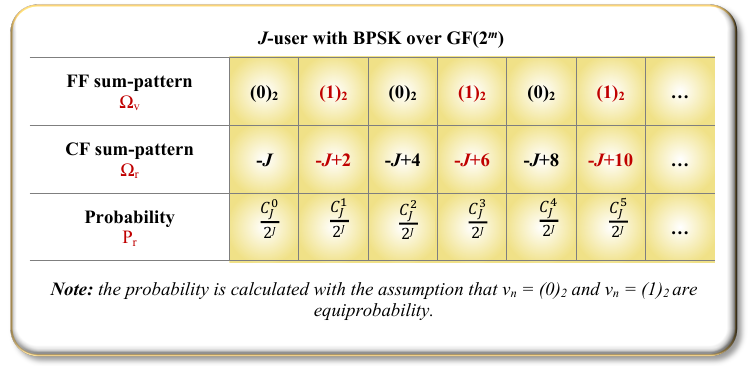}
 \caption{The relationship between CFSP sequence ${\bf r}$ and FFSP sequence ${\bf v}$ of a $J$-user FFMA system, where each user utilizes BPSK signaling.} \label{Fig_MapTable}
 \vspace{-0.25in}
\end{figure}

Based on the above facts, the transform function $\rm F_{C2F}$ maps each CFSP signal $r_{n}$ to a unique FFSP symbol $v_{n}$, i.e., ${\rm F_{C2F}}: r_{n} \mapsto v_{n}$. 
Assume that $v_{n} = (0)_2$ and $v_{n} = (1)_2$ are equally likely, 
i.e., ${\rm Pr}(v_{n}=0) = {\rm Pr}(v_{n} = 1) = 0.5$. 
Then, the probabilities of the elements in $\Omega_r$ are
 \begin{equation*} 
  {\mathcal P}_r = \left\{ {C_J^0}/{2^J}, {C_J^{1}}/{2^J},\ldots, {C_J^{J-1}}/{2^J}, {C_J^J}/{2^J} \right\}.
 \end{equation*} 
With $r_{n} \in \{-J, -J+2, \ldots, J-2, J\}$ and $v_{n} \in \{0,1\}$, 
$\rm F_{C2F}$ is a many-to-one mapping function, as summarized in Fig. \ref{Fig_MapTable}.

Next, we calculate the posterior probabilities used for decoding ${\bf y}$. 
Based on the relationship between $r_{n}$ and $v_{n}$, 
the conditional probability $v_{n}$ given by $y_{n}$ is
\begin{equation} \label{e5.11}
  P(v_{n}|y_{n}) = \frac{{\rm Pr}(v_{n})P(y_{n}|v_{n})}{P(y_{n})} ,                   
\end{equation}
where $P(y_{n})$ is the probability of $y_{n}$.
Since $y_{n}$ is determined by $r_{n}$ that is selected from the set $\Omega_r$, thus,
\begin{equation} \label{e_MAP}
  \begin{aligned}
  P(y_{n}) &= \sum_{\iota=0}^{J} {\mathcal P}_r(\iota) \cdot 
         \frac{1}{\sqrt{\pi N_0}} \exp \left\{
         - \frac{\left[y_{n} - \Omega_r(\iota) \right]^2}{N_0} 
         \right\},  \\
  \end{aligned}
\end{equation}
where ${\mathcal P}_r(\iota)$ and $\Omega_r(\iota)$ stand for the $\iota$-th element in the sets ${\mathcal P}_r$ and $\Omega_r$, respectively. 
When $v_{n} = (0)_2$, the corresponding $r_{n}$ equals to $\{-J, -J+4, -J+8, \ldots\}$. 
The posteriori probability of $v_{n} = (0)_2$ is
\begin{equation}
  P(v_{n}=0|y_{n}) = \frac{1}{{P(y_{n})}} 
  \sum_{\iota=0,\iota+2}^{\iota \le J} {\mathcal P}_r(\iota)\cdot \frac{1}{\sqrt{\pi N_0}}
  \exp \left\{
  -\frac{\left[y_{n} - \Omega_r(\iota) \right]^2}{N_0}
  \right\},
\end{equation}

Similarly, when $v_{n} = (1)_2$, it indicates $r_{n}$ belongs to $\{-J+2, -J+6, -J+10,\ldots\}$, and the posteriori probability of $v_{n} = (1)_2$ is
\begin{equation}
  P(v_{n}=1|y_{n}) = \frac{1}{{P(y_{n})}} 
  \sum_{\iota=1,\iota+2}^{\iota \le J} {\mathcal P}_r(\iota)\cdot \frac{1}{\sqrt{\pi N_0}}
  \exp \left\{
  -\frac{\left[y_{n} - \Omega_r(\iota) \right]^2}{N_0}
  \right\}.
\end{equation}

Then, $P(v_{n} = 0|y_{n})$ and $P(v_{n} = 1|y_{n})$ are used for decoding $\bf y$. 
If a binary LDPC code is utilized, we can directly calculate log-likelihood ratio (LLR) based on 
$P(v_{n} = 0|y_{n})$ and $P(v_{n} = 1|y_{n})$. 
If an NB-LDPC code is utilized, the probability mass function (pmf) can be computed based on 
$P(v_{n} = 0|y_{n})$ and $P(v_{n} = 1|y_{n})$.

When the generator matrix ${\bf G}_{gc,sym}$ in systematic form is utilized,
the decoded FFSP codeword sequence $\hat{\bf v}$ can be expressed as $\hat{\bf v} = ({\hat{\bf w}}, {\hat{\bf v}}_{\rm red})$, where ${\hat{\bf w}}$ and ${\hat{\bf v}}_{\rm red}$ are the detected FFSP sequence and parity block, respectively.

After removing the parity block ${\hat{\bf v}}_{\rm red}$ from $\hat{\bf v}$, the detected FFSP sequence ${\hat{\bf w}}$ can be divided into $K$ FFSP blocks, and each block consists of $m$ bits formed an $m$-tuple,
i.e., ${\hat{\bf w}} = ({\hat w}_0, {\hat w}_1, \ldots, {\hat w}_k,\ldots, {\hat w}_{K-1})$, 
where ${\hat w}_k = ({\hat w}_{k,0}, {\hat w}_{k,1}, \ldots, {\hat w}_{k,i}, \ldots, {\hat w}_{k,m-1})$, with $0 \le k < K$ and $0 \le i < m$.
Finally, we separate the detected FFSP block ${\hat{w}}_k$ into $J$ bits as 
$\hat{b}_{1,k}, \hat{b}_{2,k}, \ldots, \hat{b}_{j,k}, \ldots, \hat{b}_{J,k}$ by using an \textit{inverse transform function} ${\rm F}_{q2{\rm B}}$.
According to (\ref{e.w_tdma}),
the inverse transform function ${\rm F}_{q2{\rm B}}$ is given as
\begin{equation}
  {\hat b}_{j,k} = {\rm F}_{q2{\rm B}}({\hat w_k}) = {\hat w}_{k,j-1},
\end{equation}
where ${\hat w}_{k,j-1}$ is the $(j-1)$-th component of ${\hat w}_k$. 

If there is no channel code ${\mathcal C}_{gc}$, we can calculate $\Omega_r$ and ${\mathcal P}_r$ in another manner.
Without a channel code, it means that ${\bf v}_j = {\bf u}_j$.
According to (\ref{e.u_j}) and (\ref{e.x_j}),
$r_n$ can be computed as follows:
\begin{equation*}
r_{n} = \sum_{j=1}^{J} x_{j,n} = (2 b_{j-1,n} - 1) + \sum_{j' \neq j-1}^{J} (-1)
      = -J + 2 b_{j-1,n},
\end{equation*}
where $0 \le n < N$ and $1 \le j \le J$.
If $b_{j-1,n}= (0)_2$, it is able to know the corresponding CFSP signal $r_n$ is $r_n = -J$;
otherwise, if $b_{j-1,n}= (1)_2$, then $r_n$ is equal to $r_n = -J+2$. 
Hence, the value set of CFSP $r_n$ is $\Omega_r = \{-J, -J+2\}$, 
with equiprobability ${\mathcal P}_r = \{0.5, 0.5\}$.
The Euclidean distance between $-J$ and $-J+2$ is the same as that of BPSK modulation. Therefore, without channel coding, we can directly utilize BPSK demodulation to detect the CFSP signals.

Moreover, for the different forms of the generator ${\bf G}_{gc}$, we can further classify the probabilities set into two cases. 
For systematic form, we can separately compute the posteriori probabilities of the information symbols and the redundancy symbols, respectively. 
For non-systematic form, we can exploit the proposed $\Omega_r$ and ${\mathcal P}_r$, as shown in Fig. \ref{Fig_MapTable}.

\begin{figure}[t] 
  \centering
  \includegraphics[width=0.9\textwidth]{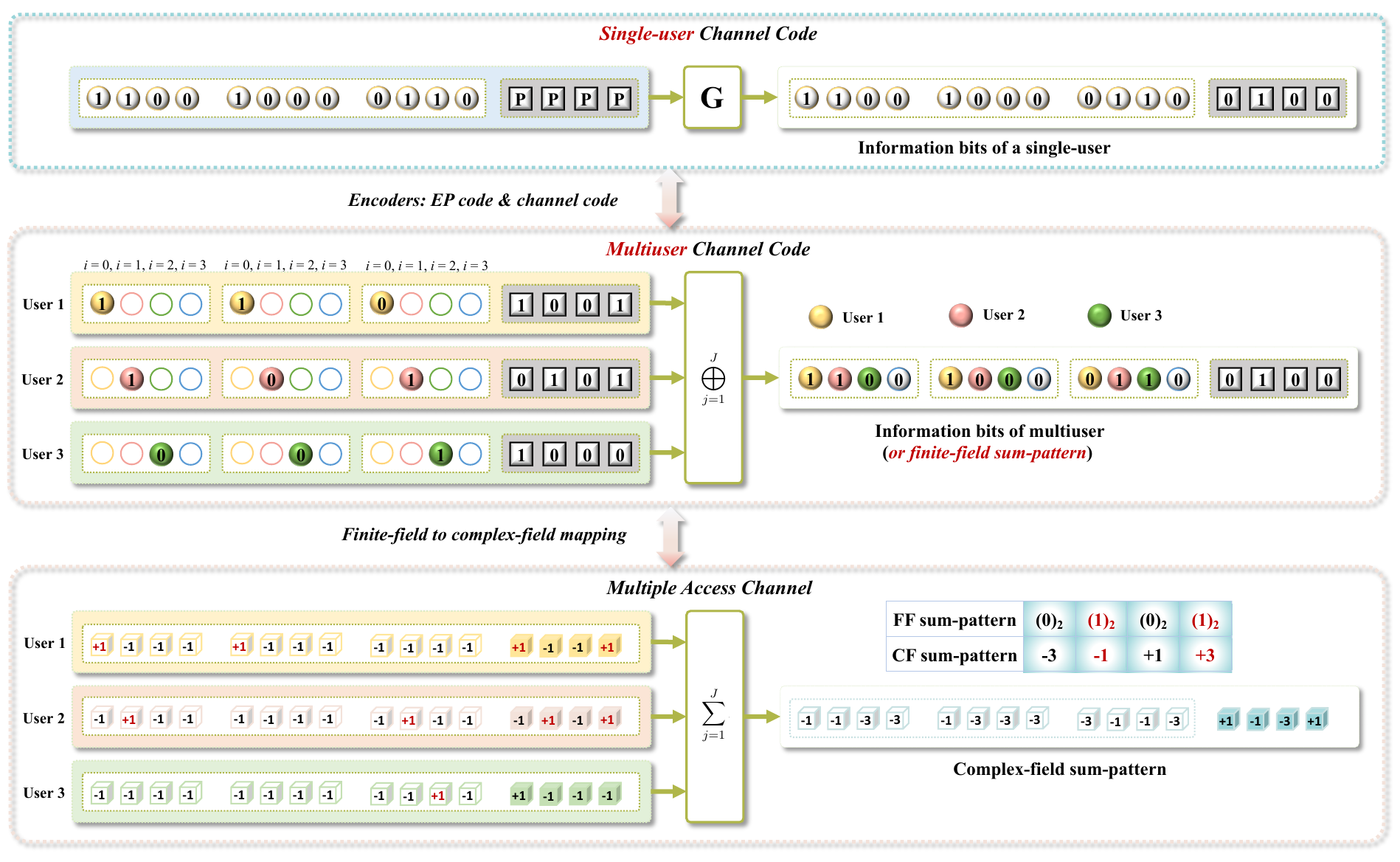}
 \caption{A diagram of Example 5. Assume there are $J = 3$ users and each user transmit $K  = 3$ bits. The bit-sequences of the 3-user are respectively ${\bf b}_1 = (1, 1, 0)$, ${\bf b}_2 = (1, 0, 1)$, and ${\bf b}_3 = (0, 0, 1)$.}
  \label{f.Eg5}
  \vspace{-0.2in}
\end{figure}

\textbf{Example 5:}
In this example, we use an FFMA system over the field GF($2^4$) to illustrate the transmission and receiving process, as shown in Fig. \ref{f.Eg5}. The field GF($2^4$) can support $4$ or fewer users. Suppose we design an FFMA system to support $J = 3$ users and each user transmits $K = 3$ bits.
The UD-EP code used is
  \(
    \Psi_{\rm o,B} = \{C_1^{\rm td}, C_2^{\rm td}, C_3^{\rm td} \},
  \)
and the UD-EP assigned to the $j$-th user is $C_j^{\rm td}$. 
Let ${\bf b}_1 = (1, 1, 0)$, ${\bf b}_2 = (1, 0, 1)$ and ${\bf b}_3 = (0, 0, 1)$ be bit-sequences of the users $1, 2$, and $3$, respectively. 
The transmission process consists of three steps, ${\rm F}_{{\rm B}2q}$ mapping (or EP encoding), channel code encoding, and ${\rm F}_{\rm F2C}$ mapping (or BPSK modulation).

In the first step of transmission, each bit $b_{j,k}$ is mapped into a symbol $u_{j,k} = {\rm F}_{{\rm B}2q}(b_{j,k})$ which is represented by a $4$-tuple over GF($2$). The transform function ${\rm F}_{{\rm B}2q}$ maps the 3 bit-sequences into three symbol-sequences over GF($2^4$) in $4$-tuple form as follows:
\begin{equation*} \label{e.Ex4_1}
  \begin{array}{c}
{\bf b}_1 = (1, 1, 0) \Rightarrow
{\bf u}_1 = (\textcolor{red}{1} 0 0 0, \textcolor{red}{1} 0 0 0, \textcolor{red}{0} 0 0 0)\\
{\bf b}_2 = (1, 0, 1) \Rightarrow
{\bf u}_2 = (0 \textcolor{red}{1} 0 0, 0 \textcolor{red}{0} 0 0, 0 \textcolor{red}{1} 0 0)\\
{\bf b}_3 = (0, 0, 1) \Rightarrow
{\bf u}_3 = (0 0 \textcolor{red}{0} 0, 0 0 \textcolor{red}{0} 0, 0 0 \textcolor{red}{1} 0)\\ 
  \end{array}.
\end{equation*}
The FFSP sequence of ${\bf u}_1$, ${\bf u}_2$, and ${\bf u}_3$ is 
\begin{equation*}
{\bf w} = \oplus_{j=1}^{3} {\bf u}_j = (1100, 1000, 0110),
\end{equation*}
which will be used at the receiving end to recover the $3$ transmitted bit-sequences.

In the second step of transmission process, each element-sequence ${\bf u}_j$ is encoded into a codeword 
${\bf v}_j = {\bf u}_j {\bf G}_{eg}$ in the channel code ${\mathcal C}_{eg}$. 
The channel code used is a $(16, 12)$ linear block code of length $16$ and dimension $12$ with the following generator matrix ${\bf G}_{eg}$ over GF($2$) in systematic form:

\begin{equation*} \label{e.G}
  \begin{array}{ll}
  {\bf G}_{eg} = 
  \left[
  \begin{array}{cccccccccccc:cccc}
   1 & 0 & 0 & 0 & 0 & 0 & 0 & 0 & 0 & 0 & 0 & 0 & 1 & 0 & 0 & 0 \\
   0 & 1 & 0 & 0 & 0 & 0 & 0 & 0 & 0 & 0 & 0 & 0 & 0 & 1 & 0 & 0 \\
   0 & 0 & 1 & 0 & 0 & 0 & 0 & 0 & 0 & 0 & 0 & 0 & 0 & 0 & 1 & 0 \\
   0 & 0 & 0 & 1 & 0 & 0 & 0 & 0 & 0 & 0 & 0 & 0 & 0 & 0 & 0 & 1 \\
   0 & 0 & 0 & 0 & 1 & 0 & 0 & 0 & 0 & 0 & 0 & 0 & 0 & 0 & 0 & 1 \\
   0 & 0 & 0 & 0 & 0 & 1 & 0 & 0 & 0 & 0 & 0 & 0 & 1 & 0 & 0 & 0 \\
   0 & 0 & 0 & 0 & 0 & 0 & 1 & 0 & 0 & 0 & 0 & 0 & 0 & 1 & 0 & 0 \\
   0 & 0 & 0 & 0 & 0 & 0 & 0 & 1 & 0 & 0 & 0 & 0 & 0 & 0 & 1 & 0 \\
   0 & 0 & 0 & 0 & 0 & 0 & 0 & 0 & 1 & 0 & 0 & 0 & 0 & 0 & 1 & 0 \\
   0 & 0 & 0 & 0 & 0 & 0 & 0 & 0 & 0 & 1 & 0 & 0 & 0 & 0 & 0 & 1 \\
   0 & 0 & 0 & 0 & 0 & 0 & 0 & 0 & 0 & 0 & 1 & 0 & 1 & 0 & 0 & 0 \\
   0 & 0 & 0 & 0 & 0 & 0 & 0 & 0 & 0 & 0 & 0 & 1 & 0 & 1 & 0 & 0 \\
  \end{array}
  \right].
  \end{array}
\end{equation*}

Hence, the three channel codewords are
\begin{equation*}
  \begin{array}{c}
{\bf v}_1 = {\bf u}_1 \cdot {\bf G}_{eg}  =
(\textcolor{red}{1} 0 0 0, \textcolor{red}{1} 0 0 0, \textcolor{red}{0} 0 0 0, \textcolor{blue}{1 0 0 1})\\
{\bf v}_2 = {\bf u}_2 \cdot {\bf G}_{eg}  =
(0 \textcolor{red}{1} 0 0, 0 \textcolor{red}{0} 0 0, 0 \textcolor{red}{1} 0 0, \textcolor{blue}{0 1 0 1})\\
{\bf v}_3 = {\bf u}_3 \cdot {\bf G}_{eg}  =
(0 0 \textcolor{red}{0} 0, 0 0 \textcolor{red}{0} 0, 0 0 \textcolor{red}{1} 0, \textcolor{blue}{1 0 0 0})
  \end{array},
\end{equation*}
and the last $4$ bits of each codeword are parity bits.

Following, each channel codeword ${\bf v}_j$ is modulated with BPSK into a signal sequence ${\bf x}_j$ as follows:
  \begin{equation*}
  \begin{array}{c}
{\bf v}_1  
\Rightarrow
{\bf x}_1 = (+1, -1, -1, -1, +1, -1, -1, -1, -1, -1, -1, -1, +1, -1, -1, +1)\\
{\bf v}_2  
\Rightarrow
{\bf x}_2 = (-1, +1, -1, -1, -1, -1, -1, -1, -1, +1, -1, -1, -1, +1, -1, +1)\\
{\bf v}_3  
\Rightarrow
{\bf x}_3 = (-1, -1, -1, -1, -1, -1, -1, -1, -1, -1, +1, -1, +1, -1, -1, -1)\\
  \end{array}.
  \end{equation*}

Then, the three modulated signal sequences ${\bf x}_1$,  ${\bf x}_2$, and  ${\bf x}_3$ are sent to a GMAC.

At the receiving end, assuming no effect of noise, the process of recovering the transmitted bit-sequences consists of $3$ steps, namely ${\rm F}_{\rm C2F}$ mapping (or demodulation), channel decoding, and ${\rm F}_{q2{\rm B}}$ mapping (or EP decoding). In the first step of recovering process, the ${\rm F}_{\rm C2F}$ function transforms the received CFSP sequence,
\begin{equation*}
  \begin{array}{c}
{\bf r} = \sum_{j=1}^{3} {\bf x}_j =   
  (-1, -1, -3, -3, -1, -3, -3, -3, -3, -1,  -1, -3,  +1, -1, -3, +1),
  \end{array}
\end{equation*}
to a FFSP codeword sequence $\hat{\bf v}$. 
For $J = 3$, we have $\Omega_r = \{-3, -1, +1, +3\}$ and $\Omega_v = \{0, 1, 0, 1\}$. 
Hence, ${\rm F_{C2F}}(-3) = (0)_2$, ${\rm F_{C2F}}(-1) = (1)_2$, ${\rm F_{C2F}}(+1) = (0)_2$, and ${\rm F_{C2F}}(+3) = (1)_2$.
This complex-field to finite-field transformation gives the following sequence:
  \begin{equation*}
  \hat{\bf v} = {\rm F_{C2F}}({\bf r}) = 
  (1, 1, 0, 0, 1, 0, 0, 0, 0, 1, 1, 0, 0, 1, 0, 0).
  \end{equation*}
In the second step of recovering process, the receiver decodes the sequence $\hat{\bf v}$ based on a designed parity check matrix ${\bf H}_{eg}$ of the channel code ${\mathcal C}_{eg}$. 
Since no noise effect is assumed, the decoded FFSP sequence is given as
\begin{equation*}
  \hat{\bf w} = (1100, 1000, 0110),
\end{equation*}
which has been divided into $K = 3$ blocks, each block consisting of $4$ bits formed an $4$-tuple.

In the third step of the receiving process, decode the FFSP sequence $\hat{\bf w}$ by using the inverse transform function ${\rm F}_{q2{\rm B}}$. Decoding recovers the transmitted bit-sequences as
 \begin{equation*} 
  \begin{array}{c}
  \hat{\bf b}_1 = (1, 1, 0), \quad
  \hat{\bf b}_2 = (1, 0, 1), \quad
  \hat{\bf b}_3 = (0, 0, 1).
  \end{array}
\end{equation*}

If transmission is affected by noise, the channel decoder has to perform error-correction process in the second step to estimate the transmitted sequences. 
$\blacktriangle \blacktriangle$

\vspace{-0.2in}
\subsection{A Diagonal-form FFMA System}

As presented early, if the generator ${\bf G}_{gc}$ is constructed over GF($2$) and in systemic form, we can obtain the \textit{sparse form codeword matrix} ${\bf V}$, as shown in (\ref{e.TM_sparse}).
For the $1 \times K$ information array ${\bf U}$ of ${\bf V}$,
each entry ${\bf U}_k$ of ${\bf U}$ is an $m \times m$ matrix, and the $m$ information bits $b_{1,k}, b_{2,k}, \ldots, b_{j,k}, \ldots, b_{m,k}$ are located on the main diagonal of $ {\bf U}_k$, 
and all off-diagonal entries are zero, where $0 \le k < K$ and $1 \le j \le J \le m$.

Permuting the columns of the $1 \times K$ array ${\bf U}$ of ${\bf V}$ and rearranging the bit sequences, we can obtain an $m \times m$ array ${\bf U}_{\rm D}$, in which ${\bf b}_1, {\bf b}_2, \ldots, {\bf b}_j,\ldots, {\bf b}_{m}$ are lying on the main diagonal of ${\bf U}_{\rm D}$ and the other entries are all $\bf 0$s, given by
\begin{equation}
  {\bf U}_{\rm D} =
  \left[
  \begin{array}{cc}
  {\bf u}_{1, \rm D} \\
  {\bf u}_{2, \rm D} \\
   \vdots   \\
  {\bf u}_{m, \rm D} 
  \end{array}
  \right]
  =
  \left[
  \begin{array}{cccc}
    {\bf b}_1 &           &        &            \\
              & {\bf b}_2 &        &            \\
              &           & \ddots &            \\
              &           &        & {\bf b}_m  \\
 
  \end{array}
  \right],
\end{equation}
where ${\bf b}_1, {\bf b}_2, \ldots, {\bf b}_j,\ldots, {\bf b}_{m}$ are $1 \times K$ vectors,
and the subscript ``D'' stands for ``diagonal''.
Let ${\bf u}_{j, \rm D}$ denote as a $1 \times mK$ information sequence of the $j$-th user, 
i.e., ${\bf u}_{j, \rm D} = ({\bf 0}, \ldots, {\bf 0}, {\bf b}_j, {\bf 0}, \ldots, {\bf 0})$,
in which the $1 \times K$ bit-sequence of the $j$-th user ${\bf b}_j$ is located at the $(j-1)$-th entry of ${\bf u}_{j, \rm D}$ and ${\bf 0}$ is a $1 \times K$ zero vector.
Note that, if $J < m$, the $(m-J)$ rightmost entries of ${\bf U}_{\rm D}$ are all zeros, 
forming an $(m-J) \times (m-J)K$ zero matrix.

Then, we encode ${\bf u}_{j, \rm D}$ by the generator matrix ${\bf G}_{gc, sym}$ in systemic form, and obtain the codeword ${\bf v}_{j,\rm D}$ of ${\bf u}_{j, \rm D}$, i.e., 
${\bf v}_{j, \rm D} = ({\bf u}_{j,\rm D}, \textcolor{blue}{{\bf v}_{j,\rm D, red}})$,
where ${\bf v}_{j,\rm D,red}$ is the parity block of ${\bf v}_{j, \rm D}$,
which is a $(N-mK)$-tuple.
The codewords of ${\bf v}_{1,\rm D}, {\bf v}_{2,\rm D}, \ldots, {\bf v}_{J,\rm D}$ together can form a \textit{diagonal form} codeword matrix ${\bf V}_{\rm D}$, given as follows:
\begin{equation}
  {\bf V}_{\rm D} =
  \left[
  \begin{array}{cc}
  {\bf v}_{1, {\rm D}} \\
  {\bf v}_{2, {\rm D}} \\
   \vdots   \\
  {\bf v}_{m, {\rm D}} 
  \end{array}
  \right]
  =
  \left[
  \begin{array}{cccc:c}
    {\bf b}_1 &           &        &             & \textcolor{blue}{{\bf v}_{1, \rm D, red}}\\
              & {\bf b}_2 &        &             & \textcolor{blue}{{\bf v}_{2, \rm D, red}}\\
              &           & \ddots &             & \vdots\\
              &           &        & {\bf b}_m   & \textcolor{blue}{{\bf v}_{m, \rm D, red}}\\
 
  \end{array}
  \right],
\end{equation}
consisting of an $m \times m$ information array ${\bf U}_{\rm D}$ and an $m \times 1$ parity array ${\bf E}_{\rm D}$, 
where ${\bf E}_{\rm D} = [{\bf v}_{1,\rm D,red}, \\ {\bf v}_{2,\rm D, red}, \ldots, {\bf v}_{m,\rm D, red}]^{\rm T}$.

From the codeword ${\bf v}_{j,\rm D}$, it is found that the useful vectors 
(i.e., ${\bf b}_j$ and ${\bf v}_{j,\rm D, red}$) 
are only located at the $(j-1)$-th entry of ${\bf u}_{j, {\rm D}}$ and the parity entry ${\bf v}_{j,\rm D, red}$, and the other entries are all zeros.
Thus, for the diagonal-form codeword ${\bf v}_{j, {\rm D}}$, we can directly modulate and transmit the shorten codeword, defined by 
${\bf v}_{j,{\rm D,S}} = ({\bf b}_j, {\bf v}_{j,\rm D, red})$, 
instead of ${\bf v}_{j,{\rm D}}$, to reduce the transmit power.
The subscript ``S'' indicates ``shorten''.
The codeword length of ${\bf v}_{j,{\rm D, S}}$ is equal to $N - (m-1)K$.
For a short packet transmission, e.g., $K$ is an extremely small number, 
the length of ${\bf v}_{j,{\rm D, S}}$ is approximately equal to the parity blocklength ${\bf v}_{j,{\rm D, red}}$.

The FFMA system based on the shorten codeword ${\bf v}_{j,{\rm D,S}}$ is referred to as \textit{diagonal-form FFMA (DF-FFMA)} system.
Note that, the permuting and rearranging operations do not affect the properties of FFMA system.
Actually, the DF-FFMA is a special case of the sparse-form FFMA.
Hence, the DF-FFMA system can be decoded like the aforementioned SF-FFMA system.
For a DF-FFMA system, the bit-sequences of $J$ users are transmitted in orthogonal mode and the $J$ parity sequences are appended to them.

\begin{figure}[t]
  \centering
  \includegraphics[width=0.5\textwidth]{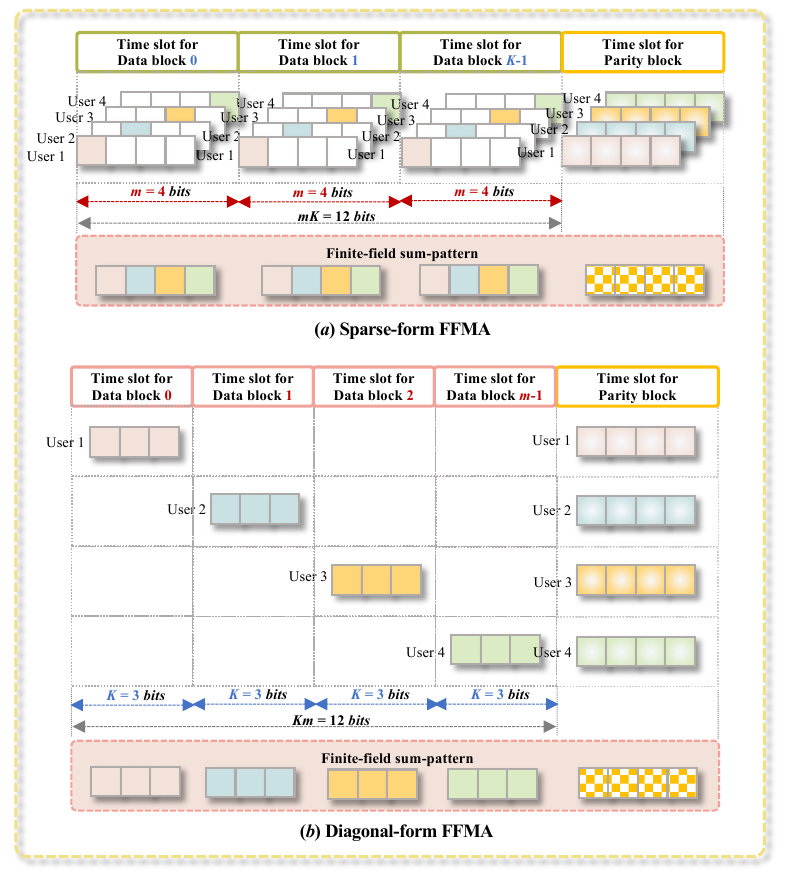}
  \caption{A diagram of the sparse-form and diagonal-form FFMA systems, where $m=4$, $K = 3$ and a $(16, 12)$ channel code ${\mathcal C}_{eg}$ used for error control.} \label{Fig_TimeSlots}
  \vspace{-0.2in}
\end{figure}

\subsection{FFMA for Sourced RA Systems}
For a sourced RA scenario, the blocklength (or DoFs) is assumed to be $N$.
There are a total of $m$ users, with each user being assigned a unique signature to denote their identity. Let the number of active users denote by $J$ where $J \le m$, and each user sends a small packet traffic, i.e., $K = 10 \sim 100$ bits.
If all the $m$ users simultaneously transmit packets, the number of total bits are equal to $mK$.
In this paper, we assume that the frame length is larger than the total bits, i.e., $N > mK$.
Hence, we can construct a binary $(N, mK)$ channel code ${\mathcal C}_{gc}$ with rate $\frac{mK}{N}$, which is utilized for error control.

In the following, we investigate both sparse-form and diagonal-form FFMA systems for  implementing sourced RA in a GMAC.

\subsubsection{SF-FFMA for Sourced RA}
When the sparse-form FFMA is employed to support sourced RA,
an EP $C_j^{\rm td}$ in the orthogonal EP code $\Psi_{\rm o,B}$ is referred to as a signature.
In this case, the orthogonal EP code $\Psi_{\rm o,B}$ is constructed over GF($2^m$),
where $\Psi_{\rm o,B} = \{C_1^{\rm td}, \ldots, C_j^{\rm td}, \ldots, C_m^{\rm td}\}$ with $1 \le j \le m$. 
If the $j$-th user is active, we can obtain an encoded codeword ${\bf v}_j$ for the $j$-th user, as presented earlier. It is noted that the active user occupies the entire frame length.

At the receiving end, we need to recover the active users' bit sequences from the received signals. In general, the detection process consists of the active user detection (AUD) and multiuser detection (MUD).
For a SF-FFMA sourced RA system, implementing the AUD is straightforward, as we only need to detect the total number of active users $J$ by using any clustering algorithms, such as $K$-means clustering.
Next, we do channel decoding algorithm as presented earlier to recover the FFSP blocks, and then extract the active users' information from the field-field.
For instance, if ${\hat b}_{j,k} = 0$ for $0 \le k < K$, it can be inferred that the $j$-th user is not active. This phenomenon is appealing for the sourced RA, since it avoids the complex AUD in complex-field.

\subsubsection{DF-FFMA for Sourced RA}
When employing the diagonal-form FFMA to support sourced RA, we can divide the entire frame into two sections: $m$ data blocks and one parity block.
The data blocklength is $K$, and the parity blocklength is $R = N-mK$.
The $m$ data blocks can be considered as $m$ time slots, similar to the structure of slotted ALOHA. The indices of data blocks range from from $1, 2, \ldots, i, \ldots, m$, and each index is denoted as a signature.
If the $j$-th user is active, its information sequence only occupies the $j$-th data block, while the other data blocks default to $0$s without transmitting signals.
Then, we can obtain a shortened codeword ${\bf v}_{j, {\rm D, S}} = ({\bf b}_j, {\bf v}_{j, \rm D, red})$ for the $j$-th user, as presented earlier, indicating that the active user only occupies a data block and the parity block.

At the receiving end, we can use the same AUD algorithm as that of slotted ALOHA to obtain the number of active users, due to their similar structures. 
Subsequently, employing the channel decoding algorithm can help in recovering the information sequences of the active users.

In summary, FFMA systems for sourced RA generally typically feature two frame structures: sparse-form and diagonal-form, each with its unique features. The diagonal-form is notably more flexible for scenarios with massive users with short data packet transmissions, as it benefits from a reduced blocklength and lower power consumption.
A diagram of the SF-FFMA and DF-FFMA systems is shown in Fig. \ref{Fig_TimeSlots}.

\section{Polarization-adjusted DF-FFMA System}
As previously discussed, users in a DF-FFMA system are characterized by a shortened blocklength and reduced power consumption. In this section, we explore a specific scenario of the DF-FFMA system, referred to as \textit{polarization-adjusted DF-FFMA (PA-DF-FFMA)} or simply \textit{PA-FFMA}.

Let $P_{avg}$ denote the average transmit power per symbol.
For an $(N, mK)$ channel code $\mathcal{C}_{gc}$, the total transmit power per user is $N \cdot P_{avg}$.
In the SF-FFMA system, the entire transmit power of $N \cdot P_{avg}$ is utilized. In contrast, the DF-FFMA system can only utilize a portion of this power, specifically $(K + R) \cdot P_{avg}$, leaving $(m-1)K \cdot P_{avg}$ unused. The proposed PA-FFMA reallocates this unused power. 
By power allocation, the channel capacity is adjusted according to the assigned power, resulting in a form of \textit{polarization}. This is why it is named \textit{polarization-adjusted (or power allocation) FFMA (PA-FFMA)}.

Note that when using LDPC codes as the channel code, the PA-FFMA effectively becomes a \textit{polarization-adjusted LDPC (PA-LDPC)} code when there is only one user, i.e., $J = 1$.

\subsection{Regular and irregular PA-FFMA}

\subsubsection{Irregular PA-FFMA}
We further analyze the shortened codeword ${\bf v}_{j, \rm D, S} = ({\bf b}_{j}, {\bf v}_{j,\rm D,red})$. Since the parity blocklength is $R = N - mK$, the codeword length of ${\bf v}_{j,\rm D,S}$ is $K+R$, or equivalently, the number of symbols is $K+R$.

In the irregular PA-FFMA scheme, different powers are allocated to different symbols of ${\bf v}_{j,\rm D,S}$.
We define the \textit{polarization-adjusted vector (PAV)} as
$\mu_{\rm pav} = (\mu_0, \mu_1, \ldots, \mu_{n_S},\ldots, \mu_{K+R-1})$,
where $0 \le n_S < K+R$. The subscript ``pav'' stands for ``polarization-adjusted vector''.
The power of each symbol of ${\bf v}_{j,\rm D,S}$ is then given by
\begin{equation*}
   P_{avg} \cdot \mu_{\rm pav} 
   = (\mu_0 P_{avg}, \mu_1 P_{avg}, \ldots, \mu_{n_S} P_{avg},\ldots, \mu_{K+R-1} P_{avg}),
\end{equation*}
where $N = \sum_{n_S=0}^{K+R-1} \mu_{n_S}$.
This ensures that the total transmit power of PA-FFMA matches that of SF-FFMA, which is $N \cdot P_{avg}$.

It is evident that irregular PA-FFMA offers significant flexibility, as each bit's power can be customized. 
However, this flexibility introduces design complexity due to the numerous parameters involved.
Therefore, irregular PA-FFMA will be addressed in future work. This paper primarily focuses on the regular PA-FFMA system, which is more straightforward to implement compared to the irregular version.

\subsubsection{Regular PA-FFMA}
Examining the structure of DF-FFMA, we see two distinct sections: the information section and the parity section (or block). 
Assume that the power assigned to the information section per symbol is $\mu_1 P_{avg}$, while the power allocated to the parity section per symbol is $\mu_2 P_{avg}$.
Then, the power allocation (or called polarization-adjusted) conditions are given as
\begin{equation}
  \begin{aligned}
  C 1&: N \cdot P_{avg} = K \cdot (\mu_1 P_{avg}) + R \cdot (\mu_2 P_{avg}),\\
  C 2&: 1  \le \left(\mu_{\rm pas} = \frac{\mu_1}{\mu_2}\right) \le m,
  \end{aligned}
\end{equation}
where $\mu_{\rm pas} = \frac{\mu_1}{\mu_2}$ is defined as \textit{polarization-adjusted scaling factor (PAS)}, and the subscript ``pas'' stands for ``polarization-adjusted scaling''.
Condition $C 1$ ensures that the total transmit power remains constant. 
Condition $C 2$ specifies that the information section receives more power than the parity section. Ideally, we aim to maximize the reliability of the information section by assigning as much power as possible to it.
Additionally, considering the unused power is $(m-1)K \cdot P_{avg}$, the maximum power that can be assigned to the information section per symbol is $m \cdot P_{avg}$, which is equivalent to repeating the information section per symbol $m$ times.

\vspace{-0.1in}
\subsection{Receiver of a PA-FFMA System}

Considering the PA-FFMA structure, the received signal sequence ${\bf y} \in {\mathbb{C}}^{1 \times N}$ can be partitioned into information and parity sections, i.e., ${\bf y} = ({\bf y}_{\rm inf}, {\bf y}_{\rm red})$. Here, ${\bf y}_{\rm inf} = (y_0, y_1, \dots, y_{mK-1})$ represents the information section, consisting of $m$ data blocks, each with a block length of $K$. 

For the information section ${\bf y}_{\rm inf}$, the received signals can be expressed as:
\begin{equation}
  y_n = \mu_1 \cdot (2 b_n - 1) + z_n,
\end{equation}
where $z_n$ is an AWGN with distribution ${\mathcal N}(0, N_0/2)$, and $0 \le n < mK$. Here, $b_n$ is the $n$-th component of the FFSP sequence ${\bf w} = (b_0, b_1, \dots, b_n, \dots, b_{N-1})$, which can also be expressed in block-level form as ${\bf w} = ({\bf b}_1, {\bf b}_2, \dots, {\bf b}_m)$, consisting of $m$ data blocks, each with block length $K$. Clearly, ${\bf y}_{\rm inf}$ represents the modulated signal, i.e., $2 b_n - 1$, that passes through an AWGN channel.

On the other hand, ${\bf y}_{\rm red}$ represents the parity section, i.e., ${\bf y}_{\rm red} = (y_{mK}, y_{mK+1}, \dots, y_{N-1})$, with length $R = N - mK$. The received signals in this section can be written as:
\begin{equation}
  y_n = \mu_2 \cdot \sum_{j = 1}^{m-1} (2 v_{j,n} - 1) + z_n,
\end{equation}
where $mK \leq n < N$. In fact, ${\bf y}_{\rm red}$ represents the superposition signal of $m$ users, which corresponds to the case where $m$ users pass through a GMAC.

\begin{figure}[t]
  \centering
  \includegraphics[width=0.78\textwidth]{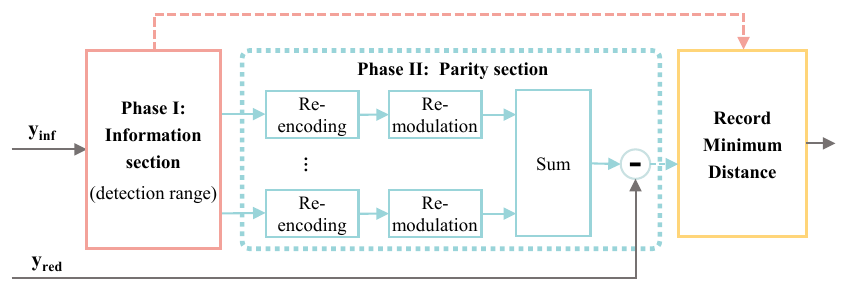}
  \caption{A diagram to show the BMD algorithm.} 
  \label{f.BMD}
  \vspace{-0.2in}
\end{figure}

\vspace{-0.1in}
\subsection{Detection Algorithm of PA-FFMA Systems}

With the increased power allocated to the information section, its reliability improves, leading to a higher degree of polarization. Building on this characteristic, we propose a \textit{bifurcated minimum distance (BMD) detection algorithm} for the PA-FFMA system. The BMD algorithm is structured into two distinct phases:
\begin{itemize}
  \item
  Phase I focuses on the information section for initial screening. This phase identifies a narrower detection range, thereby reducing the complexity of the subsequent detection process.
  \item
  Phase II concentrates on the parity section for refinement. This phase aims to fine-tune the identification process to find the closest optimal solution for the multiuser information sequences.
\end{itemize}
The proposed BMD algorithm is designed to enhance both efficiency and accuracy incrementally, and it approximates \textit{maximum a posteriori (MAP)} detection. Next, we provide a detailed description of the BMD algorithm. 

\subsubsection{\textbf{Phase I}}
In Phase I of the BMD algorithm, the input signals consist of the information section $\mathbf{y}_{\mathrm{inf}} = (y_0, y_1, \ldots, y_{mK-1})$. The \textit{log-likelihood ratio (LLR)} for each $y_n$ ($0 \leq n < mK$) is given by $2\mu_1 y_n/\sigma^2$, leading to the bit decision rule:
\begin{equation}
  \hat{b}_n = 
  \begin{cases} 
  1, & \frac{2\mu_1 y_n}{\sigma^2} \geq 0 \\
  0, & \frac{2\mu_1 y_n}{\sigma^2} < 0 
  \end{cases}.
\end{equation}
The corresponding Euclidean distances for both possible bit values are calculated as:
\begin{equation}
  D_n(\hat{b}_n) = \|y_n -\mu_1 (2\hat{b}_n-1) \|.
\end{equation}

The FFSP sequence $\mathbf{w}$ takes values from the set $\{\underbrace{00\ldots0}_{mK}, \ldots, \underbrace{11\ldots1}_{mK}\}$, comprising all $2^{mK}$ possible binary combinations. The distance metric for a detected FFSP sequence $\hat{\mathbf{w}}$ is computed as:
\begin{equation}
  D_{\mathrm{inf}, \hat{\mathbf{w}}^{\mathrm{dec}}} = \sum_{n=0}^{mK-1} D_n(\hat{b}_n),
\end{equation}
where $\hat{\mathbf{w}} = ({\hat b}_0, {\hat b}_1, \ldots, {\hat b}_n, \ldots,{\hat b}_{mK-1})$ with ${\hat b}_n \in \mathbb{B}$, and $\hat{\mathbf{w}}^{\mathrm{dec}} = {\rm F}_{\mathrm{B2D}}(\hat{\mathbf{w}})$ represents the decimal equivalent of the binary sequence, satisfying $0 \leq {\hat{\mathbf{w}}}^{\mathrm{dec}} \leq 2^{mK}-1$.

The algorithm sorts all $2^{mK}$ distance values $\{D_{\mathrm{inf},0}, \ldots, D_{\mathrm{inf}, 2^{mK}-1}\}$ in ascending order and selects the $L$ smallest distances to form the information distance set:
\begin{equation*}
  \Upsilon_{\mathrm{inf}} = \{D_{\mathrm{inf},{\hat{\mathbf{w}}}_1^{\mathrm{dec}}}, D_{\mathrm{inf},{\hat{\mathbf{w}}}_2^{\mathrm{dec}}}, \ldots, D_{\mathrm{inf},{\hat{\mathbf{w}}}_l^{\mathrm{dec}}}, \ldots, D_{\mathrm{inf},{\hat{\mathbf{w}}}_L^{\mathrm{dec}}}\},
\end{equation*}
where $D_{\mathrm{inf},{\hat{\mathbf{w}}}_1^{\mathrm{dec}}} < D_{\mathrm{inf},{\hat{\mathbf{w}}}_2^{\mathrm{dec}}} < \ldots <  D_{\mathrm{inf},{\hat{\mathbf{w}}}_l^{\mathrm{dec}}} <... < D_{\mathrm{inf},{\hat{\mathbf{w}}}_L^{\mathrm{dec}}}$ for $1 \leq l \leq L \leq 2^{mK}$.

Each distance value $D_{\mathrm{inf},\mathbf{w}_l^{\mathrm{dec}}}$ corresponds to a unique FFSP sequence obtained through the decimal-to-binary transformation:
\begin{equation}
  {\hat{\mathbf{w}}}_l = {\rm F}_{\mathrm{D2B}}({\hat{\mathbf{w}}}_l^{\mathrm{dec}}) = (\hat{\mathbf{b}}_{l,1}, \hat{\mathbf{b}}_{l,2}, \ldots, \hat{\mathbf{b}}_{l,j}, \ldots, \hat{\mathbf{b}}_{l,m}),
\end{equation}
where the binary vector ${\hat{\mathbf{w}}}_l$ is partitioned into $m$ data blocks, and the $j$-th data block ${\hat{\mathbf{b}}}_{l,j}$ is the detected bit-sequence of the $j$-th user. This establishes a one-to-one correspondence between the distance set $\Upsilon_{\mathrm{inf}}$ and the binary vector set
\(
  \Omega_{\hat{\mathbf{w}}} = \{{\hat{\mathbf{w}}}_1, {\hat{\mathbf{w}}}_2, \ldots, {\hat{\mathbf{w}}}_L\},
\)
with $\Upsilon_{\mathrm{inf}} \Leftrightarrow \Omega_{\hat{\mathbf{w}}}$. Both sets serve as inputs to Phase II of the BMD algorithm for subsequent detection processing.

\subsubsection{\textbf{Phase II}}
Phase II of the BMD algorithm utilizes the parity section for detection. Considering the $l$-th FFSP sequence $\hat{\mathbf{w}}_l \in \Upsilon_{\mathrm{inf}}$ as an example, the process applies similarly to all sequences $\hat{\mathbf{w}}_l$ for $l = 1,2,\ldots,L$.

The sequence $\hat{\mathbf{w}}_l$ is decomposed into $m$ binary blocks as 
\(
    \hat{\mathbf{w}}_l = (\hat{\mathbf{b}}_{l,1}, \hat{\mathbf{b}}_{l,2}, \ldots, \hat{\mathbf{b}}_{l,j}, \ldots, \hat{\mathbf{b}}_{l,m}),
\)
where each $\hat{\mathbf{b}}_{l,j}$ represents the $j$-th binary block. These $m$ blocks undergo parallel processing. 
For the $j$-th block $\hat{\mathbf{b}}_{l,j}$, systematic re-encoding is performed using the generator matrix $\mathbf{G}_{gc,\mathrm{sym}}$ of channel code $\mathcal{C}_{gc}$, yielding the encoded codeword:
\begin{equation}
    \hat{\mathbf{v}}_{l,j} = \hat{\mathbf{u}}_{l,j,\mathrm{D}} \cdot \mathbf{G}_{gc,\mathrm{sym}},
\end{equation}
where the input vector $\hat{\mathbf{u}}_{l,j,\mathrm{D}}$ is constructed as
\(
    \hat{\mathbf{u}}_{l,j,\mathrm{D}} = (\mathbf{0}, \ldots, \mathbf{0}, \hat{\mathbf{b}}_{l,j}, \mathbf{0}, \ldots, \mathbf{0}),
\)
where $\hat{\mathbf{b}}_{l,j}$ occupies the $l$-th data block position and all other blocks are zero vectors. 
The resulting codeword $\hat{\mathbf{v}}_{l,j}$ can be partitioned into two components\(\hat{\mathbf{v}}_{l,j} = (\hat{\mathbf{w}}_l, \hat{\mathbf{v}}_{l,j,\mathrm{red}})\), from which we extract the parity block $\hat{\mathbf{v}}_{l,j,\mathrm{red}}$ for subsequent processing.
The parity block is then modulated through the transformation ${\rm F}_{\mathrm{F2C}}$:
\begin{equation}
    \hat{\mathbf{x}}_{l,j,\mathrm{red}} = {\rm F}_{\mathrm{F2C}}(\hat{\mathbf{v}}_{l,j,\mathrm{red}}), \quad 1 \leq j \le m.
\end{equation}
The $m$ modulated signals $\hat{\mathbf{x}}_{l,1,\mathrm{red}}, \hat{\mathbf{x}}_{l,2,\mathrm{red}}, \ldots, \hat{\mathbf{x}}_{l,m,\mathrm{red}}$ are combined to form:
\begin{equation}
    \hat{\mathbf{r}}_{\mathrm{red},l} = \sum_{j=1}^{m} \hat{\mathbf{x}}_{l,j,\mathrm{red}}.
\end{equation}
The distance metric $D_{\mathrm{red},l}$ for $1 \leq l \leq L$ between this reconstructed signal and the received parity section $\mathbf{y}_{\mathrm{red}}$ is computed as:
\begin{equation}
    D_{\mathrm{red},l} = \|\mathbf{y}_{\mathrm{red}} - \hat{\mathbf{r}}_{\mathrm{red},l}\| = \left\|\mathbf{y}_{\mathrm{red}} - \sum_{j=0}^{m-1} \hat{\mathbf{x}}_{l,j,\mathrm{red}}\right\|.
\end{equation}

The total distance metric combines both information and parity sections:
\begin{equation}
    D_{\mathrm{sum},l} = D_{\mathrm{inf},l} + D_{\mathrm{red},l}.
\end{equation}

The final detected FFSP sequence corresponds to the minimum combined distance:
\begin{equation}
    \hat{\mathbf{w}} = \underset{1 \leq l \leq L}{\operatorname{argmin}} \, D_{\mathrm{sum},l}.
\end{equation}

Figure \ref{f.BMD} summarizes the complete BMD algorithm. The algorithm's error performance and computational complexity are primarily governed by the iteration parameter $L$, where increasing $L$ improves detection accuracy at the expense of higher computational load.


\section{A Network FFMA System for a Digital Network Scenario}
In this section, we present FFMA systems for a digital network scenario.
Since AIEPs consist of integers, elements from a prime field GF($p$) or powers of a primitive element from an extension field GF($p^m$) of a prime field GF($p$), they are naturally suitable for applications in a digital network.

\subsection{Virtual Resource Block}

As mentioned earlier, EPs can be regarded as virtual resources, with each EP constituting a \textit{virtual resource block (VRB)}.
Before we present network FFMA systems, we relook some of the properties of EPs developed in Sections II and III.

As shown in Section II, for a prime field GF($p$) with $p > 2$, we can construct an AIEP code ${\Psi}_{\rm s} = C_1^{\rm s} \times \ldots \times C_l^{\rm s} \ldots \times C_L^{\rm s}$ with $C_l^{\rm s} = (l, p-1)$ which is the $l$-th AIEP of the AIEP code ${\Psi}_{\rm s}$.
Each AIEP $C_l^{\rm s}$ forms a VRB and the number $L$ of VRBs in ${\Psi}_{\rm s}$ is determined by the designed FF-MUX for an AIEP code.
If ${\Psi}_{\rm s}$ is an UD-AIEP code, the number $L$ of VRBs is at most $L = \log_2(p-1)$.
If ${\Psi}_{\rm s}$ is an AIEP code without USPM structural property, then $L$ is at most $(p-1)/2$.
In this case, VRBs in an AIEP code ${\Psi}_{\rm s}$ only depends on the parameter $p$, which is called \textit{prime factor (PF)}.

\begin{figure}[t]
  \centering
  \includegraphics[width=0.45\textwidth]{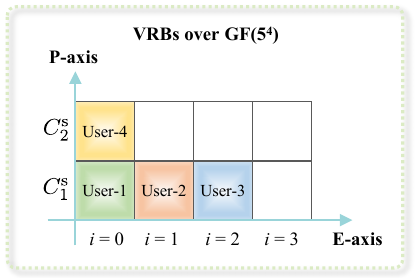}
  \caption{The VRBs over GF($5^4$), which also shows the VRB allocation of Example 6. In Example 6, user-$1$, user-$2$, user-$3$, and user-$4$ are assigned the UD-EPs $\alpha^0 \cdot C_1^{\rm s}$, $\alpha^1 \cdot C_1^{\rm s}$, $\alpha^2 \cdot C_1^{\rm s}$, and $\alpha^0 \cdot C_2^{\rm s}$, respectively.} 
  \label{Fig_Eg6}
  \vspace{-0.2in}
\end{figure}

For an extension field GF($2^m$) with $m > 1$, 
we can construct an $m$-dimension orthogonal UD-EP code ${\Psi}_{\rm o,B}$, 
i.e., ${\Psi}_{\rm o,B} = \{C_{1}^{\rm td}, \ldots, C_{i+1}^{\rm td}, \ldots, C_{m}^{\rm td}\}$, 
where $C_{i+1}^{\rm td} = \alpha^{i} \cdot C_{\rm B}$ and $0 \le i < m$.
The difference between UD-EPs in ${\Psi}_{\rm o,B}$ is determined by the location $i$.
Thus, we denote the location $i$ of $\alpha^{i} \cdot C_{\rm B}$ as a VRB.
The number of VRBs is equal to $m$, which is called \textit{extension factor (EF)}.

For an extension field GF($p^m$) with $p > 2$ and $m > 1$, 
an $m$-dimension orthogonal AIEP code $\Psi_{\rm o}$, i.e., $\Psi_{\rm o} = \{\Psi_{{\rm o},0}, \Psi_{{\rm o},1}, \ldots, \Psi_{{\rm o},i}, \ldots, \Psi_{{\rm o},m-1}\}$, is an $m$-dimension orthogonal AIEP code over GF($p$) in which each $\alpha^i \cdot C_l^{\rm s}$ of 
$\Psi_{{\rm o},i} = \{\alpha^i \cdot C_1^{\rm s}, \alpha^i \cdot C_2^{\rm s}, \ldots,  \alpha^i \cdot C_l^{\rm s},\ldots, \alpha^i \cdot C_L^{\rm s}\}$ is an AIEP over GF($p$). 
An $m$-dimension orthogonal AIEP code $\Psi_{\rm o}$ may be viewed as a two concatenated code, including both the orthogonal UD-EP code $\Psi_{\rm o, B}$ constructed over GF($2^m$) and the AIEP code ${\Psi}_{\rm s}$ constructed over GF($p$). 
Thus, an UD-AIEP $\alpha^i \cdot C_l^{\rm s}$ in $\Psi_{\rm o}$ is determined by both the location $i$ and the UD-AIEP $C_l^{\rm s}$. 
The number of VRBs of the finite-field GF($p^m$) is equal to $m \cdot L$.

Hence, for a given finite-field GF($p^m$) where $m > 2$, 
the VRBs can be represented by a Cartesian coordinate system, with the horizontal axis denoted as E-axis and the vertical axis as P-axis, as shown in Fig. \ref{Fig_Eg6}.
Actually, the proposed AIEP code $\Psi_{\rm s}$ over GF($p$) and orthogonal AIEP code $\Psi_{\rm o}$ over GF($p^m$) can be considered as constructed using the \textit{bit-to-symbol transform approach}, where each user can be assigned one or several VRB(s) for distinction.

\textbf{Example 6:} 
In Example 3, we showed that a total $8$ UD-AIEPs can be construct based on the field GF($5^4$). Suppose we want to design an FFMA system to support $J = 4$ users, and user-$1$, user-$2$, user-$3$, and user-$4$ are respectively assigned the UD-AIEPs $\alpha^0 \cdot C_1^{\rm s}$, $\alpha^1 \cdot C_1^{\rm s}$, $\alpha^2 \cdot C_1^{\rm s}$, and $\alpha^0 \cdot C_2^{\rm s}$,
as shown in \ref{Fig_Eg6}.
With the above assignment, user-$1$, user-$2$ and user-$3$ have the same AIEP $C_1^{\rm s}=\left(1,4\right)$ of $\Psi_{\rm s}$, but are with different AIEPs of $\Psi_{\rm o,B}$ (e.g., $i=0, i=1$ and $i=2$) to distinct users. 
In addition, user-$1$ and user-$4$ have the same AIEP of $\Psi_{\rm o,B}$ (e.g., $i=0$), but are with different AIEPs of $\Psi_{\rm s}$ (e.g., $C_1^{\rm s}$ and $C_2^{\rm s}$), thus user-$1$ and user-$4$ can be separated. 
$\blacktriangle \blacktriangle$

\vspace{-0.2in}
\subsection{An overview of FFMA systems} 

The conventional CFMA (Complex-Field Multiple Access) system transmitter comprises four fundamental components: 
source encoder, channel code encoder, modulation module, and CF-MUX (determined by physical resource blocks).

In contrast, our proposed FFMA system architecture features a more streamlined transmitter structure with the following components: 
element-pair (EP) encoder, optional channel encoder, finite-field to complex-field transform function (${\rm F}_{\mathrm{F2C}}$), and optional power allocation module.

The EP encoder serves multiple functions, effectively combining the roles of source encoding, channel encoding (when needed), and multiuser distinction. This multi-functional capability stems from its unique algebraic structure in finite fields.
EP codes can be implemented in two distinct forms. The \textit{symbol-wise} construction, used in the current work, implements EP codes through bit-to-symbol transformation where each output element is generated in symbol-wise form. This approach provides the fundamental building blocks for our FFMA system. 
Our ongoing research explores \textit{codeword-wise} EP construction where single or multiple bits are mapped to codewords, with each output element generated in codeword-wise form. This advanced construction enables additional features including finite-field and/or complex-field spreading, error correction capability, and enhanced spectral efficiency as a type of non-orthogonal spreading code.
The EP encoder achieves a unified framework where bits and users are treated equivalently in finite-field operations, combining source encoding and multiuser encoding into a single processing stage. This framework naturally extends to $p$-ary sources, where the EP code becomes a special case of our more general \textit{element-assemblage (EA) code}. The EA code more explicitly resembles a traditional source encoder while maintaining the advantages of finite-field processing.

Next, we employ ${\rm F}_{\mathrm{F2C}}$ transform function rather than conventional modulation due to its greater flexibility. While traditional modulation uses simple one-to-one mapping, ${\rm F}_{\mathrm{F2C}}$ supports more sophisticated mappings including both conventional modulation schemes and advanced matrix transformations. This enhanced flexibility better suits the requirements of FFMA systems, particularly in multiuser scenarios.

The optional power allocation module serves to balance error performance across users, optimize system performance, and compensate for channel variations. The design of power allocation schemes remains an important area for future research in FFMA systems, particularly when combined with the finite-field processing advantages of our approach.




\begin{figure}[t]
  \centering
  \includegraphics[width=0.9\textwidth]{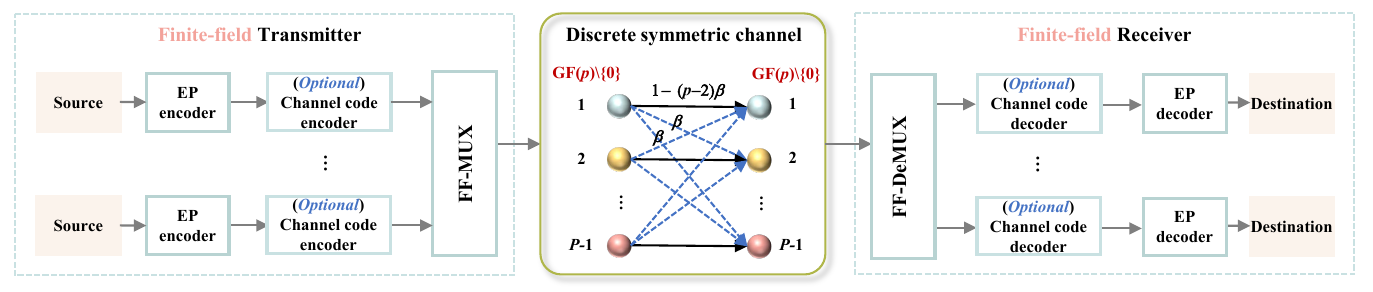}
  \caption{A diagram of a network FFMA system for a pure digital network scenario over a DSC.} \label{Fig_NetFFMA}
  \vspace{-0.2in}
\end{figure}

\vspace{-0.1in}
\subsection{A network FFMA over GF($p^m$) system}
In the following, we present a network FFMA over GF($p^m$) system for a digital network, as shown in Fig. \ref{Fig_NetFFMA}.
Suppose the system is designed to support $J$ users. 
We choose an orthogonal AIEP code $\Psi_{\rm o} = \{\Psi_{{\rm o},0}, \Psi_{{\rm o},1}, \ldots, \Psi_{{\rm o},i}, \ldots, \Psi_{{\rm o},m-1}\}$. 
Assume the $j$-th user is assigned the AIEP $\alpha^i \cdot C_l^{\rm s}$, 
i.e.,
\begin{equation*}
\alpha^i \cdot C_l^{\rm s} = \{(0, 0, \ldots, l, 0, \ldots, 0), (0, 0, \ldots, p-l, 0, \ldots, 0)\},
\end{equation*}
where $0 \le i < m$, $1 \le l \le L=(p-1)/2$ and $J \le m \cdot L$.


Let ${\bf b}_j = (b_{j,0}, b_{j,1}, \ldots, b_{j,k},\ldots, b_{j,K-1})$ be the bit-sequence at the output of user-$j$, where $b_{j,k} \in {\mathbb B}$ and $0 \le k < K$. 
Based on the orthogonal EP code $\Psi_{\rm o,B}$, we first encode the bit-sequence ${\bf b}_j$ by using the ${\rm F}_{{\rm B}2q}$ transform function,
i.e., $u_{j,k} = {\rm F}_{{\rm B}2q}(b_{j,k})$, 
into an element-sequence ${\bf u}_j$ over GF($2^m$).
Next, the element-sequence ${\bf u}_j$ is encoded by a binary linear block code ${\mathcal C}_{gc}$ into a binary codeword ${\bf v}_j = (v_{j,0}, v_{j,1},\ldots, v_{j,n}, \ldots, v_{j,N-1})$, where $v_{j,n} \in {\mathbb B}$ and $0 \le n < N$. The encoding process of $\Psi_{\rm o,B}$ and ${\mathcal C}_{gc}$ are the same as the process in Section IV.

After $\Psi_{\rm o,B}$/${\mathcal C}_{gc}$-encoding, we cluster the codewords ${\bf v}_j$ into $L$ groups.
If the users $j_1, j_2, \ldots, j_{c_l}$ are assigned to the same UD-AIEP $C_l^{\rm s}$ in $\Psi_{\rm s}$ but at different locations in $\Psi_{\rm o,B}$, i.e., $\alpha^i \cdot C_l^{\rm s}$ for $0 \le i < m$, the codewords of the users $j_1, j_2, \ldots, j_{c_l}$ are clustered into one group, i.e., 
${\bf V}_{c_l} = \{ {\bf v}_{j_1}, {\bf v}_{j_2}, \ldots, {\bf v}_{j_{c_l}}\}$.
The FFSP of the codewords in ${\bf V}_{c_l}$ is 
\begin{equation*}
{\bf v}_{c_l} = \bigoplus_{j \in {\bf V}_{c_l}} {\bf v}_j 
= {\bf v}_{j_1} \oplus {\bf v}_{j_2} \oplus \ldots \oplus {\bf v}_{j_{c_l}},
\end{equation*}
where the subscript $l$ of ``$c_l$'' is ranged from $1 \le l \le L$,
and ${\bf v}_{c_l} = (v_{c_l,0}, v_{c_l,1}, \ldots, v_{c_l,n}, \ldots, v_{c_l,N-1})$.

The $L$ FFSP codewords ${\bf v}_{c_1}, {\bf v}_{c_2}, \ldots, {\bf v}_{c_L}$ correspond to the $L$ AIEPs in $\Psi_{\rm s}$.
Then, each component $v_{c_l,n}$ in ${\bf v}_{c_l}$ is mapped into an element $s_{c_l,n}$ in $C_l$, i.e., $s_{c_l,n} = {\rm F}_{{\rm B}2q}(v_{c_l,n})$, given by
\begin{equation}
  s_{c_l,n} = \left\{
    \begin{matrix}
      (l)_p,   & v_{c_l,n} = (0)_2 \\
      (p-l)_p, & v_{c_l,n} = (1)_2 \\
    \end{matrix}
    \right.
\end{equation}
where $C_l^{\rm s} = (l, p-l)$ is an AIEP of $\Psi_{\rm s}$.
The mapping ${\rm F}_{{\rm B}2q}$ of the components of ${\bf v}_{c_l}$ results in a sequence
${\bf s}_{c_l} = (s_{c_l,0}, s_{c_l,1}, \ldots, s_{c_l,n}, \ldots, s_{c_l,N-1})$ over GF($p$).

Next, the $L$ sequences ${\bf s}_{c_1}, {\bf s}_{c_2}, \ldots, {\bf s}_{c_L}$ 
are multiplexed into a sequence ${\bf x} = (x_0, x_1, \ldots, x_{N-1})$ over GF($p$) by the FF-MUX ${\mathcal A}_{\rm M}$. 
If ${\Psi}_{\rm s}$ is an UD-AIEP code, ${\mathcal A}_{\rm M}$ is a finite-field addition operation, 
where $x_n = \bigoplus_{l=1}^{L} s_{c_l, n}$ and $x_n \in {\rm GF}(p)\backslash \{0\}$. 
If ${\Psi}_{\rm s}$ is not an UD-AIEP code, we should design ${\mathcal A}_{\rm M}$ for different finite-fields, e.g., C-UDEPs method.

The multiplexed sequence $\bf x$ is a sequence of FFSPs of the orthogonal AIEP code $\Psi_{\rm o}$ over GF($p^m$), which are transmitted over a discrete symmetric channel (DSC). Consequently, both the input and output discrete digits of the DSC are drawn from ${\rm GF}(p)\backslash \{0\} = \{1, 2, \ldots, p-1\}$.

Let ${\bf y} = (y_0, y_1, \ldots, y_n, \ldots, y_{N-1})$ be the received sequence, where $y_n \in {\rm GF}(p)\backslash \{0\}$. The transition probability of the DSC is defined as
\begin{equation}
  P(y_n|x_n) = \left\{
  \begin{matrix}
    \beta,           & y_n \neq x_n\\
    1 - (p-2)\beta,  & y_n = x_n\\
  \end{matrix},
  \right.
\end{equation}
where $\beta$ is decided by the practical physical channel.

To recover the transmit bit-sequences, we first recover the $L$ AIEP coded sequences $\hat {\bf s}_{c_1}, \hat {\bf s}_{c_2}, \ldots, \hat {\bf s}_{c_L}$ from ${\bf y}$ by lookup FFSP decoding tables, as illustrated in Figs. 1 and 2. 
Next, the $L$ AIEP coded sequences $\hat {\bf s}_{c_1}, \hat {\bf s}_{c_2}, \ldots, \hat {\bf s}_{c_L}$ are decoded through the inverse function of ${\rm F}_{{\rm B}2q}$, denoted as ${\rm F}_{q2{\rm B}}$, i.e., $\hat {\bf v}_{c_l} = {\rm F}_{q2{\rm B}}(\hat {\bf s}_{c_l})$, to obtain the binary codewords $\hat {\bf v}_{c_1}, \hat {\bf v}_{c_2}, \ldots, \hat {\bf v}_{c_L}$.
Subsequently, channel decoding is performed on the codewords $\hat {\bf v}_{c_1}, \hat {\bf v}_{c_2}, \ldots, \hat {\bf v}_{c_L}$, to obtain the sequences $\hat {\bf u}_{c_1}, \hat {\bf u}_{c_2}, \ldots, \hat {\bf u}_{c_L}$.
Finally, by decoding the UD-AIEP code $\Psi_{\rm o,B}$, we recover all the transmitted bit-sequences $\hat {\bf b}_j$. 
The decoding process follows a methodology similar to that presented in Section IV.

\textbf{Example 7:} 
For a given finite-field GF($5^4$), we design a network FFMA system to support $J = 4$ users, and the VRB allocation scheme of the $4$ users has been given in Example 6.
We use the $(16, 12)$ linear block code ${\mathcal C}_{eg}$ whose generator matrix ${\bf G}_{eg}$ has been given in Example 5. 
Suppose each user transmits $K = 3$ at a time.
Let ${\bf b}_1 = (1, 1, 0)$, ${\bf b}_2 = (1, 0, 1)$, ${\bf b}_3 = (0, 0, 1)$, and ${\bf b}_4 = (0, 1, 0)$ be bit-sequences to be transmitted by the $4$ users, respectively. Transmission of these $4$ bit-sequences consists of $4$ steps.

In the first step, each bit $b_{j,k}$ is encoded by EP code $\Psi_{\rm o,B}$ into a symbol $u_{j,k} = {\rm F}_{{\rm B}2q}(b_{j,k})$. Then, encode the element-sequence ${\bf u}_j$ into a codeword in the channel code ${\mathcal C}_{eg}$, i.e., ${\bf v}_{j} = {\bf u}_j \cdot {\bf G}_{eg}$ as shown below.

  \begin{equation*} \label{e.Ex7_1}
  \begin{array}{ll}
  \alpha^0 \cdot C_1^{\rm s}: {\bf b}_1 = (1, 1, 0) &\Rightarrow
  {\bf u}_1 = (\textcolor{red}{1} 0 0 0, \textcolor{red}{1} 0 0 0, \textcolor{red}{0} 0 0 0) \\
  &\Rightarrow 
  {\bf v}_1 = {\bf u}_1 \cdot {\bf G}_{eg}  =
  (\textcolor{red}{1} 0 0 0, \textcolor{red}{1} 0 0 0, \textcolor{red}{0} 0 0 0, \textcolor{blue}{1 0 0 1})
  \\
  \alpha^1 \cdot C_1^{\rm s}: {\bf b}_2 = (1, 0, 1) &\Rightarrow
  {\bf u}_2 = (0 \textcolor{red}{1} 0 0, 0 \textcolor{red}{0} 0 0, 0 \textcolor{red}{1} 0 0) \\
  &\Rightarrow 
  {\bf v}_2 = {\bf u}_2 \cdot {\bf G}_{eg}  =
  (0 \textcolor{red}{1} 0 0, 0 \textcolor{red}{0} 0 0, 0 \textcolor{red}{1} 0 0, \textcolor{blue}{0 1 0 1})
  \\
  \alpha^2 \cdot C_1^{\rm s}: {\bf b}_3 = (0, 0, 1) &\Rightarrow
  {\bf u}_3 = (0 0 \textcolor{red}{0} 0, 0 0 \textcolor{red}{0} 0, 0 0 \textcolor{red}{1} 0) \\
  &\Rightarrow 
  {\bf v}_3 = {\bf u}_3 \cdot {\bf G}_{eg}  =
  (0 0 \textcolor{red}{0} 0, 0 0 \textcolor{red}{0} 0, 0 0 \textcolor{red}{1} 0, \textcolor{blue}{1 0 0 0})
  \\
  \hdashline
  \alpha^0 \cdot C_2^{\rm s}: {\bf b}_4 = (0, 1, 0) &\Rightarrow
  {\bf u}_4 = (\textcolor{red}{0} 0 0 0, \textcolor{red}{1} 0 0 0, \textcolor{red}{0} 0 0 0) \\
  &\Rightarrow 
   {\bf v}_4 = {\bf u}_4 \cdot {\bf G}_{eg}  =
  (\textcolor{red}{0} 0 0 0, \textcolor{red}{1} 0 0 0, \textcolor{red}{0} 0 0 0, \textcolor{blue}{0 0 0 1})
  \\  
  \end{array}.
  \end{equation*}
 
In the second step, we cluster the codewords ${\bf v}_j$ into 
${\bf V}_{c_1} =\{{\bf v}_1, {\bf v}_2, {\bf v}_3 \}$ and 
${\bf V}_{c_2} =\left\{ {\bf v}_4 \right\}$, 
and compute the FFSPs ${\bf v}_{c_1}$ and ${\bf v}_{c_2}$ of the codewords in ${\bf V}_{c_1}$ and ${\bf V}_{c_2}$, as follows:
  \begin{equation*}
   \begin{aligned}
   {\bf v}_{c_1} &= \bigoplus_{j=1}^{3} {\bf v}_j = (1 1 0 0, 1 0 0 0, 0 1 1 0, 0 1 0 0), \\
   {\bf v}_{c_2} &= {\bf v}_4 = ({0} 0 0 0, {1} 0 0 0, {0} 0 0 0, {0 0 0 1}).\\
    \end{aligned}
  \end{equation*}

Next, we encode ${\bf v}_{c_1}$ and ${\bf v}_{c_2}$ by the AIEPs $C_1^{\rm s}$ and $C_2^{\rm s}$, i.e., ${\bf s}_{c_1} = {\rm F}_{{\rm B}2q}({\bf v}_{c_1})$ and ${\bf s}_{c_2} = {\rm F}_{{\rm B}2q}({\bf v}_{c_2})$ as shown below.
  \begin{equation*}
   \begin{aligned}
  {\bf s}_{c_1} = (4 4 1 1, 4 1 1 1, 1 4 4 1, 1 4 1 1)_5,\\
  {\bf s}_{c_2} = (2 2 2 2, 3 2 2 2, 2 2 2 2, 2 2 2 3)_5.\\
    \end{aligned}
  \end{equation*}
Following, we multiplex ${\bf s}_{c_1}$ and ${\bf s}_{c_2}$ into the sequence ${\bf x}$,
  \begin{equation*}
    {\bf x} = {\bf s}_{c_1} \oplus {\bf s}_{c_2} = (1 1 3 3, 2 3 3 3, 3 1 1 3, 3 1 3 4)_5.
  \end{equation*}
Then, the multiplexed sequence ${\bf x}$ is transmitted.

At the receiving end, through the FFSP decoding table given by Fig. 1, we can recover ${\bf s}_{c_1}$ and ${\bf s}_{c_2}$ from ${\bf x}$. Then, do opposite operations of the transmitter, the bit-sequences can be recovered. $\blacktriangle  \blacktriangle$

\section{Simulation results}

In this section, we show the error performance of our proposed FFMA systems.
As presented in Sect. IV, the transmitter consists of an EP code $\Psi_{\rm o,B}$, a channel code ${\mathcal C}_{gc}$, and a transform function ${\rm F}_{\rm F2C}$. The channel code can be either a non-binary (NB) channel code or a binary channel code. Additionally, the generator matrix of the channel code ${\mathcal C}_{gc}$ can take both systematic form ${\bf G}_{gc, sym}$ and non-systematic form ${\bf G}_{gc, nonsym}$. 
Hence, we construct an NB-LDPC code ${\mathcal C}_{gc, q}$ over GF($q$) and two binary LDPC codes ${\mathcal C}_{gc, b1}$ and ${\mathcal C}_{gc, b2}$ for the proposed FFMA systems.

We begin by investigating the effects of systematic and non-systematic generator matrices for the channel code ${\mathcal C}_{gc}$. 
Next, we will present the error performance of each configuration of FFMA systems and compare it with that of slotted ALOHA and IDMA systems over a GMAC.
Finally, we will evaluate the error performance of network FFMA systems over a DSC.


\begin{figure}[t]
  \centering
  \includegraphics[width=0.5\textwidth]{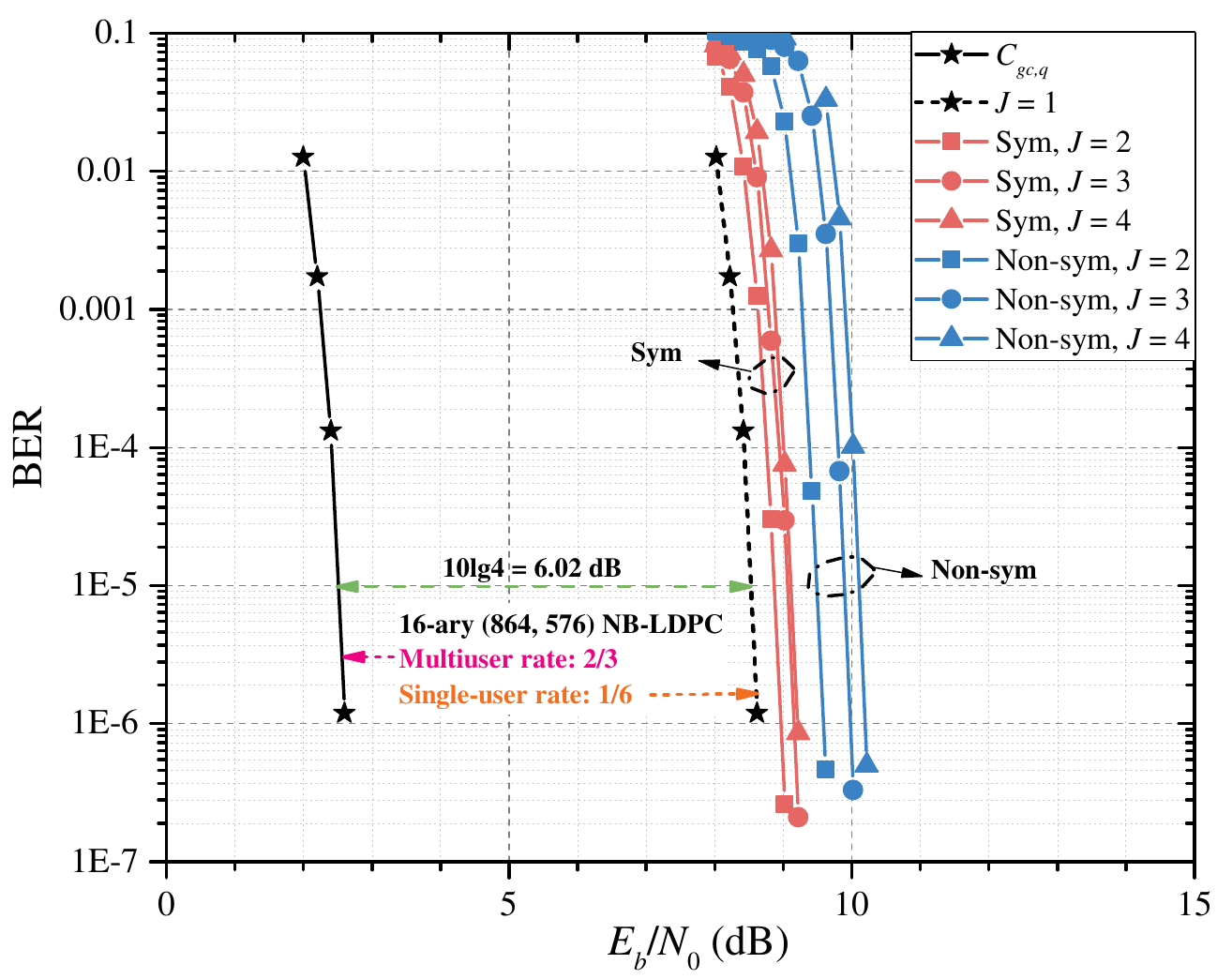}
  \caption{The BER performance of the SF-FFMA over GF($2^4$) in a GMAC, where $J = 1, 2, 3, 4$ and a $16$-ary $(864, 576)$ NB-LDPC code ${\mathcal C}_{gc, q}$ is used for error control.}
  \label{f.NB}
  \vspace{-0.25in}
\end{figure}

\subsection{Systematic and Non-systematic Generator Matrices}
First, we examine the effects of systematic and non-systematic generator matrices when an NB-LDPC code ${\mathcal C}_{gc, q}$ is used for error control.
Fig. \ref{f.NB} shows the BER performance of the SF-FFMA system in a GMAC, 
which includes both systematic and non-systematic forms of the generator matrix of ${\mathcal C}_{gc, q}$.

An orthogonal UD-EP code $\Psi_{\rm o,B}$ is constructed over GF($2^4$), so that the system can support up to $4$ users. 
We set $Q = q = 2^4$ and construct a $16$-ary $(864, 576)$ NB-LDPC code ${\mathcal C}_{gc, q}$ over GF($2^4$) with a rate of $R_c = \frac{2}{3}$. Since the system can support a maximum of four users, we have $R_c = R_{\rm MU} = 4 R_{\rm SU}$ and $R_{\rm SU} = \frac{1}{6}$.
In simulating the error performance of the system, decoding is carried out with $50$ iterations of the FFT-QSPA \cite{LinBook3}.


From Fig. \ref{f.NB}, we see that the system in systematic form performs better than the system in non-systematic form, because more accurate probabilities of the elements $\Omega_r$ are used.
In fact, the BER performance very much depends on the probabilities of the elements in $\Omega_r$. Two different sets of probabilities of the elements in $\Omega_r$ may results in a big gap in decoding of the error control code ${\mathcal C}_{gc, q}$. To show this, we consider both the non-systematic form and systematic form with $J = 4$. 
In non-systematic form, we directly exploit the given $\Omega_r$ and ${\mathcal P}_r$, i.e., $\Omega_r=\{-4,-2,0,+2,+4\}$ and ${\mathcal P}_r= \{0.0625,0.25,0.375,0.25,0.0625\}$.
For the systematic form, we can set $\Omega_r = \{-4,-2\}$ and ${\mathcal P}_r = \{0.5,0.5\}$ to the first $K_Q = 576$ received information symbols; and $\Omega_r = \{-4,-2,0,+2,+4\}$ and ${\mathcal P}_r=\{0.0625,0.25,0.375,0.25,0.0625\}$ to the next $N_Q-K_Q=288$ received parity symbols. Hence, the generator matrix of the channel code in systematic form is preferred for our proposed FFMA systems.
Additionally, it was observed that the BER performance slightly deteriorates as the number of users increases.

When $J = 1$, the required $E_b/N_0$ for the SF-FFMA system in a GMAC is $10 \cdot \lg 4 = 6.02$ dB higher than that of the NB-LDPC code ${\mathcal C}_{gc, q}$ in an AWGN channel to achieve the same BER. This increase in $E_b/N_0$ is attributed to the varying data rate, specifically from $2/3$ to $1/6$.
However, according to Shannon's theory, a channel code with a lower rate generally requires a lower $E_b/N_0$ to achieve the same error performance. This phenomenon further highlights the conflict between designing multiuser channel codes and single-user channel codes.


In an FFMA system utilizing an NB-LDPC code for error control, supporting a large number of simultaneous users can be challenging due to the high decoding complexity. Consequently, a binary LDPC code is often preferred in practice. Therefore, in the following discussion, we will focus exclusively on binary LDPC codes for error control.

\begin{figure}[t]
  \centering
  \includegraphics[width=0.5\textwidth]{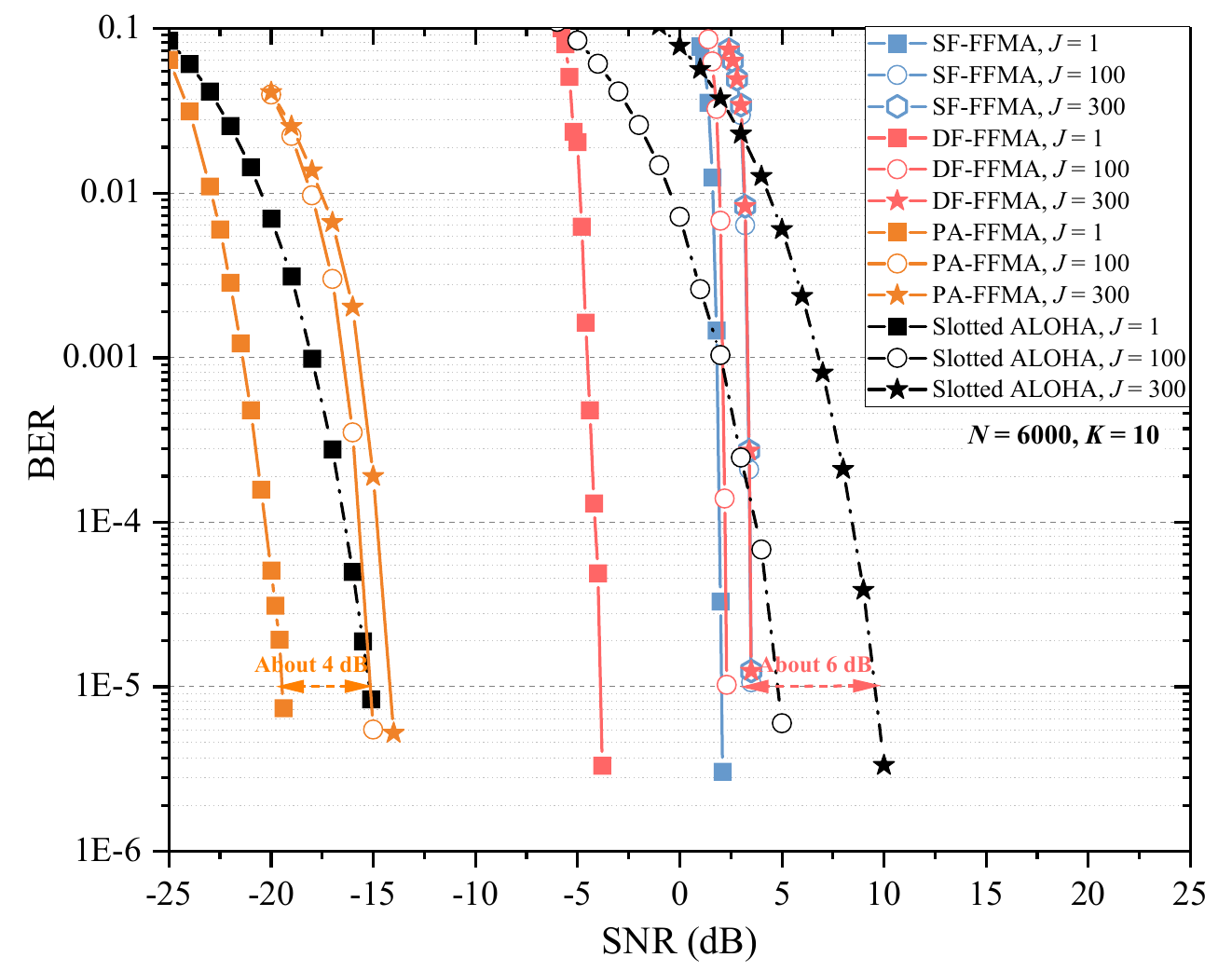}
  \caption{The BER performances of different FFMA systems in a GMAC. The FFMA systems are used a binary $(6000, 3000)$ LDPC code ${\mathcal C}_{gc, b1}$ in systematic form for error control, and the slotted ALOHA system utilizes repetition code for error control, where $N = 6000$, $K = 10$ bits, and $J = 1, 100, 300$ for comparisons.}
  \label{f.TDMA}
  \vspace{-0.25in}
\end{figure}

\subsection{Error Performance of FFMA Systems over a GMAC}
\begin{figure*}[t!]
  \centering
  \subfloat[${\mathcal C}_{gc,b1}$ with $R_c = 0.5$.]{\includegraphics[width=0.5\textwidth]{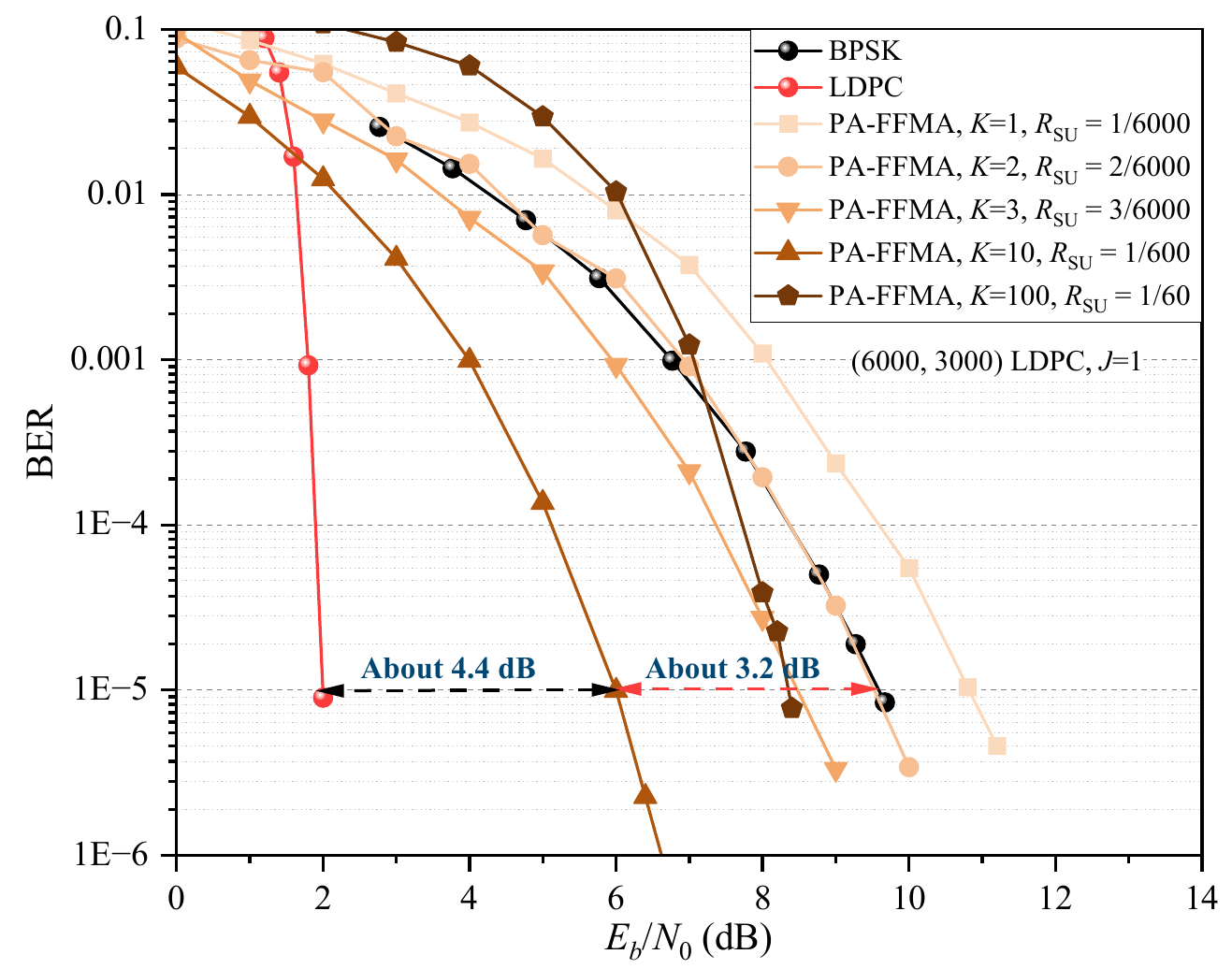}%
  \label{LDPC_sub1}}
  \subfloat[${\mathcal C}_{gc,b2}$ with $R_c = 0.84$.]{\includegraphics[width=0.5\textwidth]{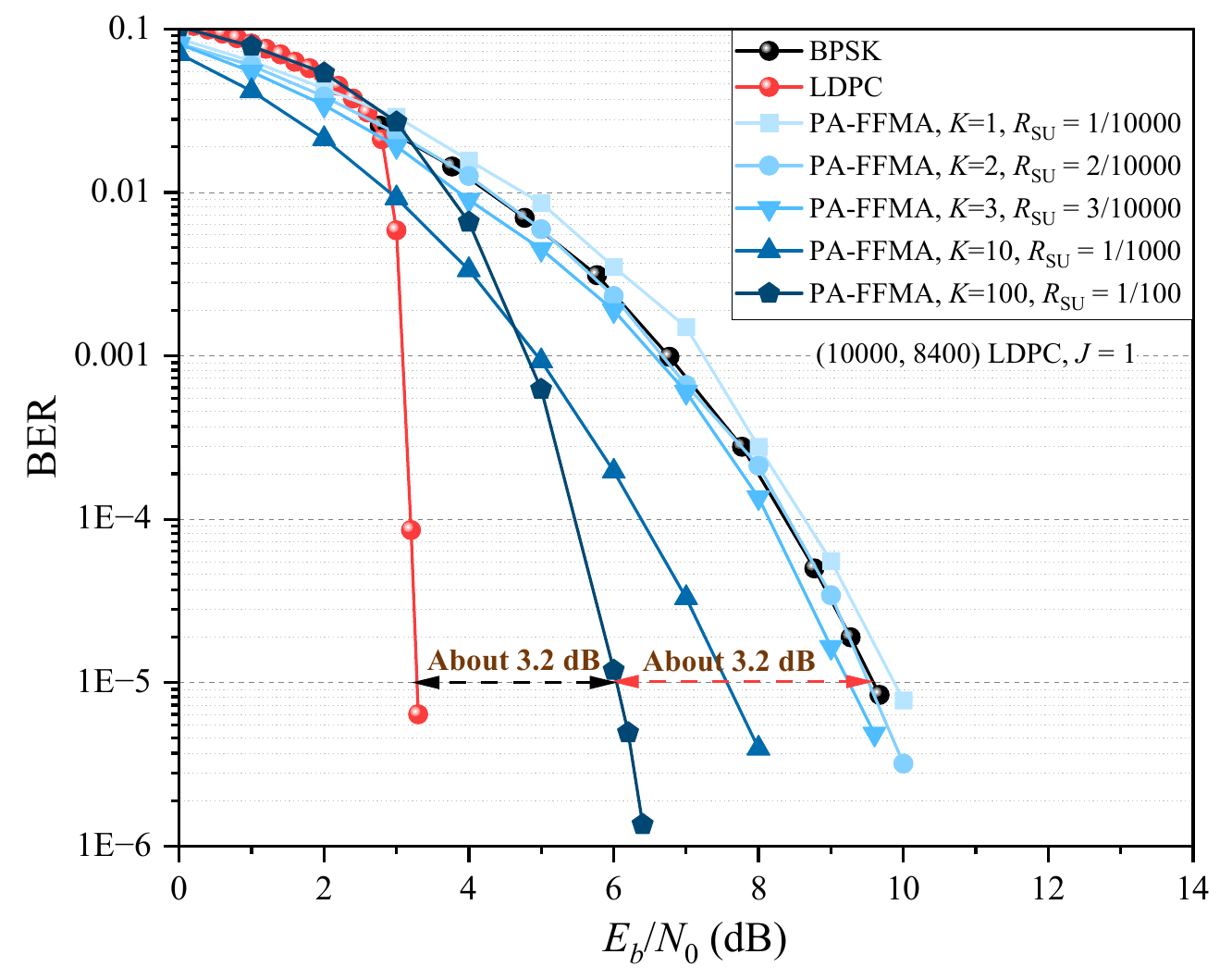}%
  \label{LDPC_sub2}}
  \caption{BER performances of low-rate channel codes constructed based on PA-FFMA.During the simulations, we set $J =1$ and $K = 1, 2, 3, 10, 100$ bits/user for comparison.}
  \label{f.LDPC}
  \vspace{-0.2cm}
\end{figure*}

In the following sub-section, we present the error performances of our proposed FFMA systems in a GMAC, as illustrated in Figs. \ref{f.TDMA}, \ref{f.LDPC} and \ref{f.IDMA}.

\subsubsection{Comparisons between slotted ALOHA and FFMA sourced RA systems}

First, we assume a total number of resources $N = 6000$, with each user transmitting $K = 10$ bits. 
The maximum number of serving users is set to $m = 300$, and the number of arriving users is $J = 1, 100,$ and $300$ where $J \le m$. 
For the FFMA systems, we construct a binary $(6000, 3000)$ LDPC code ${\mathcal C}_{gc,b1}$ for error control, with the generator matrix of ${\mathcal C}_{gc,b1}$ in systematic form.

Additionally, we assume that the AUD can successfully recover the number of arriving users, allowing us to ignore the effects of AUD. This assumption is reasonable since AUD is applied uniformly across all sourced MA systems; thus, the error performances caused by AUD are consistent for all MA systems.
The other simulation conditions for various MA systems are outlined as follows:
\begin{itemize}
  \item
  For the SF-FFMA systems, we construct an orthogonal UD-EP code $\Psi_{\rm o,B}$ over GF($2^{300}$), where $m = 300$. This code consists of $300$ orthogonal EPs, capable of accommodating up to $300$ users. Each user is assigned a unique EP code as their identity, occupying the entire frame length $N = 6000$. Decoding is performed at the receiving end using $50$ iterations of the min-sum algorithm (MSA).
  \item
  In the DF-FFMA systems, the frame comprises $m = 300$ data blocks and one parity block. The data blocklength and parity blocklength are $K = 10$ and $R = 3000$, respectively. Each user is assigned a unique data index as their identity, occupying both a data block and a parity block, i.e., ${\bf v}_{j,\rm D, S} = ({\bf b}_{j}, {\bf v}_{j, \rm D, red})$, with a total length of $K + R = 3010$. Decoding at the receiving end is also conducted with $50$ iterations of the MSA.
  \item
  The PA-FFMA systems share the same structural parameters as the DF-FFMA systems, with the polarization-adjusted scaling factor set to $\mu_{\rm pas} = 300$. At the receiving end, the BMD algorithm is employed to decode the multiuser data sequences.
  \item
  For the slotted ALOHA systems, we assume that the number of time slots equals the total number of users, employing a repetition code for error control. After AUD processing, the slotted ALOHA system can be considered a type of TDMA system. Consequently, the blocklength of each time slot is $\frac{N}{J}$, with each bit repeated $\frac{N}{JK}$ times.
  The repetition code can also be viewed as a spreading technique, with its spreading factor (SF) equal to $\frac{N}{JK}$. Therefore, the signal processing gain of the repetition code is approximately $10 \log_{10}(\frac{N}{JK})$. 
  MAP detection is employed at the receiving end to recover the transmitted information sequences.
\end{itemize}

From Fig. \ref{f.TDMA}, we see that the BER of the system goes down as the number of users increases. For the SF-FFMA systems, the BERs of the $J = 100$ and $300$ cases are almost the same. When the BER is $P_b = 10^{-5}$, the gap between the BERs of $J=1$ and $300$ is only $1.5$ dB, verifying the efficiency of the proposed FFMA system.
Then, we investigate the BER performance between SF-FFMA and DF-FFMA systems.
Under the same simulation conditions, the DF-FFMA provides better BER than that of the SF-FFMA, since the default bits are $0$s which are available at the receiving end. 
However, with the increased number of users, the differences between the sparse-form and diagonal-form systems become smaller. 
For $J = 300$, both systems exhibit the same BER.
Additionally, it is found that the PA-FFMA system provides significantly better BER performance than both the SF-FFMA and DF-FFMA systems. This improvement is attributed to a polarization gain of approximately $10 \log_{10}(\mu_{\rm pas}) \approx 24.77$ dB.

Following, we compare our proposed FFMA systems with the slotted ALOHA system. 
When $J = 1$, slotted ALOHA with repetition coding outperforms both the SF-FFMA and DF-FFMA systems in terms of BER due to a coding gain of approximately $10 \log_{10}(600) \approx 27.78$ dB. 
However, the proposed PA-FFMA system still offers an additional coding gain of approximately $4$ dB over the slotted ALOHA system, as it combines both polarization gain from power and coding gain from the entire DoFs.
For a large number of users, i.e., $J \ge 100$, all three configurations of the proposed FFMA systems can offer much better BER performance than the slotted ALOHA system. Specifically, at $P_b = 10^{-5}$ and $J = 300$, the SF-FFMA (or DF-FFMA) systems provide a coding gain of approximately $6$ dB over the slotted ALOHA system, further validating the ability of the proposed FFMA systems to enhance BER performance in scenarios with massive user counts.

Table 1 compares the slotted ALOHA and FFMA-based RA systems. As the number of users $J$ increases, the total transmit power of the slotted ALOHA system decreases. Consequently, to achieve the same BER performance, the required SNRs for $J = 1$ and $J = 300$ range from $-15.3$ dB to $9.6$ dB. From Fig.~\ref{f.TDMA}, when $J = 300$, the DF-FFMA system achieves the same BER performance as the SF-FFMA system, but with only $\frac{3010}{6000}$ of the total power required by the SF-FFMA system. This indicates that the DF-FFMA system requires less power to achieve the same BER performance.

Next, we analyze the decoding algorithms across different FFMA systems. The SF-FFMA and DF-FFMA systems, which use the MSA decoding algorithm, exhibit significantly lower decoding complexity. In contrast, the PA-FFMA system, which uses the BMD algorithm, has decoding complexity that depends heavily on the number of iterations $L$. Although a larger number of iterations generally improves BER performance, it also increases the complexity considerably. Therefore, selecting appropriate parameters to balance error performance and complexity is crucial.

\begin{small}
\begin{table}[t]
\caption{Comparison of slotted ALOHA and FFMA sourced RA systems.}
\label{table_example}
\centering
\begin{tabular}{p{2.2cm}|p{2.2cm}|p{2.2cm}|p{2.2cm}|p{2.2cm}}
\toprule
    & \textbf{Slotted ALOHA} & \textbf{SF-FFMA} & \textbf{DF-FFMA} & \textbf{PA-FFMA} \\
\hline
 \textbf{Transmitter} 
    & Channel encoder $\Rightarrow$ CF-MUX 
    & FF-MUX $\Rightarrow$ channel encoder 
    & FF-MUX $\Rightarrow$ channel encoder 
    & FF-MUX $\Rightarrow$ channel encoder \\
\hline
   \textbf{Receiver} 
    & MAP 
    & MSA  
    & MSA  
    & BMD \\
\hline
    \textbf{Total power} 
    & $\frac{N}{J} \cdot P_{avg}$ 
    & $N \cdot P_{avg}$ 
    & $(K+R) \cdot P_{avg}$ 
    & $N \cdot P_{avg}$ \\
\bottomrule
\end{tabular}
\end{table}
\end{small}

\subsubsection{BER performance of low-rate channel codes based on PA-FFMA}

As mentioned previously, when $J = 1$, the proposed PA-FFMA system can be viewed as a type of LDPC code. In this case, due to the small number of information bits (e.g., $K = 1, 2, 3, 10, 100$) and the large block length (e.g., $N = 10000$), the PA-FFMA system with $J = 1$ essentially becomes a low-rate channel code (e.g., $R_c = 1/10000, 2/10000, 3/10000, 1/1000, 1/100$).
In this subsection, we investigate the BER performance of low-rate channel codes constructed using the PA-FFMA framework.

Since $ J = 1 $, the BER performance of the PA-FFMA system is primarily determined by the $(N, mK)$ channel codes $ {\mathcal C}_{gc} $. We consider two cases: in the first case, the channel code $ {\mathcal C}_{gc} $ is a binary $(6000, 3000)$ LDPC code $ {\mathcal C}_{gc, b1} $ with rate of $0.5$; in the second case, $ {\mathcal C}_{gc} $ is a binary $(10000, 8400)$ LDPC code $ {\mathcal C}_{gc, b2} $ with rate of $0.84$.
It is important to note that we only consider channel codes with rates greater than or equal to $0.5$. This is because the BMD algorithm heavily relies on the reliability of the information section. Therefore, we restrict our analysis to channel codes that satisfy $ R_c \ge 0.5 $.

When $J = 1$ and $K = 1, 2, 3, 10, 100$ bits/user, the BER performances of the PA-FFMA systems with the channel codes $\mathcal{C}_{gc, b1}$ (rate $0.5$) and $\mathcal{C}_{gc, b2}$ (rate $0.84$) are shown in Fig.~\ref{f.LDPC}.

As observed in Fig.~\ref{LDPC_sub1}, as the number of $K$ increases, the BER of the PA-FFMA system with $R_c = 0.5$ initially improves (i.e., from $K = 1$ to $K = 10$), then degrades (i.e., from $K = 10$ to $K = 100$). In the range of $K = 10$ to $K = 100$, a local extremum appears. However, the exact value of $K$ that reaches this extremum is unknown, as all results are based on computer simulations. Nevertheless, it is known that the typical traffic for a single user is around $K = 10$ to $K = 100$. Therefore, we conclude that $K = 10$ achieves better BER performance compared to $K = 100$.
Specifically, when $K = 10$, the BER of the proposed PA-FFMA system with $R_c = 0.5$ shows a gain of approximately $3.2$ dB compared to the uncoded case. In contrast, there remains a gap of about $4.4$ dB relative to the performance of the original LDPC code $\mathcal{C}_{gc, b1}$.

In contrast to the BER behavior of the PA-FFMA with $R_c = 0.5$, the BER of the PA-FFMA system with the channel code $\mathcal{C}_{gc, b2}$ (rate 0.84) improves monotonically with the increased number of $K$, i.e., from $K = 1$ to $K = 100$.
When $K = 100$, the BER of the proposed PA-FFMA system with $R_c = 0.84$ shows a gain of approximately $3.2$ dB compared to the uncoded case. In contrast, it still exhibits a gap of about $3.2$ dB relative to the performance of the original LDPC code $\mathcal{C}_{gc, b2}$.



\begin{figure}[t]
  \centering
  \includegraphics[width=0.5\textwidth]{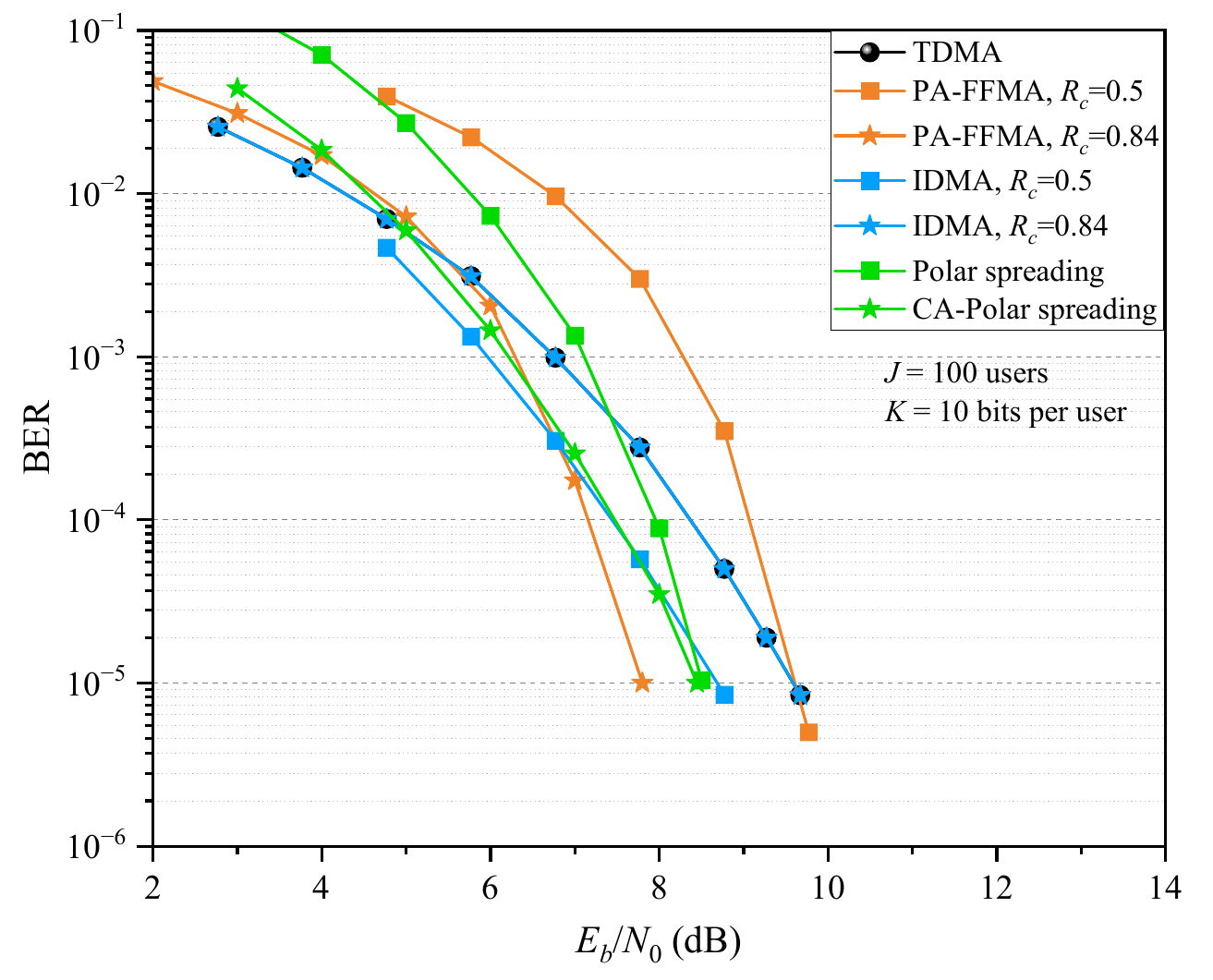}
  \caption{BER performances of various MA systems in a GMAC. During the simulations, we set $K =10$ bits/user and $J = 100$ for comparison.} 
  \label{f.IDMA}
  \vspace{-0.22in}
\end{figure}

\subsubsection{Comparisons among different MA systems}

Next, we examine the BER performance of PA-FFMA systems, which function as multiuser channel codes, and compare them with the IDMA \cite{IDMA} and polar spreading \cite{Polar_4} systems.

In Fig.~\ref{f.IDMA}, we compare our proposed PA-FFMA system with IDMA system, and polar spreading system \cite{Polar_4}.
We set the nubmer of bits is $K = 10$ and the number of users is $J = 100$. 
The PA-FFMA system employs two binary LDPC codes for error correction. One is a binary $(6000, 3000)$ LDPC code ${\mathcal C}_{gc, b1}$, and the other is a binary $(10000, 8400)$ LDPC code ${\mathcal C}_{gc, b2}$ with a rate of $0.84$.

For fair comparison with the $\mathcal{C}_{gc,b1}$ case, the IDMA benchmark system allocates $N = 6000$ resources. The IDMA transmitter chain comprises a rate $1/2$ convolutional encoder producing $20$-bit codewords, followed by a repetition code with spreading factor $300$, and user-specific interleaving. This configuration maintains the same $N = 6000$ frame length as PA-FFMA, with turbo iterative processing at the receiver. When comparing with the $\mathcal{C}_{gc,b2}$ case, the IDMA system adapts to $N = 10000$ resources using a rate $0.84$ convolutional encoder that outputs $12$-bit codewords and increases the spreading factor to $840$.
The higher-rate IDMA configuration ($R_c = 0.84$) demonstrates limited error correction capability due to the extremely short $12$-bit codeword length. Consequently, its bit error rate performance degenerates to that of uncoded BPSK transmission.

In the polar spreading system, we utilize two types of polar codes: a $(32,10)$ polar code and a CA-polar code with a CRC length of $2$. The spreading sequence is generated using a Gaussian random sequence, with a SF set to 312. As a result, the total frame length for both the polar and CA-polar spreading systems is $9984$, which is slightly shorter than that of the PA-FFMA and IDMA systems, both of which have a frame length of $10,000$.
The iterative decoding process of the polar spreading system combines successive interference cancellation (SIC), MMSE estimation, and single-user decoding for both the polar and CA-polar codes. Specifically, the successive cancellation list (SCL) decoding algorithm is employed, with the number of decoding paths set to 64. It is worth noting that, due to the source RA (Random Access), energy detection (ED) is not required during the decoding process.

From Fig.~\ref{f.IDMA}, it can be seen that the PA-FFMA system with $R_c = 0.84$ provides a coding gain of approximately $2.1$ dB compared to the IDMA system (or the uncoded case), and a $0.65$ dB coding gain compared to the CA-polar spreading system. This demonstrates that the proposed PA-FFMA system is capable of achieving a significant coding gain, even with a large number of users, such as $J = 100$. These results highlight the effectiveness of our proposed FFMA system. However, the PA-FFMA system with $R_c = 0.5$ provides worse BER performance, since the BMD algorithm heavily relies on the reliability of the information section. The result is the same as the aforementioned result.

Consequently, the differences in BER are primarily attributed to the lengths of the channel codewords. To maintain the same frame length for both the FFMA and IDMA systems (e.g., $N = 8400$), the parity blocklength in the IDMA system is $2$, as channel encoding must occur before the CF-MUX. In contrast, our proposed PA-FFMA system performs channel encoding after the FF-MUX, allowing it to utilize a parity blocklength of $1600$.
According to Shannon's theory, longer channel codes can provide better coding gains. Therefore, the BER performance of our PA-FFMA system is superior to that of the IDMA system.

\begin{figure}[t]
  \centering
  \includegraphics[width=0.5\textwidth]{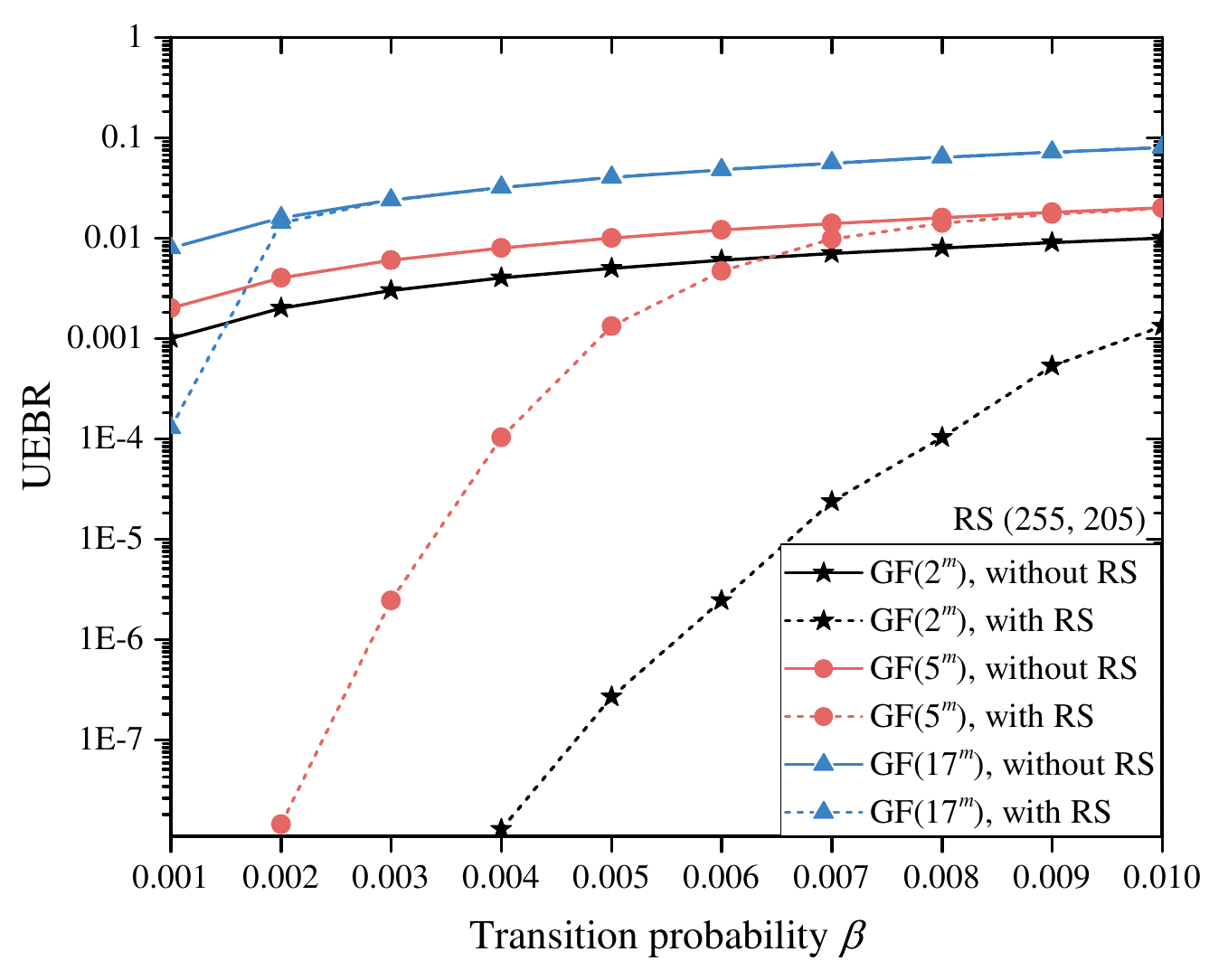}
  \caption{The UEBR performance of a network FFMA in a DSC, where the $(255, 205)$ RS code is used for error control. } 
  \label{f.network}
  \vspace{-0.22in}
\end{figure}

\subsection{Error Performance of Network FFMA Systems over a DSC}
Finally, we investigate the error performance of a network FFMA system over a DSC, utilizing the $(255, 205)$ Reed-Solomon (RS) code for error control with hard-decision decoding. The horizontal axis represents the transition probability $\beta$, while the vertical axis depicts the unresolved erasure bit rate (UEBR).

The UD-AIEP code is performed in three different fields GF($2^m$), GF($5^m$) and GF($17^m$) to support $m$, $2m$, and $4m$ users, respectively. 
The UEBR performances of the systems using $3$ different base prime fields GF($2$), GF($5$) and GF($17$) are shown in Fig. \ref{f.network}. From the figure, we see that the UEBR performances become poorer as the PF $p$ increases. This is due to the fact, for a given transition probability $\beta$, as the PF $p$ of the prime base field increases, the correct receiving probability, $1-(p-2)\beta$, becomes smaller. 
However, using a large PF $p$, more users can be simultaneously multiplexed, and the \textit{multiplex efficiency} is equal to $\log_2 (p-1)$.
This provides a tradeoff between the multiplexed number of users and the UEBR performance. 
Fig. \ref{f.network} shows that the RS code improves the UEBR performance significantly.

\section{Conclusion and Remarks}

In this paper, we proposed an FFMA technique to support massive users with short packet traffic, to solve the FBL of multiuser reliability transmission problem. To achieve this objection, each user is assigned to a unique EP over a prime field or its extension field. 
The Cartesian product of $J$ distinct EPs can form an EP code.
If the EP code has USPM structural property, we say it is an UD-EP code.
In this paper, we have constructed symbol-wise UD-EP codes which are based on prime fields and their extension fields, and their encoding and decoding in conjunction with a channel error-control code are used for error control.

Based on the orthogonal UD-EP code $\Psi_{\rm o, B}$ constructed over GF($2^m$), we proposed both sparse-form and diagonal-form FFMA systems. 
Additionally, through power allocation, we introduced a special configuration of the DF-FFMA system, referred to as PA-FFMA.
The proposed PA-FFMA can be viewed as a type of multiuser channel code, thereby obtaining both power and coding gains from the entire DoF. Hence, PA-FFMA can provide well-behaved BER than that of the current CFMA systems.
It is noted that, when $J = 1$, the proposed PA-FFMA is equivalent to a polarization ajusted LDPC code, which can support low-rate transmission.

In an FFMA system, an EP is considered a VRB for MA communications. The number of VRBs is determined by the extension field $m$ and the prime field $p$ of a given finite field GF($p^m$).
Furthermore, we proposed a network FFMA system over the extension field GF($p^m$) derived from a prime field GF($p$), designed for pure digital communication scenarios. In this setup, different VRBs are assigned to different users, allowing for user separation without ambiguity.
Simulation results demonstrated that the proposed FFMA system effectively supports a large number of users while maintaining favorable error performance in a GMAC. For network FFMA systems, increased user multiplexing can be achieved by utilizing a larger prime field $p$, with the multiplexing efficiency equal to $\log_2(p-1)$.

There are many unsolved works left. For example, developing FFMA systems for fading channel is desirable. It is also appealing to design FFMA based unsourced RA systems.



\vfill

\begin{thebibliography}{99}


\bibitem{FAdachi1}
F. Adachi, D. Garg, S. Takaoka, and K. Takeda, ``Broadband CDMA Techniques," \textit{IEEE Wireless Communications}, vol. 12, no. 2, pp. 8-18, April 2005.




\bibitem{YChen_2018}
Y. Chen et al., ``Toward the Standardization of Non-Orthogonal Multiple Access for Next Generation Wireless Networks," \textit{IEEE Communications Magazine}, vol. 56, no. 3, pp. 19-27, March 2018.


\bibitem{6G}
C. -X. Wang et al., ``On the Road to 6G: Visions, Requirements, Key Technologies, and Testbeds,'' \textit{IEEE Communications Surveys \& Tutorials}, vol. 25, no. 2, pp. 905-974, Secondquarter 2023.

\bibitem{6G_white}
``6G typical scenarios and key capabilities,'' IMT-2030 (6G) Promotion Group, White Paper, Jul. 2022. [Online]. 


\bibitem{UMA_2022}
Y. Li et al., ``Unsourced multiple access for 6G massive machine type communications,'' \textit{China Communications}, vol. 19, no. 3, pp. 70-87, March 2022.

\bibitem{UMA_PZ}
P. Fan et al., ``Random access for massive Internet of things: current status, challenges and opportunities,'' \textit{Journal on Communications}, vol. 42, no. 4, pp. 1-21, April 2021.






\bibitem{MIT_2017}
Y. Polyanskiy, ``A perspective on massive random-access,'' \textit{IEEE International Symposium on Information Theory-Proceedings, ISIT 2017}, 2017, pp. 2523–2527. 


\bibitem{SourcedRA_1}
M. Ke, Z. Gao, M. Zhou, D. Zheng, D. W. K. Ng and H. V. Poor, ``Next-Generation URLLC With Massive Devices: A Unified Semi-Blind Detection Framework for Sourced and Unsourced Random Access,'' \textit{IEEE Journal on Selected Areas in Communications}, vol. 41, no. 7, pp. 2223-2244, July 2023.


\bibitem{Yu_UDAS}
Qi-yue Yu, and Ke-xun Song, ``Uniquely Decodable Multi-Amplitude Sequence for Grant-Free Multiple-Access Adder Channels,'' \textit{IEEE Transactions on Wireless communications}, vol. 22, no. 12, pp. 8999-9012, Dec. 2023.



\bibitem{SourcedRA_2}
X. Shao, X. Chen, D. W. K. Ng, C. Zhong and Z. Zhang, ``Cooperative Activity Detection: Sourced and Unsourced Massive Random Access Paradigms,'' \textit{IEEE Transactions on Signal Processing}, vol. 68, pp. 6578-6593, 2020.


\bibitem{SourcedRA_3}
P. Agostini, Z. Utkovski, A. Decurninge, M. Guillaud and S. Stańczak, ``Constant Weight Codes With Gabor Dictionaries and Bayesian Decoding for Massive Random Access,'' \textit{IEEE Transactions on Wireless Communications}, vol. 22, no. 5, pp. 2917-2931, May 2023.


\bibitem{SourcedRA_4}
Z. Liang and J. Zheng, ``Mixed Massive Random Access,'' \textit{IEEE 20th International Conference on Communication Technology (ICCT)}, Nanning, China, 2020, pp. 311-316.


\bibitem{ALOHA_1}
F. Schoute, ``Dynamic Frame Length ALOHA,'' \textit{IEEE Transactions on Communications}, vol. 31, no. 4, pp. 565-568, April 1983.


\bibitem{Capacity_GMC_2017}
X. Chen, T.-Y. Chen and D. Guo, ``Capacity of Gaussian many-access channels,'' \textit{IEEE Trans. Information Theory}, vol. 63, no. 6, pp. 3516-3539, Jun. 2017.


\bibitem{Capacity_GMC_2020}
J. Gao, Y. Wu and W. Zhang, ``Energy-efficiency of Massive Random Access with Individual Codebook,'' \textit{IEEE Global Communications Conference, Globecom 2020}, Taipei, Taiwan, 2020, pp. 1-6.

\bibitem{Capacity_GMC_2021}
R. C. Yavas, V. Kostina and M. Effros, ``Gaussian Multiple and Random Access Channels: Finite-Blocklength Analysis,'' \textit{IEEE Trans. Information Theory}, vol. 67, no. 11, pp. 6983-7009, Nov. 2021.

\bibitem{Capacity_GMC_Yury}
S. S. Kowshik and Y. Polyanskiy, ``Fundamental Limits of Many-User MAC With Finite Payloads and Fading,'' \textit{IEEE Trans. Information Theory}, vol. 67, no. 9, pp. 5853-5884, Sept. 2021.



\bibitem{MIT_2017_2}
O. Ordentlich and Y. Polyanskiy, ``Low complexity schemes for the random access Gaussian channel,'' \textit{IEEE International Symposium on Information Theory-Proceedings, ISIT 2017}, 2017, pp. 2528–2532. 


\bibitem{BCH_MA}
I. Bar-David, E. Plotnik, et al., ``Forward Collision Resolution A Technique for Random Multiple-Access to the Adder Channel,'' \textit{IEEE Trans. Information Theory}, vol. 39, no. 5, 1993, pp. 1671–1675. 




\bibitem{CCS_1}
V. K. Amalladinne, G. S. Member, et al., ``A Coded Compressed Sensing Scheme for Unsourced Multiple Access,'' \textit{IEEE Trans. on Information Theory}, vol. 66, no. 10, 2020, pp. 6509–6533.

\bibitem{CCS_2}
R. Calderbank and A. Thompson, ``CHIRRUP: A practical algorithm for unsourced multiple access,'' \textit{Information and Inference}, vol. 9, no. 4, 2020. 


\bibitem{Polar_1}
J. Dai, K. Niu, et al., ``Polar-Coded Non-Orthogonal Multiple Access,'' \textit{IEEE Trans. Signal Processing}, vol. 66, no. 5, 2018, pp. 1374–1389. 

\bibitem{Polar_2}
E. Abbe and E. Telatar, ``Polar Codes for The $m$-user Multiple Access Channel,'' \textit{IEEE Trans.  Information Theory}, vol. 58, no. 8, 2012. 


\bibitem{Polar_3}
M. Zheng, Y. Wu, et al., ``Polar Coding and Sparse Spreading for Massive Unsourced Random Access,'' \textit{IEEE Vehicular Technology Conference, VTC 2020}, Nov. 2020, pp. 8–13.

\bibitem{Polar_4}
A. K. Pradhan, V. K. Amalladinne, K. R. Narayanan and J. -F. Chamberland, ``Polar Coding and Random Spreading for Unsourced Multiple Access," 2020 IEEE International Conference on Communications (ICC), Dublin, Ireland, 2020, pp. 1-6.

\bibitem{IDMA}
P. Li, L. Liu, K. Wu and W. K. Leung, ``Interleave division multiple-access,'' \textit{IEEE Transactions on Wireless Communications}, vol. 5, no. 4, pp. 938-947, April 2006.

\bibitem{IDMA_1}
A. K. Pradhan, V. K. Amalladinne, et al., ``Sparse IDMA: A Joint Graph-Based Coding Scheme for Unsourced Random Access,'' \textit{IEEE Trans. Communications}, vol. 70, no. 11, pp. 7124-7133, Nov. 2022.


\bibitem{SVC_1}
H. Ji, S. Park, and B. Shim, ``Sparse Vector Coding for Ultra Reliable and Low Latency Communications,'' \textit{IEEE Trans. Wireless Communications}, vol. 17, no. 10, pp. 6693-6706, Oct. 2018.


\bibitem{SVC_2}
L. Yang and P. Fan, ``Improved Sparse Vector Code Based on Optimized Spreading Matrix for Short-Packet in URLLC,'' \textit{IEEE Wireless Communications Letters}, vol. 12, no. 4, pp. 728-732, April 2023.












\bibitem{PAC_1}
E. Arıkan, ``On the Origin of Polar Coding,'' \textit{IEEE Journal on Selected Areas in Communications}, vol. 34, no. 2, pp. 209-223, Feb. 2016.

\bibitem{PAC_2}
E. Arıkan, ``From sequential decoding to channel polarization and back again,'' 
https://arxiv.org/abs/1908.09594.

\bibitem{PAC_3}
M. Rowshan, A. Burg and E. Viterbo, ``Polarization-Adjusted Convolutional (PAC) Codes: Sequential Decoding vs List Decoding,'' \textit{IEEE Transactions on Vehicular Technology}, vol. 70, no. 2, pp. 1434-1447, Feb. 2021.



\bibitem{Liao1972}
H. H. J. Liao, ``Multiple access channels," Ph.D dissertation, Dept. Electrical Engineering, Univ. Hawaii, Honolulu, HI, 1972.

\bibitem{Kasami1976}
T. Kasami, and Shu Lin, ``Coding for a Multiple-Access Channel," \textit{IEEE Trans. Information Theory}, vol. IT-22, no. 2, pp. 129-137, March 1976.



\bibitem{Kasami1978}
T. Kasami, and Shu Lin, ``Bounds on the Achievable Rates of Block Coding for a Memoryless Multiple-Access Channel," \textit{IEEE Trans. Information Theory}, vol. IT-24, no. 2, pp. 187-197, March 1978.






\bibitem{Kasami1983}
T. Kasami, Shu Lin, Victor K. Wei, and Saburo Yamamura, ``Graph Theoretic Approaches to the Code Construction for the Two-User Multiple-Access Binary Adder Channel," \textit{IEEE Trans. Information Theory}, vol. IT-29, no. 1, pp. 114-130, 1983.







\bibitem{Fan1995}
P. Fan, M. Darnell, and B. Honary, ``Superimposed codes for the multi-access binary adder channel'', \textit{IEEE Trans. Information Theory}, vol. 41, no. 4, pp. 1178-1182, 1995.















\bibitem{R3}
R. Ahlswede, Ning Cai, S. Li and R. W. Yeung, ``Network information flow,'' \textit{IEEE Trans. on Information Theory}, vol. 46, no. 4, pp. 1204-1216, July 2000.






\bibitem{Polar2009}
Erdal Arikan, ``Channel Polarization: A Method for Constructing Capacity-Achieving Codes for Symmetric Binary-Input Memoryless Channels,'' \textit{IEEE Trans. Information Theory}, vol. 55, no. 7, pp. 3051-3073, July 2009.

\bibitem{JDai2016}
J. Dai, K. Niu, Z. Si, and J. Lin, ``Polar coded non-orthogonal multiple access," \textit{IEEE International Symposium on Information Theory-Proceedings, ISIT 2016}, Barcelona, pp. 988-992, June 2016.











\bibitem{Shu2009}
William E. Ryan and Shu Lin, Channel Codes classical and Modern, Cambridge University Press, 2009.


\bibitem{LinBook3}
J. Li, S. Lin, K. Abdel-Ghaffar, W. E. Ryan, and D. J. Costello, ``LDPC Code Designs, Constructions, and Unification," Cambridge University Press, 2017.

\bibitem{Thomas}
Thomas M. Cover, and Joy A. Thomas, ``Elements of Information Theory,'' Tsinghua University Press, 2010.



\end{thebibliography}
\end{document}